\documentclass[twocolumn]{aastex631}

\usepackage{textcomp}
\usepackage{subfigure}
\usepackage{appendix}

\usepackage{comment}
\usepackage{rotating}
\usepackage{amsmath}
\usepackage{amssymb}
\usepackage{amsfonts}

\newcommand{\x}{\mbox{$\times$}} 
\newcommand{\Msun}{\mbox{$M_{\odot}$}}

\shortauthors{Morii et al.}

\begin{document}

\title{The ALMA Survey of 70 $\mu$m Dark High-mass Clumps in Early Stages (ASHES).\\ IX. Physical Properties and Spatial Distribution of Cores in IRDCs}

\author{Kaho Morii}
\affil{Department of Astronomy, Graduate School of Science, The University of Tokyo, 7-3-1 Hongo, Bunkyo-ku, Tokyo 113-0033, Japan email: kaho.morii@grad.nao.ac.jp}
\affil{National Astronomical Observatory of Japan, National Institutes of Natural Sciences, 2-21-1 Osawa, Mitaka, Tokyo 181-8588, Japan}

\author{Patricio Sanhueza}
\affil{National Astronomical Observatory of Japan, National Institutes of Natural Sciences, 2-21-1 Osawa, Mitaka, Tokyo 181-8588, Japan}
\affil{Department of Astronomical Science, SOKENDAI (The Graduate University for Advanced Studies), 2-21-1 Osawa, Mitaka, Tokyo 181-8588, Japan}

\author{Fumitaka Nakamura}
\affil{National Astronomical Observatory of Japan, National Institutes of Natural Sciences, 2-21-1 Osawa, Mitaka, Tokyo 181-8588, Japan}
\affil{Department of Astronomical Science, SOKENDAI (The Graduate University for Advanced Studies), 2-21-1 Osawa, Mitaka, Tokyo 181-8588, Japan}
\affil{Department of Astronomy, Graduate School of Science, The University of Tokyo, 7-3-1 Hongo, Bunkyo-ku, Tokyo 113-0033, Japan}

\author{Qizhou Zhang}
\affiliation{Center for Astrophysics $|$ Harvard \& Smithsonian, 60 Garden Street, Cambridge, MA 02138, USA}

\author{Giovanni Sabatini}
\affil{INAF - Istituto di Radioastronomia - Italian node of the ALMA Regional Centre (It-ARC), Via Gobetti 101, I-40129 Bologna, Italy}

\author{Henrik Beuther}
\affil{Max Planck Institute for Astronomy, Konigstuhl 17, 69117 Heidelberg, Germany}

\author{Xing Lu}
\affil{Shanghai Astronomical Observatory, Chinese Academy of Sciences, 80 Nandan Road, Shanghai 200030, People’s Republic of China}

\author{Shanghuo Li}
\affil{Max Planck Institute for Astronomy, K\"{o}nigstuhl 17, D-69117 Heidelberg, Germany}

\author{Guido Garay}
\affil{Departamento de Astronomía, Universidad de Chile, Las Condes, Santiago 7550000, Chile}

\author{James M. Jackson}
\affil{Green Bank Observatory, 155 Observatory Rd, Green Bank, WV 24944, USA}

\author{Fernando A. Olguin}
\affil{Institute of Astronomy and Department of physics, National Tsing Hua University, Hsinchu 30013, Taiwan}

\author{Daniel Tafoya}
\affil{Department of Space, Earth and Environment, Chalmers University of Technology, Onsala Space Observatory, 439~92 Onsala, Sweden}

\author{Ken'ichi Tatematsu}
\affil{National Astronomical Observatory of Japan, National Institutes of Natural Sciences, 2-21-1 Osawa, Mitaka, Tokyo 181-8588, Japan}
\affil{Department of Astronomical Science, SOKENDAI (The Graduate University for Advanced Studies), 2-21-1 Osawa, Mitaka, Tokyo 181-8588, Japan}

\author{Natsuko Izumi}
\affil{Academia Sinica Institute of Astronomy and Astrophysics, 11F of AS/NTU Astronomy-Mathematics Building, No.1, Section 4, Roosevelt Road, Taipei 10617, Taiwan}

\author{Takeshi Sakai}
\affil{Graduate School of Informatics and Engineering, The University of Electro-Communications, Chofu, Tokyo 182-8585, Japan}

\author{Andrea Silva}
\affil{National Astronomical Observatory of Japan, National Institutes of Natural Sciences, 2-21-1 Osawa, Mitaka, Tokyo 181-8588, Japan}

\begin{abstract}
The initial conditions found in infrared dark clouds (IRDCs) provide insights on how high-mass stars and stellar clusters form. 
We have conducted high-angular resolution and high-sensitivity observations toward thirty-nine massive IRDC clumps, which have been mosaicked using the 12m and 7m arrays from the Atacama Large Millimeter/submillimeter Array (ALMA).  
The targets are 70\,$\mu$m dark massive (220--4900\,$M_\odot$), dense ($>$10$^4$\,cm$^{-3}$), and cold ($\sim$10--20\,K) clumps located at distances between 2 and 6 kpc. 
We identify an unprecedented number of 839 cores, with masses between 0.05 and 81\,$M_\odot$ using 1.3 mm dust continuum emission. 
About 55\% of the cores are low-mass ($<$1\,$M_\odot$),  
whereas $\lesssim$1\% (7/839) are high-mass ($\gtrsim$27\,$M_\odot$). 
We detect no high-mass prestellar cores. 
The most massive cores (MMC) identified within individual clumps lack sufficient mass to form high-mass stars without additional mass feeding.  
We find that the mass of the MMCs is correlated with the clump surface density, implying denser clumps produce more massive cores and a larger number of cores.  
There is no significant mass segregation except for a few tentative detections. 
In contrast, most clumps show segregation once the clump density is considered instead of mass. 
Although the dust continuum emission resolves clumps in a network of filaments, some of which consist of hub-filament systems, the majority of the MMCs are not found in the hubs. 
Our analysis shows that high-mass cores and MMCs have no preferred location with respect to low-mass cores at the earliest stages of high-mass star formation. 
\end{abstract}


\section{Introduction} \label{sec:intro}

The study of cores embedded in massive prestellar clumps is expected to provide important information of the early phases of high-mass stars $>8\,M_\odot$ and cluster formation. For instance, 
the spatial distribution of cores, the fragmentation properties, and core masses would help to understand the very early phase of star formation from the clump to core scale. 
Do high-mass cores (e.g., $\gtrsim 30\,M_\odot$) form in the early phase or do only low-mass cores form at early times? Is there any primordial mass segregation? 
Answers to these questions are important to constrain high-mass star formation scenarios.

A long debate of core accretion \citep[e.g.,][]{McKeeTan02, McKeeTan03} versus competitive accretion \citep[e.g.,][]{Bonnell01, Bonnell04} has not been settled over twenty years \citep[e.g.,][]{Tan14}. 
Recently, more scenarios have been proposed, such as the global hierarchical collapse \citep[]{vazquez19} and the inertial inflow model \citep[]{Padoan20, Pelkonen21}.
Hub-filaments, sites where multiple filaments converge, have also attracted attention as the birthplace of high-mass stars \citep[]{Myers09, Peretto14, Kumar20}.  
The turbulent core accretion model predicts a massive core as an initial condition in which a high-mass star forms via monolithic collapse. 
Other models (clump-fed scenarios) predict continuous mass feeding toward initially low-mass cores that eventually accrete sufficient mass to form high-mass stars. 
The competitive accretion \citep[][]{Bonnell01} and the global hierarchical collapse \citep[]{vazquez19} models describe core growth in a cluster environment. 
A cloud fragments into cores with a mass near the thermal Jeans mass, forming an initial cluster of low-mass (proto-)stars. They grow in mass by additional mass feeding. Especially, protostars near the center of the cluster, or at the bottom of the gravitational potential, efficiently gain mass and have the largest probability of forming high-mass stars. Simulations made by \citet[][]{Wang10} suggest that not even the presence of magnetic fields can prevent mass accretion or core mass growth. The inertial-inflow model proposed by \citet{Padoan20} explains the process of acquiring additional mass from converging flows.  

To study the initial conditions of high-mass star formation, infrared dark clouds (IRDCs) are considered to be the best targets \citep[][]{Rathborne06, Chambers09, Sanhueza12, Sanhueza19}. 
They are identified by {\it Spitzer} and {\it Herschel} surveys such as the Galactic Legacy Infrared Mid-Plane Survey Extraordinaire \citep[GLIMPSE,][]{Benjamin03-GLIMPSE}, the survey of the inner Galactic plane using the Multiband Infrared Photometer for {\it Spitzer} \citep[MIPSGAL,][]{Carey09}, and the {\it Herschel} Infrared GALactic plane survey \citep[Hi-GAL,][]{Molinari10}, and studied in millimeter/submillimeter surveys such as the Atacama Pathfinder EXperiment \citep[APEX,][]{Gusten06} Telescope Large Area Survey of the Galaxy \citep[ATLASGAL,][]{Schuller09}, the Bolocam Galactic Plane Survey \citep[BGPS,][]{Aguirre11}, the millimeter Astronomy Legacy Team 90 GHz (MALT90) survey \citep[][]{Foster11, Foster13, Jackson13}, and the Radio Ammonia Mid-plane Survey \citep[RAMPS,][]{Hogge18}.  
Recent observations, using sub/millimeter arrays such as SMA (Submillimeter Array), CARMA (Combined Array for Research in Millimeter-wave Astronomy), NOEMA (NOrthern Extended Millimeter Array), and ALMA (Atacama Large Millimeter/submillimeter Array), have achieved high spatial resolution and high sensitivity, resolving dense cores embedded in IRDCs even though they are at far distances from the Sun (over a few kpc).  
Most case studies have investigated which physical processes affect the fragmentation \citep[e.g.,][]{Zhang09, Wang11, Zhang14, Ohashi16, Csengeri17, Li19a, Sanhueza19, Li21a, Liu21, Zhang21}, whether high-mass prestellar cores exist \citep[e.g.,][]{Zhang11, Lu15, Sanhueza17, Pillai19, Svoboda19, Louvet19, Barnes21}, if low-mass cores form before, coevally, or after the formation of high-mass cores \citep[e.g.,][]{zhang15, Pillai19, Svoboda19, Li21a}, if cores acquire additional mass from their surroundings \citep{Schneider10, Henshaw14, Contreras18, Olguin21, Redaelli22}, and their chemistry \citep{Sanhueza12,Sanhueza13,Sakai15,Sakai18, Sakai22,Liu20Chem,Feng20,Sabatini22}, for example. 
However, the small sample size and possible environmental effects make difficult to reach general conclusions. A systematic study of a statistically significant sample of IRDCs is desired to characterize the core properties, especially in the very early prestellar phase. 

We have conducted the ALMA Survey of 70 $\mu$m Dark High-mass Clumps in Early Stages (ASHES). 
The angular resolution of $\sim$1\farcs2 allowed us to resolve cores embedded in clumps. 
In a pilot survey, we mosaicked twelve candidate massive prestellar clumps with ALMA in dust continuum and molecular line emission at $\sim$224\,GHz \citep[][]{Sanhueza19}. 
We investigated fragmentation process \citep[][]{Sanhueza19}, outflows \citep[][]{Li20}, chemistry \citep[][]{Sabatini22, Li22-arxiv}, and dynamical properties (Li et al. 2022, submitted). 
We also conducted some case studies about a peculiar outflow \citep[][]{Tafoya21}, active star formation signatures \citep[][]{Morii21}, and deuterated chemistry \citep[][]{Sakai22}.

Here, we present the dust continuum emission of the complete sample toward thirty-nine targets, and study core physical properties such as core mass and spatial distribution, and, in particular, the properties of the most massive cores in each clump. 
The paper is organized as follows: Section~\ref{sec:sample} describes the sample selection. Section~\ref{sec:obs} summarizes the observation setups and data analysis. Results are presented in Section~\ref{sec:result}, and we discuss core mass, a correlation between the mass and the spatial distribution of the most massive cores in Section~\ref{sec:core_mass} and \ref{sec:mmc}, respectively. Finally, we discuss the high-mass star formation picture based on our finding in Section~\ref{sec:hmsf}, and Section~\ref{sec:conclusion} concludes the present study. 

\section{Sample Selection}
\label{sec:sample}
\subsection{Selection Strategy}

\begin{figure*}
    \centering
    \includegraphics[width=18.2cm]{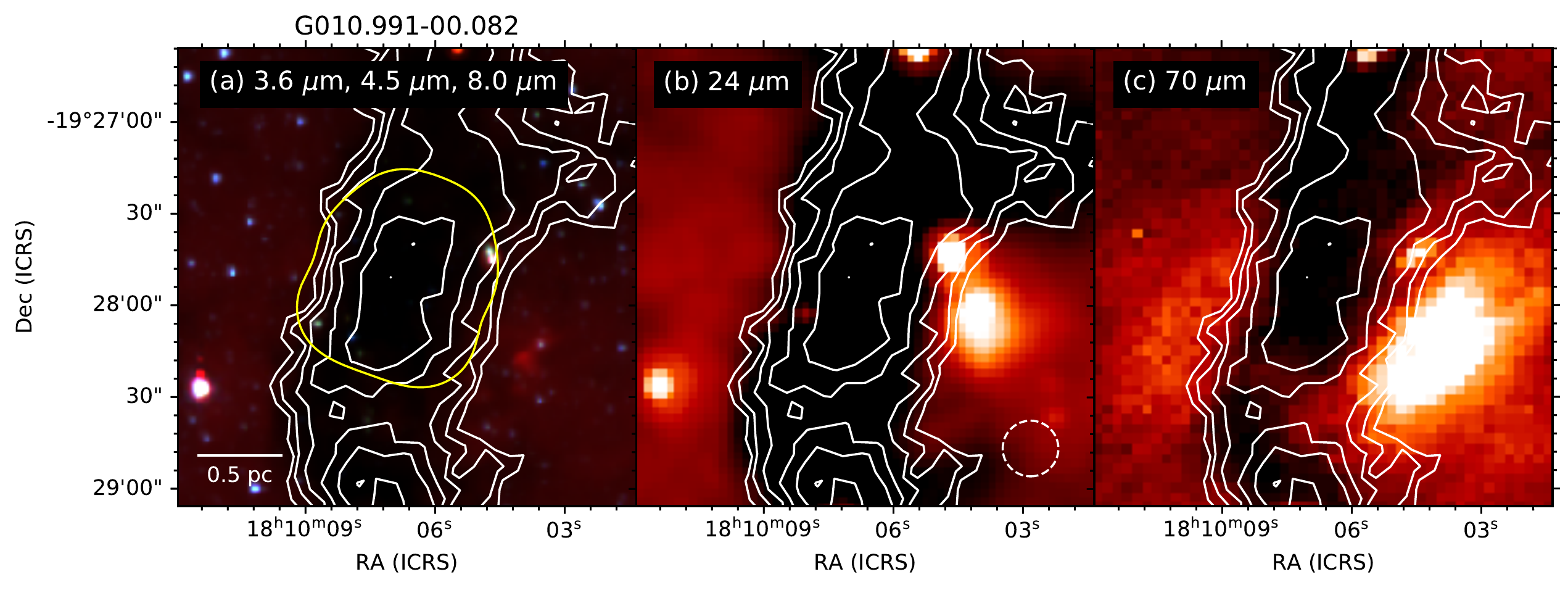}
    \caption{{\it Spitzer} and {\it Herschel} infrared images for G010.991--00.082. 
 (a) {\it Spitzer}/IRAC three-color (3.6 $\mu$m in blue, 4.5 $\mu$m in green, and 8.0 $\mu$m in red) image. The yellow contour represents the area mosaicked with ALMA in ASHES. 
 The white contours are 870 $\mu$m dust continuum emission from the ATLASGAL survey.
 Contour levels for the 870 $\mu$m dust continuum emission are (3, 4.2, 6, 8.5, 12, and 17)$\times \sigma$, where $\sigma=71$ mJy\,beam$^{-1}$ is the RMS noise level. 
 (b) {\it Spitzer}/MIPS 24 $\mu$m image and (c) {\it Herschel}/PACS 70 $\mu$m image.
  A white dashed circle on the bottom right in panel (b) shows the beam size ($\sim18\farcs2$) of the ATLASGAL survey. The complete figure set (39 images) is available after Appendix. }
    \label{fig:IR_G10}
\end{figure*}

We selected massive prestellar clump candidates without bright infrared sources to reveal the very early phase of high-mass star formation  from the 1st and 4th quadrants following \citet{Traficante15} and the MALT90 survey \citep{Foster11,Foster13,Jackson13}, respectively. 
In the MALT90 survey, \citet{Jackson13} used the ATLASGAL 870 $\mu$m survey to select a sample of 3246 high-mass clumps, almost all in the 4th quadrant, that were observed in molecular lines. 
Combining {\it Herschel} and ATLASGAL dust continuum emission, \citet{Guzman15} derived column density and dust temperature maps for the whole sample. \citet{Whitaker17} derived kinematic distances, and \citet{Contreras17} calculated the mass, density, and luminosities of the clumps. 
\citet{Guzman15} classified clumps that lack from 3.6 to 70\,$\mu$m ({\it Spitzer}/{\it Herschel}) compact emission as quiescent, which are the best prestellar clump candidates. 

The mean temperature of these clumps is $\sim$15 K, with a range from 9 to 23\,K, supporting the idea that these clumps host the early stages of star formation. 
We impose additional selection criteria for clump mass and density to ensure the selection of the best prestellar candidates with the potential to form high-mass stars. 
The clump mass, the mass surface density, and the volume density should be larger than 500\,$M_\odot$, $\sim$0.1 g cm$^{-2}$, and $5\times10^3$ cm$^{-3}$, respectively. 
In addition, targets are limited to within a distance of 6 kpc to ensure good spatial resolutions. At this distance, the angular resolution is comparable to the size of low-mass cores \citep[e.g., $\sim$7000 au]{Kirk06}. 
For some clumps that are not included in the analysis of \citet[][]{Contreras17} \footnote{ Clumps not included in \citet[][]{Contreras17} are G010.991--00.082, G014.492--00.139, G331.372--00.116, G333.481--00.224, G340.222--00.167, and G340.232--00.146.}, we estimate clump mass by using the column density and radius from \citet[][]{Guzman15} and the distance from  \citet[][]{Whitaker17}. 
Finally, we selected 18 clumps only in the 4th quadrant satisfying the conditions above, from which 11 were presented in the pilot survey \citep[][]{Sanhueza19}.  

We selected additional sources from the 1st quadrant, using the work by \citet{Traficante15}, who studied 3493 clumps using dust emission from {\it Herschel} and $^{13}$CO ($J=1-0$) emission from the Galactic Ring Survey \citep[GRS]{Jackson06}, and identified 667 starless clump candidates with no counterpart(s) at 70 $\mu$m,  without checking for {\it Spitzer} point sources.  We used the physical properties in \citet{Traficante15} for the sample selection. We followed the same procedure for clumps designated as starless, and selected 20 prestellar, high-mass clump candidates (one presented in the pilot survey) that we visually inspected to verify a lack of emission from compact {\it Spitzer} sources. We finally included one  70\,$\mu$m-dark prestellar clump candidate, G023.477+00.114, that has a potential to form high-mass stars satisfying all conditions above that had been previously studied by \citet[][]{Beuther13, Beuther15}. A detailed study of this source was presented using ASHES data in \citet[][]{Morii21}. 
In summary, the entire ASHES sample is composed of 39 IRDC clumps (see Table~\ref{tab:clump}). 

Figure~\ref{fig:IR_G10} shows the {\it Spitzer} and {\it Herschel} images
for G010.991--00.082, one of the targets. 
The left panel shows the three-color composite image (3.6\,$\mu$m in blue, 4.5\,$\mu$m in green, and 8\,$\mu$m in red) taken in the GLIMPSE survey \citep[][]{Benjamin03-GLIMPSE}. 
For comparison, the center and right panels display the 24 and 70\,$\mu$m emission from the MIPSGAL \citep[][]{Carey09} and Hi-GAL \citep[][]{Molinari10} surveys, respectively, with contours of 870\,$\mu$m continuum emission \citep[][]{Schuller09}. 

\subsection{Potential for High-mass Star Formation} 
\label{sec:capability}
Molecular cloud surveys suggest as empirical thresholds for high-mass star formation at a surface density $> 0.05$\,g\,cm$^{-2}$ \citep[e.g., ATLASGAL][]{urquhart14, He15}.  
Another threshold that has been proposed by \citet{KauffmannPillai10}, which after being scaled following \citet[][]{Sanhueza19}, is $M>580\,R^{1.33}$ ($M$ and $R$ are the mass and radius in units of $M_\odot$ and pc, respectively).   

The mass and radius of clumps depend on the definition used in each study. 
\citet{Traficante15} estimated the clump mass from the spectral energy distribution (SED) fitting of far-infrared wavelengths, while the radius of clumps is defined from the fitting of the 250 $\mu$m emission. 
In the MALT90 survey, \citet{Contreras17} used the deconvolved radius from the two-dimension Gaussian fitting as the physical radius, and they estimated the clump mass from the radius and the column density, which was computed from SED fitting. 
\citet{Urquhart18} estimated the clump mass using ATLASGAL 870 $\mu$m continuum emission.
Additionally, the temperature and distances adopted are also different among catalogs. 

The clump size is difficult to define because clumps are sometimes adjoining, and their borders are unclear. The definition and the measuring method are different from reference to reference. 
Typically, clump mass is estimated from the column density obtained from SED fitting or from the ATLASGAL continuum flux with temperature information. 
One of the caveats for using the column density from the {\it Herschel} SED fitting is the low resolution of 35$''$ at 500\,$\mu$m, which barely resolves clumps. 
Therefore, we decide to estimate clump mass and radius using the ATLASGAL continuum emission, which has an angular resolution of 18.2$''$. 
We measured flux density ($F_\nu$) and the radius of clumps ($R_\mathrm{cl}$) from a two-dimension Gaussian fitting of 870 $\mu$m ATLASGAL continuum images using CASA viewer \citep[]{McMullin07}. 
The radius corresponds to half of the full width at half maximum (FWHM) of the best-fit Gaussian deconvolved with the beam size, estimated by using the CASA task $\mathtt{imfit}$. 
Using the flux density of 870 $\mu$m continuum emission from the ATLASGAL survey, the clump mass can be estimated assuming optically thin conditions as: 
\begin{equation}
    M_\mathrm{cl} = \mathbb{R}\frac{d^2 F_\nu}{\kappa_\nu B_\nu (T_\mathrm{dust})},
\label{equ:Mass}
\end{equation}
where $\mathbb{R} =$ 100 is the gas-to-dust mass ratio, $\kappa_\mathrm{0.87\,mm} = $1.72\,cm$^2$\,g$^{-1}$ \citep[][]{Schuller09} is the dust absorption coefficient, $d$ is the distance to the source, and $B_\nu$ is the Planck function for a dust temperature $T_\mathrm{dust}$. 
The dust absorption coefficient ($\kappa_\mathrm{0.87\,mm}$) is calculated from $\kappa_\mathrm{1.3\,mm}$ assuming a dust emissivity spectral index ($\beta$) of 1.5, where $\kappa_\mathrm{1.3\,mm}= 0.9$\,cm$^2$\,g$^{-1}$ is from \citet[][]{Ossenkopf94} (see Section~\ref{sec:core-phy}). 
As for the dust temperature, we adopted the temperatures ($T_\mathrm{cl} \sim$10--20\,K) from the SED fitting between 160 $\mu$m and 870 $\mu$m from Hi-GAL and ATLASGAL survey at the continuum peak position \citep[][]{Guzman15}. 
We adopted the distances of clumps from \citet{Whitaker17}, ranging from 2.4--6\,kpc.  
The different assumption that can be made on $\mathbb{R}$ is that it varies depending on the galactocentric distance. Comparing the variation of $\mathbb{R}$ with its uncertainties and with what is found in \citet[][]{Sabatini22}, $\mathbb{R} = 100$ is a reasonable approximation. 
The uniformly re-calculated clumps properties, mass ($M_\mathrm{cl}$), radius ($R_\mathrm{cl}$), surface density ($\Sigma_\mathrm{cl} = M_\mathrm{cl} /(\pi R^2_\mathrm{cl})$), and volume density ($n(\mathrm{H_2})_\mathrm{cl} = M_\mathrm{cl} / \Bar{m}_\mathrm{H_2}(4\pi R^3_\mathrm{cl}/3)$), are summarized in Table~\ref{tab:clump}. Here, $\Bar{m}_\mathrm{H_2}$ is the mean molecular mass per hydrogen molecule, and we adopt $\Bar{m}_\mathrm{H_2}=2.8\,m_\mathrm{H}$ \citep[][]{Kauffmann08}. Throughout this work, these are the clump properties that will be used in the analysis. 

\begin{figure}
 \centering
 \includegraphics[width=9.6cm]{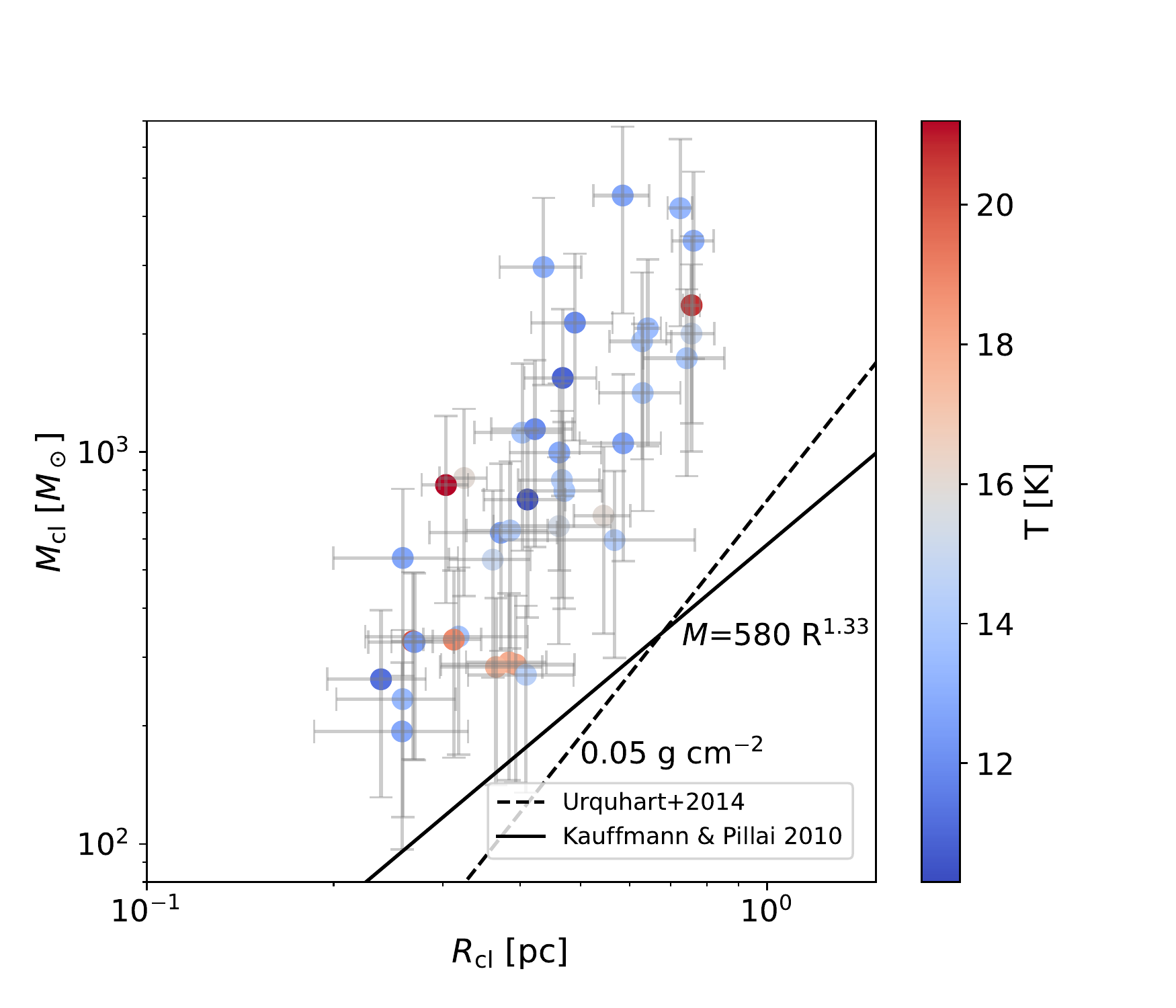}
 \caption{Mass-Radius relation colored by dust temperature. Physical properties were derived from 2-d Gaussian fitting
 to 870 $\mu$m continuum images. The black lines represent empirical thresholds for high-mass star formation ($M=580\,R^{1.33}$). Dashed line corresponds  to $\Sigma_\mathrm{cl} = 0.05$ g\,cm$^{-2}$ \citep[][]{urquhart14}. The relationship proposed by \citet{KauffmannPillai10} is shown as a solid black line.}
 \label{fig:MRcl}
\end{figure}

Figure~\ref{fig:MRcl} shows the mass as a function of clump radius colored by dust temperature. We overlaid the widely adopted thresholds for high-mass star formation as black lines. 
The solid one shows the high-mass star formation relation proposed by \citet{KauffmannPillai10}, as described above.
The dashed line represents $\Sigma_\mathrm{cl}>0.05$\,g\,cm$^{-2}$ suggested by \citet{urquhart14} and \citet{He15}. 
All targets from the ASHES sample satisfy these thresholds and are therefore thought to be capable of forming high-mass stars. 

Additionally, we can estimate a possible maximum stellar mass formed in a clump using the clump mass following \citet{Sanhueza17, Sanhueza19}. 
\citet{Larson03} obtained an empirical relation between the total stellar mass of a cluster ($M_\mathrm{cluster}$) and the maximum stellar mass in the cluster ($m^*_\mathrm{max}$) as
\begin{eqnarray}
    m^*_\mathrm{max} &=& 1.2\left( \frac{M_\mathrm{cluster}}{M_\odot}\right)^{0.45}\,M_\odot \\
    &=& 15.6\left( \frac{M_\mathrm{clump}}{10^3\,M_\odot} \frac{\varepsilon_\mathrm{SFE}}{0.3} \right)^{0.45}\,M_\odot,
    \label{equ:larson}
\end{eqnarray}
\\ \noindent
where the star formation efficiency, $\epsilon_\mathrm{SFE}$, is evaluated as $\varepsilon_\mathrm{SFE}=0.1-0.3$ for nearby embedded clusters \citep{Lada03}.
More recently, \citet{Sanhueza19} derive another relation for the maximum stellar mass that could be formed in a clump following the Kroupa's initial mass function (IMF; \citealt[][]{Kroupa01}) as
\begin{equation}
\label{equ:sanhueza}
    { m^*_\mathrm{max}=\left(\frac{0.3}{\varepsilon_\mathrm{SFE}}\frac{21.0}{M_\mathrm{clump}/M_\odot} + 1.5 \times 10^{-3}\right)^{-0.77}\,M_\odot. }
\end{equation}
Here we assumed that the relation of $M_\mathrm{cluster}= \varepsilon_\mathrm{SFE} M_\mathrm{clump}$, with $M_\mathrm{clump}=M_\mathrm{cl}$ from Table~\ref{tab:clump}. 
The calculated maximum stellar masses are in the range of $\sim$8--53\,$M_\odot$, with an assumption of $\varepsilon_\mathrm{SFE}=0.3$ for Equations~\ref{equ:larson} and \ref{equ:sanhueza}. 

Thus, the ASHES clumps fulfill known conditions for high-mass star formation and they are therefore considered high-mass prestellar clump candidates in this work.

\begin{deluxetable*}{lccccccccccc}
\label{tab:clump}
\tabletypesize{\footnotesize}
\tablecaption{Physical Properties of ASHES Clumps}
\tablewidth{0pt}
\tablehead{
\colhead{Clump Name}  & \colhead{R.A.} & \colhead{Decl.}  & \colhead{$\sigma$}  & \colhead{$d$} & \colhead{$T_\mathrm{cl}$}& \colhead{$M_\mathrm{ref}$} & \colhead{$R_\mathrm{cl}$} & \colhead{$M_\mathrm{cl}$} & \colhead{$\Sigma_\mathrm{cl}$} & \colhead{$n(\mathrm{H_2})_\mathrm{cl}$}\\ 
\colhead{} & \colhead{(ICRS)} & \colhead{(ICRS)} & \colhead{(km\,s$^{-1}$)} & \colhead{(kpc)} & \colhead{(K)} & \colhead{($M_\odot$)} & \colhead{(pc)} & \colhead{($M_\odot$)} & \colhead{(g\,cm$^{-2}$)} & \colhead{(10$^4$\,cm$^{-3}$)} \\
\colhead{(1)} & \colhead{(2)} & \colhead{(3)} & \colhead{(4)} & \colhead{(5)} & \colhead{(6)} & \colhead{(7)} & \colhead{(8)} & \colhead{(9)}  & \colhead{(10)} & \colhead{(11)}} 
    \startdata
    G010.991-00.082&18:10:06.65&-19:27:50.7&1.1&3.7&12.0&2230&0.49&2300&0.64&6.74\\
    G014.492-00.139&18:17:22.03&-16:25:01.9&1.8&3.9&13.0&5200&0.44&3200&1.11&13.30\\
    G015.203-00.441&18:19:52.56&-15:56:00.2&1.0&2.4&20.1&930&0.27&400&0.33&7.14\\
    G016.974-00.222&18:22:31.99&-14:16:02.3&1.2&3.6&12.8&1378&0.26&200&0.21&4.02\\
    G018.801-00.297&18:26:19.27&-12:41:17.3&1.3&4.7&13.3&2809&0.72&4500&0.57&4.08\\
    G018.931-00.029&18:25:35.70&-12:26:53.7&1.5&3.6&20.7&423&0.76&2500&0.30&2.00\\
    G022.253+00.032&18:31:39.60&-09:28:39.5&1.0&5.2&13.8&3010&0.32&400&0.24&4.29\\
    G022.692-00.452&18:34:13.78&-09:18:42.2&1.2&4.9&18.0&1426&0.39&300&0.13&1.70\\
    G023.477+00.114&18:33:39.53&-08:21:09.6&1.3&5.3&13.9&1000&0.38&1300&0.61&8.48\\
    G024.010+00.489&18:33:18.40&-07:42:28.1&1.0&5.7&12.7&3529&0.26&600&0.57&11.97\\
    G024.524-00.139&18:36:30.82&-07:32:26.6&1.5&5.5&13.1&1555&0.76&3700&0.43&2.89\\
    G025.163-00.304&18:38:17.20&-07:02:58.3&1.2&4.2&12.9&7245&0.46&1100&0.33&3.84\\
    G028.273-00.167&18:43:31.32&-04:13:19.5&1.6&5.0&10.9&1722&0.47&1700&0.50&5.71\\
    G028.541-00.237&18:44:15.80&-04:00:49.8&1.4&5.3&13.9&6028&0.63&2100&0.35&2.91\\
    G028.564-00.236&18:44:18.09&-03:59:33.5&1.9&5.3&12.7&15276&0.59&4900&0.94&8.43\\
    G028.927+00.394&18:42:43.18&-03:22:56.3&1.0&5.9&15.4&1616&0.46&700&0.22&2.45\\
    G030.704+00.104&18:47:00.12&-01:56:03.9&1.5&5.9&15.0&1749&0.76&2200&0.25&1.76\\
    G030.913+00.719&18:45:11.58&-01:28:08.2&0.9&3.5&12.4&1074&0.27&400&0.32&7.00\\
    G033.331-00.531&18:54:03.20&+00:06:53.4&2.0&6.1&14.3&1065&0.57&600&0.13&1.13\\
    G034.133+00.076&18:53:21.47&+01:06:12.1&1.2&3.8&16.0&579&0.55&700&0.17&1.49\\
    G034.169+00.089&18:53:22.57&+01:08:29.7&1.0&3.8&17.9&596&0.38&300&0.14&1.83\\
    G034.739-00.119&18:55:09.83&+01:33:14.5&1.2&5.3&12.7&2833&0.37&700&0.32&4.69\\
    G036.666-00.114&18:58:39.77&+03:16:16.5&0.9&3.6&13.4&881&0.26&300&0.25&6.00\\
    G305.794-00.096&13:16:33.40&-62:49:42.1&1.3&5.0&16.0&1560&0.32&900&0.58&9.06\\
    G327.116-00.294&15:50:57.18&-54:30:33.6&1.4&3.9&14.3&580&0.39&700&0.30&4.23\\
    G331.372-00.116&16:11:34.10&-51:35:00.1&1.8&5.4&14.0&1640&0.63&1500&0.25&2.07\\
    G332.969-00.029&16:18:31.61&-50:25:03.1&1.1&4.3&12.6&1170&0.59&1100&0.22&1.88\\
    G333.016-00.751&16:21:56.39&-50:53:45.2&2.1&3.7&17.6&690&0.37&300&0.15&2.12\\
    G333.481-00.224&16:21:39.97&-50:11:44.8&1.2&3.5&18.9&593&0.31&400&0.24&4.52\\
    G333.524-00.269&16:22:03.39&-50:11:47.2&1.5&3.5&21.2&2400&0.30&900&0.64&11.10\\
    G337.342-00.119&16:37:21.00&-47:19:25.3&2.8&4.7&14.5&460&0.41&300&0.12&1.52\\
    G337.541-00.082&16:37:58.48&-47:09:05.1&1.1&4.0&12.0&1180&0.42&1200&0.46&5.49\\
    G340.179-00.242&16:48:40.88&-45:16:01.1&1.9&4.1&14.0&1470&0.74&1900&0.23&1.60\\
    G340.222-00.167&16:48:30.83&-45:11:05.8&1.0&4.0&15.0&760&0.36&600&0.29&4.39\\
    G340.232-00.146&16:48:27.56&-45:09:51.9&2.1&3.9&14.0&710&0.47&900&0.26&2.97\\
    G340.398-00.396&16:50:08.85&-45:11:47.9&1.8&3.7&13.5&1690&0.64&2200&0.36&2.86\\
    G341.039-00.114&16:51:14.11&-44:31:27.2&1.1&3.6&14.3&1070&0.47&900&0.28&3.05\\
    G343.489-00.416&17:01:01.19&-42:48:11.0&1.0&2.9&10.3&810&0.41&800&0.32&3.98\\
    G345.114-00.199&17:05:26.26&-41:22:55.4&1.1&2.9&11.2&1350&0.24&300&0.33&7.63\\
    \enddata
    \tablenotetext{}{By replacing G for AGAL, the source name Column (1) matches with the ATLASGAL simple names \citep[][]{Schuller09, urquhart14, Urquhart18}. Column (4) present the observed velocity dispersion, which was obtained by the fitting of the line profile of C$^{18}$O ($J=2-1$) averaged within the clump with a 1-d Gaussian, which was observed by Total Power (TP). The distance from the Sun (5) is from \citet{Whitaker17}, and the dust temperature (6) is from \citet{Guzman15}.  Mass (7) is cited from \citet{Traficante15} and \citet{Contreras17} for clumps in the 1st and 4th quadrant, respectively. Properties in columns (8) (9) (10) and (11) are fitting results of 2-d Gaussian for 870 $\mu$m continuum images.} 
\end{deluxetable*}

\section{Observations and Data Reduction} \label{sec:obs}
Observations of the 39 ASHES clumps were carried out with ALMA in Band 6 ($\sim$224 GHz; $\sim$1.34 mm).
The data were acquired through three cycles: Cycle 3 (2015.1.01539.S, PI: P. Sanhueza), Cycle 5 (2017.1.00716.S, PI: P. Sanhueza), and Cycle 6 (2018.1.00192.S, PI: P. Sanhueza). 
The observations were taken with the main 12 m array and the Atacama Compact Array (ACA), including both the 7 m array and Total Power (TP).
The whole IRDC clumps were covered by Nyquist-sampled ten-pointing and three-pointing mosaics with the 12 m array and the 7 m array, respectively. 
A ten-pointing mosaic corresponds to 0.97 arcmin$^2$ within the 20\% power point, equivalent to the effective field-of-view (FOV) of $\sim$1$'$ per target. 
The mosaicked observations enable us to observe a large area of clumps as defined by single-dish continuum observations. The yellow contour in Figure~\ref{fig:IR_G10} represents the FOV of the combined data (i.e., 12 + 7 m).  
The total on-source time of the 12 m array observations per mosaic was $\sim$16 minutes, while sources observed in multiple executions have a total time of $\sim$25 minutes per mosaic. 
As for the 7 m array observations, the total on-source time was $\sim$90--100 minutes for the first thirteen sources except for G023.477+00.114 ($\sim$30 minutes) and $\sim$50--70 minutes for the remaining sources in Table~\ref{tab:obs-ashes}. 
Sources were observed in slightly different configurations through different ALMA Cycles, resulting in slightly different angular resolutions. 
The observations are sensitive to structures with an angular scale smaller than $\sim$11$''$ and $\sim$19$''$ for 12 m array and 7 m array, respectively.
The detailed observation setups and the synthesized beam sizes for all sources are summarized in Table~\ref{tab:obs-ashes}.

Data reduction was carried out using CASA software package versions 4.5.3, 4.6, 4.7, and 5.4.0 for calibration and 5.4.0 and 5.6.0 for imaging \citep[]{McMullin07}.
Continuum images were produced by averaging line-free channels. The effective bandwidth for continuum emission was $\sim$3.7 GHz. 
After subtracting continuum emission, we combined the 12 m array data with the 7 m array data using the CASA task $\mathtt{concat}$, and then the combined visibility data were Fourier transformed and cleaned together.  
In this work, we only used TP data of C$^{18}$O ($J=2-1$) line to estimate the velocity dispersion of the target clumps because TP antennas do not provide continuum emission. 

All images have 512 $\times$ 512 pixels with a pixel size of 0$\farcs$2. We used $\mathtt{TCLEAN}$ with Brigg's robust weighting of 0.5 to the visibilities and an imaging option of MULTISCALE with scales of 0, 5, 15, and 25 times the pixel size, considering the spatially extended nature of the emission in IRDCs. 
Average 1$\sigma$ RMS noise level is $\sim$0.094 mJy beam$^{-1}$ with a beam size of $\sim$1$\farcs$2. The RMS noise levels are also summarized in Table~\ref{tab:obs-ashes} for all targets.  
All images shown in this paper are the ALMA 12 and 7 m combined, before the primary beam correction, while all measured fluxes are derived from the combined data corrected for the primary beam attenuation. 

\begin{deluxetable*}{lccccccc}
\label{tab:obs-ashes}
\tabletypesize{\footnotesize}
\tablecaption{Observational Parameters and Information of Continuum Images}
\tablewidth{0pt}
\tablehead{
\colhead{Clump Name}  & \colhead{Baselines} & \colhead{ Configuration} & \colhead{Number of}& \colhead{Beam Size} & \colhead{$F_\mathrm{12m+7m}/F_\mathrm{12m}$} & \colhead{$F_\mathrm{recov}$} & \colhead{RMS Noise} \\ 
\colhead{} & \colhead{ (m) } & \colhead{} & \colhead{Antennas}& \colhead{($''\,\times\,''$)} & \colhead{} & \colhead{(\%)}&\colhead{(mJy beam$^{-1}$)} } 
    \startdata
    G010.991--00.082 & 15--330 & C36--1 & 41 (9--10)&1.29$\times$0.86 & 2.5 & 12&0.12\\ 
    G014.492--00.139 & 15--330 & C36--1 & 41 (9--10)&1.29$\times$0.85 &2.5 & 29 &0.17\\ 
    G015.203--00.441 & 15--314 & C43--2 & 42--45 (9--12)&1.48$\times$1.06 &1.5 & 27&0.11  \\
    G016.974--00.222 & 15--314 & C43--2 & 42--45 (9--12)&1.48$\times$1.07 & 1.4 & 15&0.08  \\
    G018.801--00.297 & 15--314 & C43--2 & 42--45 (9--12)&1.48$\times$1.07 &6.1 & 31&0.15  \\
    G018.931--00.029 & 15--314 & C43--2 & 42--45 (9--12)&1.48$\times$1.07 &1.5 & 12&0.12  \\
    G022.253+00.032 & 15--314 & C43--2 & 42--45 (9--12)&1.50$\times$1.09 &1.4 & 15&0.08  \\
    G022.692--00.452 &15--314 & C43--2 & 42--45 (9--12)&1.50$\times$1.08 & 2.3 & 23&0.09  \\
    G023.477+00.114 & 15--314 & C43--1 & 45 (10--11)&1.36$\times$1.08 &1.1 & 20 &0.09 \\
    G024.010+00.489 & 15--314 & C43--2 & 42--45 (9--12)&1.49$\times$1.08 &1.5 & 23 &0.12 \\
    G024.524--00.139 & 15--314 & C43--2 & 42--45 (9--12) &1.51$\times$1.08 &1.3 & 17&0.09  \\
    G025.163--00.304 & 15--314 & C43--2 & 42--45 (9--12)&1.50$\times$1.08 &1.7 & 15&0.10  \\
    G028.273--00.167 & 15--314 & C43--2 & 45--46 (10--12)&1.48$\times$1.07 &1.5 & 12&0.08  \\
    G028.541--00.237 & 15--314 & C43--2 & 45--46 (10--12)&1.50$\times$1.07 &1.2 & 7&0.08 \\
    G028.564--00.236 & 15--314 & C43--2 & 45--46 (10--12)&1.50$\times$1.06 & 1.4 & 15 &0.14 \\
    G028.927+00.394 & 15--314 & C43--2 & 45--46 (10--12)&1.52$\times$1.06 &1.1 & 14&0.08  \\
    G030.704+00.104 & 15--314 & C43--2 & 45--46 (10--12)&1.50$\times$1.07 &1.1 & 11&0.09  \\
    G030.913+00.719 & 15--314 & C43--2 & 45--46 (10--12)&1.49$\times$1.07 &1.2 & 10 &0.07 \\
    G033.331--00.531 & 15--314 & C43--2 & 45--46 (10--12)&1.49$\times$1.10 &1.2 & 14 &0.07 \\
    G034.133+00.076 & 15--314 & C43--2 & 45--46 (10--12)&1.48$\times$1.10 &1.2 & 19&0.07  \\
    G034.169+00.089 & 15--314 & C43--2 & 45--46 (10--12)&1.47$\times$1.09 &1.4 & 12 &0.07 \\
    G034.739--00.119 & 15--314 & C43--2 & 45--46 (10--12)&1.49$\times$1.09 &1.3 & 18 &0.08 \\
    G036.666--00.114 & 15--314 & C43--2 & 45--46 (10--12)&1.49$\times$1.10 &1.3 & 18 &0.07 \\
    G305.794--00.096 & 15--455 & C43--2 & 45 (10–-11)&1.24$\times$0.98 & 1.5 & 25 &0.10 \\
    G327.116--00.294 & 15--330 & C36--1 & 48 (8)&1.32$\times$1.11 &1.3 & 17&0.09  \\
    G331.372--00.116 & 15--330 & C36--1 & 48 (8)&1.34$\times$1.09 &1.8 & 21 &0.08 \\
    G332.969--00.029 & 15--330 & C36--1 & 48 (8)&1.35$\times$1.08 &1.7 & 11 &0.08 \\
    G333.016--00.751 & 15--314 & C43--2 & 46--48 (10--12)&1.49$\times$1.11 &1.6 & 11&0.08  \\
    G333.481--00.224 & 15--314 & C43--2 & 46--48 (10--12)&1.48$\times$1.11 &1.6 & 20&0.08  \\
    G333.524--00.269 & 15--314 & C43--2 & 46--48 (10--12)&1.49$\times$1.10 &1.5 & 17 &0.12 \\
    G337.342--00.119 & 15--314 & C43--2 & 46--48 (10--12)&1.43$\times$1.10 & 1.4 & 7 &0.07\\
    G337.541--00.082 & 15--639 & C36--2/3--C40-1 & 41--43 (8--9)&1.29$\times$1.18 &1.4 & 15 &0.07 \\
    G340.179--00.242 & 15--704 & C36--2/3--C40-4 & 36--41 (8--9)&1.41$\times$1.29 &2.7 & 9 &0.09 \\ 
    G340.222--00.167 & 15--704 & C36--2/3--C40-4 & 36--41 (8--9)&1.40$\times$1.28 &2.3 & 22 &0.11 \\
    G340.232--00.146 & 15--704 & C36--2/3--C40-4 & 36--41 (8--9)&1.39$\times$1.26 &3.9 & 45&0.14 \\
    G340.398--00.396 & 15--314 & C43--2 & 46--48 (10--12)&1.41$\times$1.09 &1.5 & 10 &0.08 \\
    G341.039--00.114 & 15--639 & C36--2/3--C40-1 & 41--43 (8--9)&1.30$\times$1.18 & 1.4 & 23&0.07  \\
    G343.489--00.416 & 15--639 & C36--2/3--C40-1 & 41--43 (8--9)&1.30$\times$1.18 &1.5 & 15&0.07  \\
    G345.114--00.199 & 15--314 & C43--2 & 46 (10--11)&1.25$\times$1.12 &2.0 & 12 &0.08 \enddata
    \tablenotetext{}{Baselines and configurations are observational setups for 12 m array. For 7 m array, the baselines range from 8 to 49 m. The number (or range) of antennas of 7 m are shown within parenthesis. When observations were carried out in the multiple execution blocks, and a different number of antennas were used, the variations were shown as ranges. The method to calculate the flux ratio and the recoverable fraction is described in Section~\ref{sec:cont-image}. $F_\mathrm{recov}$ is recovered flux by ALMA with respect to single dish observations. Continuum sensitivity and synthesized beam in the obtained 12 m + 7 m combined images.}
\end{deluxetable*}

\section{Results}
\label{sec:result}
\subsection{Dust Continuum Emission}
\label{sec:cont-image}

\begin{figure*}
\gridline{\fig{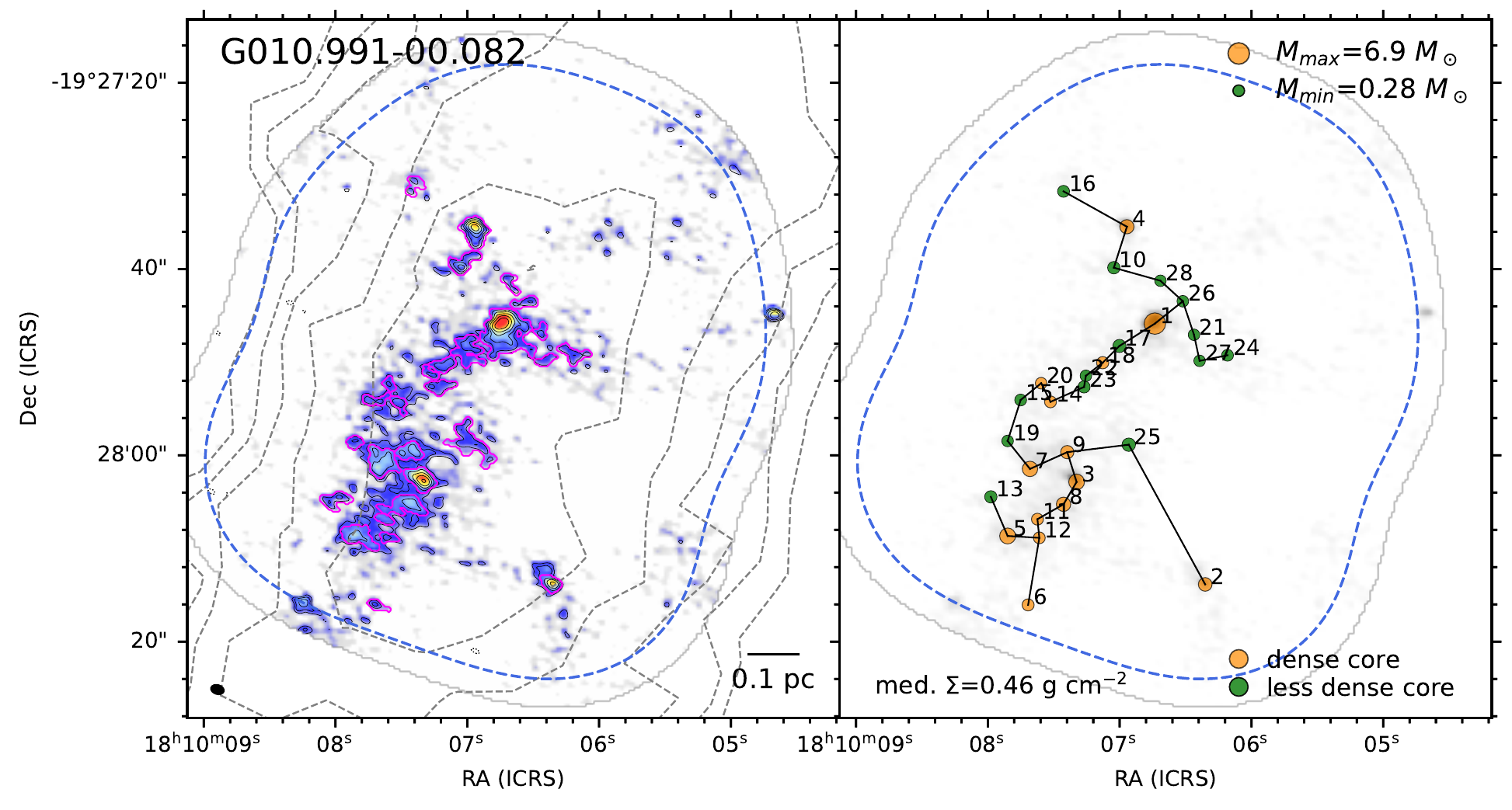}{0.9\textwidth}{}}
\vspace{-8mm}
\gridline{\fig{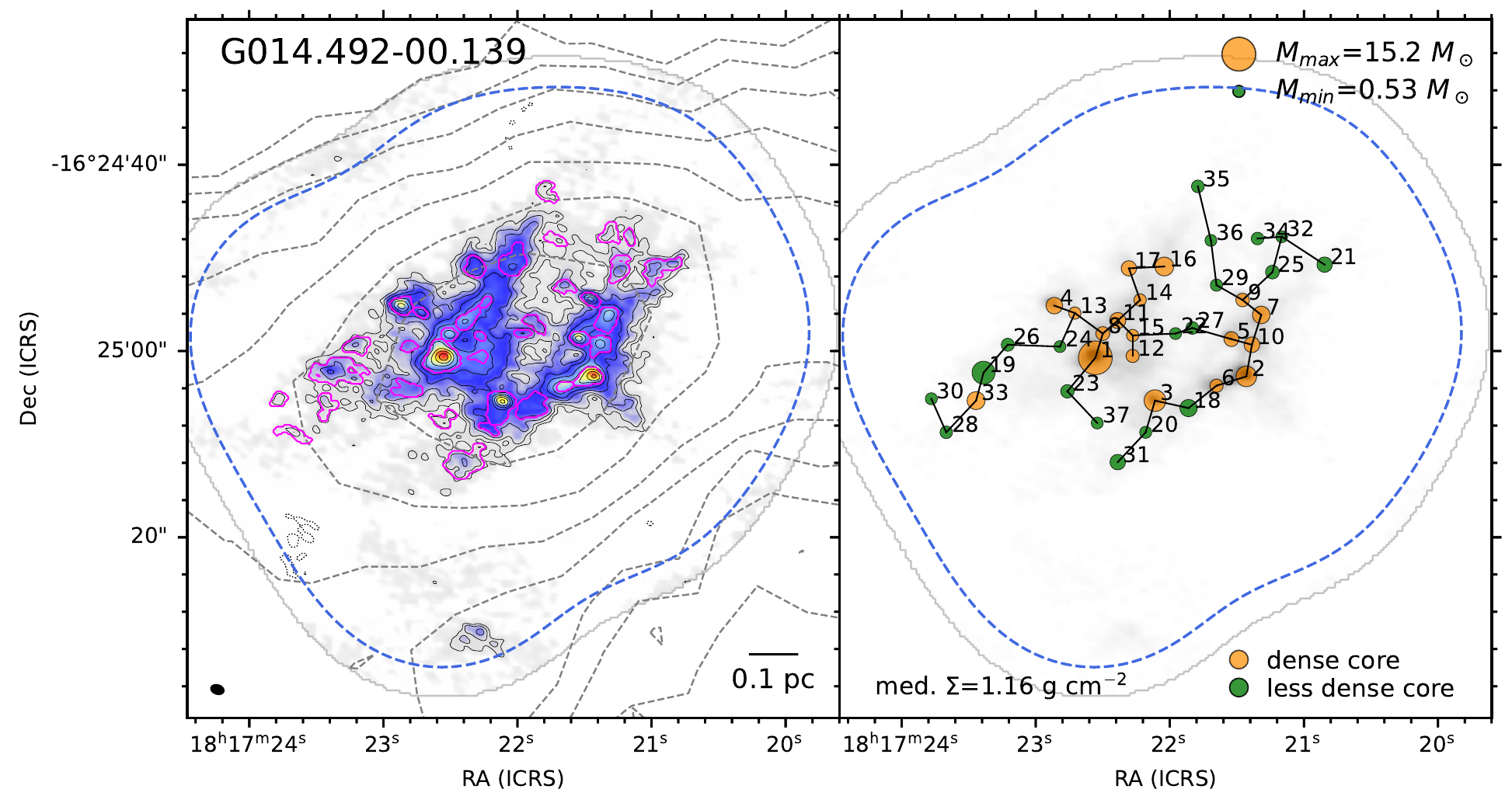}{0.9\textwidth}{}}
\vspace{-10mm}
\caption{Left: ALMA 1.3 mm continuum image for two IRDCs:  G010.991--00.82 and G014.492--00.139. Magenta thick contour represents leaf structures identified by the dendrogram algorithm (Section~\ref{sec:core-identify}). Black-solid contours represent $3 \times 2^\mathrm{n}\sigma$ (n=1, 2, 3, ...), where $\sigma$ is the continuum image RMS noise levels summarized in Table~\ref{tab:obs-ashes}. The dotted contours show the negative components (-4$\sigma$). The black-dashed contours show the 870\,$\mu$m continuum emission from the ATLASGAL survey. Contour levels are same with solid contours but with $\sigma =71$ mJy\,beam$^{-1}$ for G010.991--00.082, and $\sigma =83$ mJy\,beam$^{-1}$ for G014.492--00.139. The black ellipse in the bottom left corner represents the synthesized beam size. The black line indicates the spatial scale in the bottom right corner. The blue dashed contours correspond to a primary beam response of 30\%. 
Right: Circles represent core properties. The size corresponds to the estimated mass, and the position is centered at the continuum peak of each core. 
Orange circles represent cores with a surface density larger than the median of the cores' surface density of the corresponding clump, which are denoted on the bottom left (0.46 g\,cm$^{-2}$ and 1.16 g\,cm$^{-2}$ for G010.99 and G014.49, respectively). The remaining less dense cores are highlighted as green circles. 
Black segments show the outcome from the minimum spanning tree, which corresponds to the set of straight lines that connect cores in a way that minimizes the sum of the lengths. The digits represent the core-id named in order of peak intensity. The complete figure set (39 images) is shown after Appendix. }
\label{fig:ashes_cont_1}
\end{figure*}

The left panels of Figure~\ref{fig:ashes_cont_1} present the 1.3 mm continuum images for G010.991--00.082 and G014.492--00.139. We overlaid the single-dish 870 $\mu$m continuum emission obtained by ATLASGAL observations as black-dashed contours. Compared with the ATLASGAL observations, ALMA observations succeeded in resolving the area around the emission peak of massive clumps. 
As reported by the pilot survey \citep[]{Sanhueza19}, we confirmed that the internal structures of these clumps vary from region to region: some are filamentary (e.g., G023.47 and G025.16), and some are clumpy (e.g., G014.49 and G333.52). We also resolved some clumps (e.g., G024.01 and G024.52) in networks of hub filaments (see Section~\ref{sec:hub}). 

We combined the 12 m array and the 7 m array data to mitigate the missing flux. Indeed, the combined data have $\sim$1.7 times more flux than the 12 m-only data for the entire FOV.  To estimate how much flux is recovered in the combined images, we have scaled the 870 $\mu$m emission assuming a dust emissivity spectral index ($\beta$) of 1.5 as $F_\mathrm{recov} = F_\mathrm{1.3\,mm, ALMA}$/$F_\mathrm{1.3\,mm, exp}$, where $F_\mathrm{1.3\,mm, ALMA}$ is the 1.3 mm flux density obtained by ALMA in the FOV 
and $F_\mathrm{1.3\,mm, exp}$ is estimated as $F_\mathrm{1.3\,mm, exp}=F_\mathrm{0.87\,mm}(1.3/0.87)^{-(1.5+2)}$.  
$F_\mathrm{0.87\,mm}$ was measured within the area corresponding to the ALMA FOV (blue contours in  Figures \ref{fig:ashes_cont_1}).  
We integrated pixels where the intensity is more than twice the RMS noise level to calculate $F_\mathrm{1.3\,mm, ALMA}$. 
The estimated recoverable flux was between 7 and 45\%. The estimated values ($F_\mathrm{12m+7m}/F_\mathrm{12m}$ and $F_\mathrm{recov}$) are summarized in Table~\ref{tab:obs-ashes}.
These are consistent with SMA and ALMA observations in other IRDC studies \citep[e.g.,][]{Sanhueza17, Liu18}.

\subsection{Core Identification}
\label{sec:core-identify}
\begin{deluxetable*}{lcccccc}
\label{tab:dendro_G10}
\tabletypesize{\footnotesize}
\tablecaption{Core Properties Obtained from Dendrogram}
\tablewidth{0pt}
\tablehead{
\colhead{Clump Name}  & \colhead{Core Name}  & \colhead{R.A.} & \colhead{Decl.} & \colhead{FWHM$_\mathrm{maj} \times$ FWHM$_\mathrm{min}$}  &\colhead{Peak intensity} & \colhead{Flux Density} \\ 
\colhead{} & \colhead{} & \colhead{(ICRS)} & \colhead{(ICRS)} & \colhead{($''$\,$\times$\,$''$)} &  \colhead{(mJy beam$^{-1}$)} & \colhead{(mJy)}} 
    \startdata
    G010.991--00.082&ALMA1&18:10:06.73&--19.27.45.85&2.81$\times$2.59&2.70&12.63\\
    G010.991--00.082&ALMA2&18:10:06.35&--19.28.13.85&1.22$\times$0.80&2.32&2.80\\
    G010.991--00.082&ALMA3&18:10:07.32&--19.28.02.85&1.78$\times$1.12&2.27&4.91\\
    G010.991--00.082&ALMA4&18:10:06.94&--19.27.35.45&1.78$\times$1.34&1.91&4.04\\
    G010.991--00.082&ALMA5&18:10:07.85&--19.28.08.65&2.41$\times$1.15&1.41&4.33\\
    G010.991--00.082&ALMA6&18:10:07.69&--19.28.16.05&1.29$\times$0.49&1.34&1.08\\
    G010.991--00.082&ALMA7&18:10:07.68&--19.28.01.45&1.95$\times$1.69&1.09&4.60\\
    G010.991--00.082&ALMA8&18:10:07.42&--19.28.05.25&2.87$\times$1.19&1.01&3.53\\
    G010.991--00.082&ALMA9&18:10:07.39&--19.27.59.65&1.67$\times$1.00&0.88&2.12\\
    G010.991--00.082&ALMA10&18:10:07.04&--19.27.39.85&2.12$\times$1.01&0.86&1.60\enddata
\tablecomments{Table~\ref{tab:dendro_G10} is published in its entirety in machine-readable format. A portion is shown here for guidance regarding its form and content.}
\end{deluxetable*} 
We identified cores using the obtained 1.3 mm continuum images. 
In this paper, following the definitions of \citet[][]{Sanhueza19}, we use the term “core” to describe a compact, dense object within a clump with a size of $\sim$0.01--0.1 pc, a mass of $\sim$10$^{-1}-10^2$ $M_\odot$, and a volume density of $\sim$10$^5$ cm$^{-3}$ that will likely form a single star or a small multiple system. 
We adopt the $\mathtt{dendrogram}$ technique \citep{Rosolowsky08} to extract cores, which is implemented in the astrodendro python package.
The $\mathtt{dendrogram}$ technique classifies the hierarchical structure. 
The principal parameters are $F_\mathrm{min}$, $\delta$, and $S_\mathrm{min}$. 
$F_\mathrm{min}$ sets the minimum value above which we define structures, and $\delta$ sets a minimum significance to separate them.  
$S_\mathrm{min}$ is the minimum number of pixels to be contained in the smallest individual structure (defined as a $\mathtt{leaf}$ in the $\mathtt{dendrogram}$ technique). 
The $\mathtt{dendrogram}$ algorithm classifies the hierarchical structures as $\mathtt{leaf}$, $\mathtt{branch}$, and $\mathtt{trunk}$.
A $\mathtt{leaf}$ is a structure that has no substructure, and a $\mathtt{trunk}$ is the largest structure. 
The remaining structure is $\mathtt{branch}$, which has leaves as internal structures.

We set $F_\mathrm{min}$, $\delta$, and $S_\mathrm{min}$ as 2.5$\sigma$, 1.0$\sigma$, and the half-pixel numbers of the beam area following the pilot survey \citep{Sanhueza19}. 
Since these parameters are optimistic values, we applied the additional constraint to the flux density to exclude  suspicious structures. We have excluded cores with a flux density smaller than 3.5$\sigma$. 
Additionally, we eliminated cores at the edge of FOV. 
As a result of the $\mathtt{dendrogram}$ technique, 839 $\mathtt{leaf}$ structures were identified in total. The identified $\mathtt{leaf}$ structures are indicated as magenta contours in the left panels of Figure~\ref{fig:ashes_cont_1}. 
We define a $\mathtt{leaf}$ as a core.  
The number of cores in each region ranges from 8 to 39 (median of 20). 
We named cores ALMA1, ALMA2, ... in order of the peak intensity. 
Table~\ref{tab:dendro_G10} gives peak position, size (major and minor FWHM), peak intensity, and flux density of cores identified by the dendrogram algorithm (the properties for all cores are summarized in a machine-readable table). 
As an example, the right panels in Figure~\ref{fig:ashes_cont_1} show the identified cores in G010.991--00.082 and G014.492--00.139 as circles. Each circle position corresponds to each core's continuum peak position, and the digits on the right panel show the core-id, indicating the order of the peak intensity in each region. 

We confirmed only a weak correlation between the number of detected cores and the sensitivity, and between the number of detected cores and the clump distance with Spearman's rank coefficient of $r_\mathrm{s}$ =-0.15 and -0.22, and $p$--value is 0.36 and 0.17, respectively. 
Although the spatial resolution differs among the different clumps in the sample, this effect seems weak for the core identification. 

\subsection{Core Physical Parameters}
\label{sec:core-phy}
\begin{figure*}
    \centering
    \includegraphics[width=16cm]{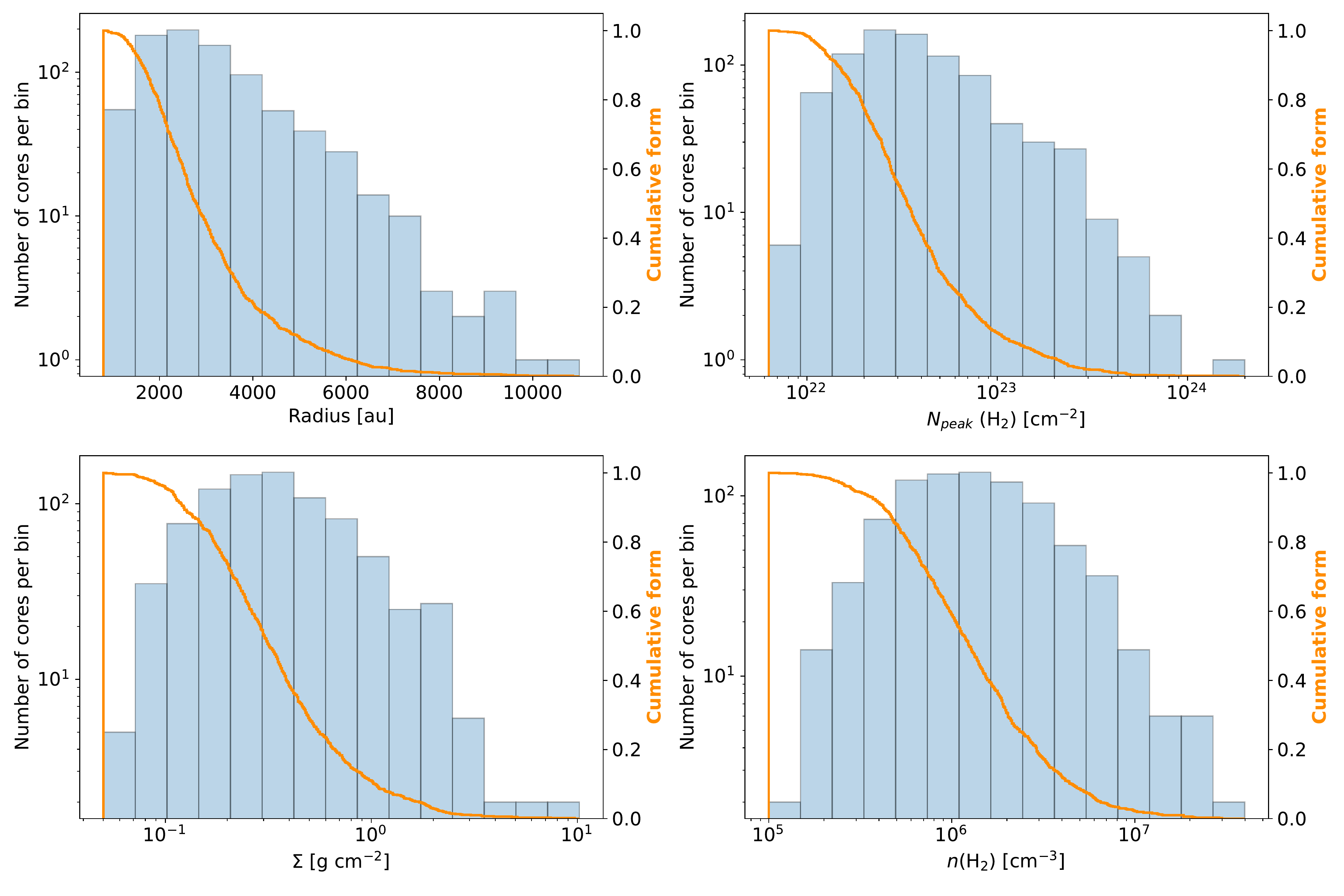}
    \caption{Core density properties such as core radius (au), peak column density ($N_\mathrm{peak}(\mathrm{H_2})$), surface density ($\Sigma$), and volume density ($n(\mathrm{H_2})$). Orange line in each panel represents the inverse-cumulative density distribution.}
    \label{fig:phy_para}
\end{figure*}

\begin{deluxetable*}{lccccccc}
\label{tab:core_phy}
\tabletypesize{\footnotesize}
\tablecaption{Core physical parameters}
\tablewidth{0pt}
\tablehead{
\colhead{Clump Name} & \colhead{Core Name} & \colhead{$T$} & \colhead{$M_\mathrm{core}$} & \colhead{$N_\mathrm{peak}(\mathrm{H_2})$} & \colhead{Radius} & \colhead{$\Sigma$} & \colhead{$n(\mathrm{H_2})$} \\ 
\colhead{} & \colhead{} & \colhead{(K)} & \colhead{($M_\odot$)} & \colhead{($\times10^{23}$\,cm$^{-2}$)} & \colhead{(au)} & \colhead{(g\,cm$^{-2}$)} & \colhead{($\times10^6$\,cm$^{-3}$)}} 
    \startdata
    G010.991--00.082&ALMA1&13.4&6.90&1.62&4950&0.78&1.72\\
    G010.991--00.082&ALMA2&12.5&1.68&1.53&1860&1.41&7.95\\
    G010.991--00.082&ALMA3&12.1&3.10&1.57&2680&1.27&4.87\\
    G010.991--00.082&ALMA4&14.1&2.05&1.06&2890&0.70&2.58\\
    G010.991--00.082&ALMA5&10.8&3.24&1.16&3090&0.96&3.31\\
    G010.991--00.082&ALMA6&13.2&0.60&0.82&1440&0.78&6.03\\
    G010.991--00.082&ALMA7&12.2&2.87&0.75&3300&0.72&2.42\\
    G010.991--00.082&ALMA8&12.3&2.17&0.68&3510&0.52&1.52\\
    G010.991--00.082&ALMA9&11.5&1.44&0.66&2480&0.71&2.87\\
    G010.991--00.082&ALMA10&12.8&0.92&0.55&2680&0.35&1.44\enddata
\tablecomments{ $T$ is the temperature used for mass calculation. The radius is calculated from the geometric mean of the FWHM divided by 2. The full table is published in machine-readable format. A portion is shown here for guidance regarding its form and content.}
\end{deluxetable*} 
Core masses can be estimated from the dust continuum emission, assuming the dust emission is optically thin, using Equation~\ref{equ:Mass} following previous ASHES papers \citep[][]{Sanhueza19, Morii21}. 
We adopt a gas-to-dust mass ratio of 100, and a dust opacity, $\kappa_\mathrm{1.3\,mm} =0.9$\,cm$^2$\,g$^{-1}$ computed by \citet{Ossenkopf94} for the dust grains with the ice mantles at a volume density of $10^6$ cm$^{-3}$. For the mass estimation, the fluxes obtained from dendrograms were used, without removing any additional background emission other than the inherently removed by the interferometer.  
In the case the background emission is important, the core masses could be considered to be upper limits. 
For cores in eleven ASHES clumps, we used the rotational temperatures derived from NH$_3$ observations at $\sim$5\arcsec \citep[][Allingham et al., 2022 in prep.]{Li22-arxiv}. For the remaining twenty-eight clumps, we adopted the clump temperature for the core mass calculation. Column 3 in Table~\ref{tab:core_phy} shows the temperature used for each core in the mass estimation. We confirmed that the dust emission is optically thin  by using Equation (B.2) from \cite{pouteau22}. We find that 99\% of cores have an optical depth smaller than 0.04 and the maximum value is $\sim$0.1.

The core mass ($M_\mathrm{core}$) given in column 4 of Table~\ref{tab:core_phy} ranges from 0.05 to 81 $M_\odot$. In the right panels in Figure~\ref{fig:ashes_cont_1}, the size of circles indicates the core mass.  
More than half ($\sim$55\%) of cores have a mass smaller than 1 $M_\odot$ and only 3.5\% (29/839) of cores have $M_\mathrm{core}>10\,M_\odot$. 
Assuming a core-to-star formation efficiency of 30\%, 27\,$M_\odot$ of core masses are necessary to form a high-mass star. In total, seven cores satisfy this mass threshold. 
Compared with the pilot survey, we revealed six additional high-mass cores thanks to the larger number of clumps. 
More discussion about core masses will be presented in  Section~\ref{sec:discussion}. 

By summing over the mass of all cores in each clump, we can estimate the core formation efficiency (CFE). We define the CFE as the ratio of the sum of core masses to the clump mass, $M_\mathrm{cl}$ \citep[e.g.,][]{Louvet14}.  
It ranges from 0.6\% to 16\%, indicating that most of the clump mass has not been assembled into cores at the early phase traced in these 70 $\mu$m dark IRDCs.  
Table~\ref{tab:core_summary} summarize the core properties in each clump such as  the total number of cores identified in each clump ($N$(core)), the maximum core mass ($M_\mathrm{max}$), the median of core mass ($M_\mathrm{med}$), and the minimum core mass ($M_\mathrm{min}$) as well as the CFE. 
Table~\ref{tab:core_summary} also shows $M_{3.5\sigma}$, the mass sensitivity in each clump, which corresponds to the mass when $F_\mathrm{1.3\,mm}$ is 3.5$\sigma$.

\begin{deluxetable*}{lcccccc}
\label{tab:core_summary}
\tabletypesize{\footnotesize}
\tablecaption{Properties of Cores in Each Clump}
\tablewidth{0pt}
\tablehead{
\colhead{Clump Name}  & \colhead{$N$(core)\tablenotemark{a}} & \colhead{$M_\mathrm{max}$} & \colhead{$M_\mathrm{med}$} & \colhead{$M_\mathrm{min}$} & \colhead{$M_{3.5\sigma}$} & \colhead{CFE\tablenotemark{b}} \\ 
\colhead{} & \colhead{}  & \colhead{($M_\odot$)}  & \colhead{($M_\odot$)}& \colhead{($M_\odot$)} & \colhead{($M_\odot$)} & \colhead{(\%)}} 
    \startdata
    G010.991--00.082&28&6.9&0.67&0.28&0.26&1.5\\
    G014.492--00.139&37&15.4&1.64&0.54&0.37&3.2\\
    G015.203--00.441&26&5.2&0.30&0.06&0.05&5.3\\
    G016.974--00.222&13&5.2&0.60&0.32&0.16&6.4\\
    G018.801--00.297&8&10.0&2.78&1.25&0.46&0.7\\
    G018.931--00.029&15&9.0&0.65&0.16&0.12&1.1\\
    G022.253+00.032&16&6.2&1.14&0.32&0.29&8.3\\
    G022.692--00.452&12&11.4&1.28&0.28&0.20&13.1\\
    G023.477+00.114&19&22.6&1.44&0.41&0.34&6.7\\
    G024.010+00.489&16&38.4&2.77&1.31&0.56&16.4\\
    G024.524--00.139&23&13.1&1.74&0.53&0.41&1.7\\
    G025.163--00.304&18&19.0&1.18&0.43&0.25&5.1\\
    G028.273--00.167&19&8.4&2.78&0.67&0.39&3.5\\
    G028.541--00.237&18&13.4&2.42&0.41&0.31&3.0\\
    G028.564--00.236&35&81.1&3.83&0.75&0.58&6.7\\
    G028.927+00.394&9&14.1&1.62&0.57&0.33&4.3\\
    G030.704+00.104&21&15.7&1.02&0.44&0.35&2.3\\
    G030.913+00.719&12&11.3&0.49&0.15&0.14&6.1\\
    G033.331--00.531&9&28.9&0.61&0.38&0.34&6.1\\
    G034.133+00.076&15&5.0&0.33&0.13&0.11&2.3\\
    G034.169+00.089&22&1.7&0.25&0.11&0.10&2.6\\
    G034.739--00.119&24&9.7&1.42&0.46&0.32&8.6\\
    G036.666--00.114&13&10.4&0.55&0.19&0.13&9.8\\
    G305.794--00.096&34&9.1&1.50&0.29&0.26&8.6\\
    G327.116--00.294&21&10.8&0.51&0.11&0.17&5.0\\
    G331.372--00.116&39&8.0&0.74&0.25&0.31&3.0\\
    G332.969--00.029&20&4.1&0.47&0.26&0.22&1.5\\
    G333.016--00.751&27&1.9&0.24&0.13&0.10&3.3\\
    G333.481--00.224&25&6.7&0.42&0.12&0.09&8.3\\
    G333.524--00.269&38&8.4&1.06&0.24&0.11&8.4\\
    G337.342--00.119&16&2.6&0.56&0.22&0.20&4.2\\
    G337.541--00.082&19&9.9&1.06&0.16&0.18&2.8\\
    G340.179--00.242&16&1.9&0.63&0.16&0.21&0.6\\
    G340.222--00.167&21&8.4&1.10&0.31&0.21&7.1\\
    G340.232--00.146&16&31.7&1.06&0.37&0.27&6.2\\
    G340.398--00.396&29&5.0&0.55&0.23&0.15&1.4\\
    G341.039--00.114&35&7.3&0.60&0.13&0.11&5.5\\
    G343.489--00.416&29&7.3&0.23&0.05&0.12&2.2\\
    G345.114--00.199&26&2.1&0.37&0.13&0.12&4.4\\
    \enddata
    \tablenotetext{a}{Total number of cores identified in each clump}
    \tablenotetext{b}{Core formation efficiency, a ratio of total core mass to clump mass.}
\end{deluxetable*} 

The major source of uncertainty in the mass calculation is the gas-to-dust mass ratio and the dust opacity. 
Assuming that all possible values of $\mathbb{R}$ and $\kappa_\mathrm{1.3\,mm}$ are distributed uniformly between the extreme values, $70<\mathbb{R}<150$ and $0.7<\kappa_\mathrm{1.3\,mm}<1.05$ \citep[e.g.,][]{Devereux90,Ossenkopf94, Vuong03, Sabatini19, Sabatini22}, the standard deviation can be estimated \citep{Sanhueza17}. We adopt the uncertainties derived by \citet{Sanhueza17} of 23\% for the gas-to-dust mass ratio and  28\% for the dust opacity, for the adopted values of 100 and 0.9 cm$^2$\,g$^{-1}$, respectively. 
In addition, considering an absolute flux uncertainty of 10\% for ALMA observations in band 6, a temperature uncertainty of $\sim$20\%, and a distance uncertainty of $\sim$10\%, we estimate the uncertainties of core mass, volume density, and a surface density of $\sim$50\% \citep[see][for more details]{Sanhueza17,Sanhueza19}. 
For the protostellar cores, the actual dust temperature can be higher than the value derived at a coarser angular resolution using single-dish observations \citep[][]{Li21a, Morii21}. However, we confirm that the NH$_3$ kinetic temperatures derived  at core scales in ASHES clumps with available NH$_3$ observations are lower than 23 K \citep[][]{Lu15, Li22-arxiv}.  A similar range of NH$_3$ temperatures was also reported in other IRDC cores \citep{Wang14}.

The core radius was defined as half of the geometric mean of the FWHM (Table~\ref{tab:clump}) provided by astrodendro; see additional details in the astrodendro website\footnote{https://dendrograms.readthedocs.io/en/stable/}. 
The core radius ranges from $\sim$5$\times 10^{-3}$ pc to $\sim$5$\times 10^{-2}$ pc, corresponding to $\sim$10$^3-10^4$ au, and its median is 2680\,au. 
The core size is only $\lesssim$5\% of the maximum recoverable scale, and the missing flux contribution for core mass estimation is expected to be small. 
The surface density, $\Sigma$, and the molecular volume density, $n(\mathrm{H_2})$, were estimated assuming a spherical core as:  $\Sigma=M_\mathrm{core}/\pi r^2_\mathrm{core}$ and  $n(\mathrm{H_2})=M_\mathrm{core} / \Bar{m}_\mathrm{H_2}(4\pi r^3_\mathrm{core}/3)$, where $r_\mathrm{core}$ is the core radius.  Figure~\ref{fig:phy_para} shows the estimated radius and densities of cores. The orange lines are the cumulative histogram, indicating which percent of cores have higher density than a certain value. 
The estimated surface density ranges from $\sim$0.05--10\,g\,cm$^{-2}$, and the volume density ranges from  $\sim$10$^{5}-3\times10^{7}$\,cm$^{-3}$. 
We found that $\sim$10\% of cores (91/839) have a surface density higher than unity, which is the condition suggested by \citet{KrumholzMcKee08} as a threshold for high-mass star formation. They concluded that 1 g\,cm$^{-2}$ is the minimum necessary to halt excessive fragmentation and allow formation of high-mass stars. 
Actually, all seven high-mass cores ($M_\mathrm{core} > 27\,M_\odot$) have a surface density higher than this threshold.

The peak column density, $N_\mathrm{peak}(\mathrm{H_2}$), was derived from 
\begin{equation}
    N_\mathrm{peak}(\mathrm{H_2}) = \mathbb{R}\,\frac{I_\mathrm{1.3\,mm, peak}}{\Omega\, \Bar{m}_\mathrm{H_2}\,\kappa_\mathrm{1.3\,mm}\,B_\mathrm{1.3\,mm}(T_\mathrm{dust})},
\end{equation}
where $I_\mathrm{1.3\,mm, peak}$ is the peak intensity measured at the continuum peak, and $\Omega$ is the beam solid angle. 
In the top-right panel of Figure~\ref{fig:phy_para}, the histogram of the peak column density is shown. It ranges from 10$^{22}$--10$^{24}$\,cm$^{-2}$ with a peak around $3\times10^{22}$\,cm$^{-2}$. The orange line indicates less than 20\% of cores have a peak column density higher than $10^{23}$\,cm$^{-2}$. 
Physical parameters of cores are summarized in Table~\ref{tab:core_phy} (the full table can be accessed in machine-readable form). 

\subsection{Identification of Hub-filament System}
\label{sec:filament}
In addition to the compact structures (cores), the continuum emission also resolved clumps in a network of filaments, some of which consist of hub-filament systems. 
A hub is the convergence of multiple filaments and recently caught more attention as the possible birthplace of high-mass stars \citep[e.g., ][]{Motte18}.  

We applied the publicly available filament finding package $\mathtt{FilFinder}$ \citep[][]{Koch15} to the ASHES continuum images to identify filamentary structures. 
It firstly creates a mask of a filamentary structure by adopting an intensity threshold, and each structure within the identified mask is reduced to a skeleton. These skeletonized structures are pruned by imposing thresholds for their lengths or sizes prior to obtain the extracted filaments. We exclude very short structures and identified the prominent ones (see Appendix~\ref{sec:Appendix_A} for more details). 

\begin{figure}
    \centering
    \includegraphics[width=8.7cm]{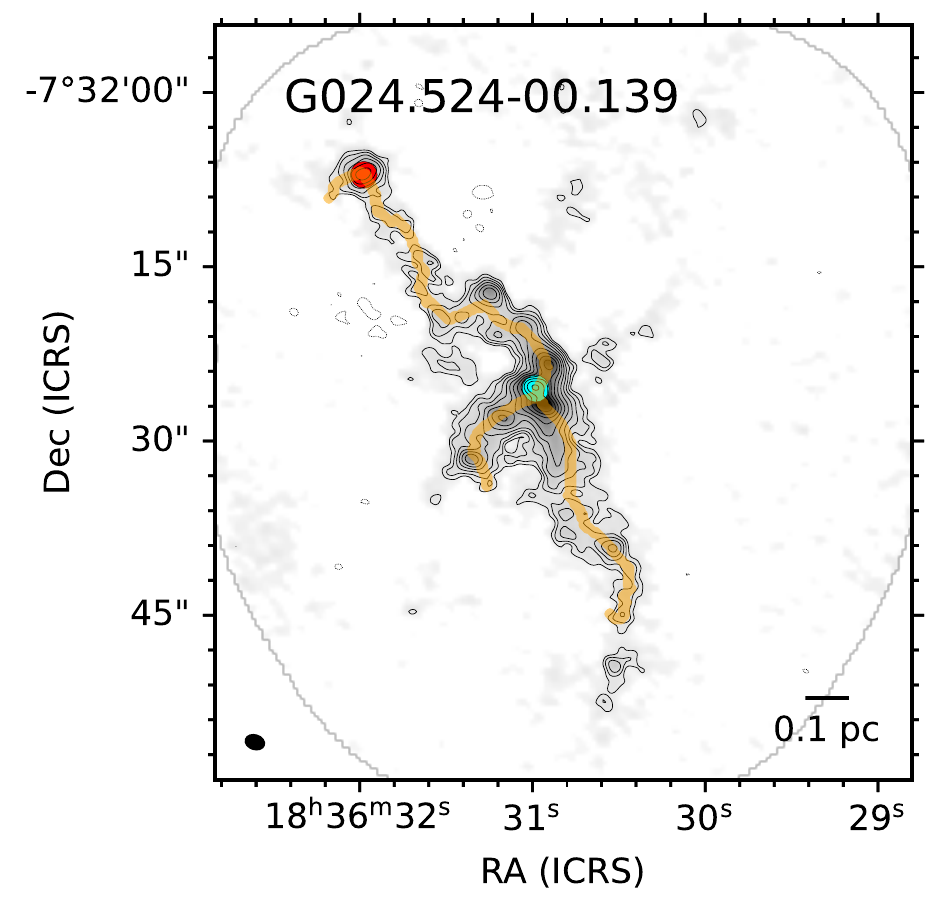}
    \caption{An example of identified filament in G025.524--00.139. Background image and contours are 1.3 mm continuum emission, the same as the right panel of Figure~\ref{fig:ashes_cont_1}. Orange lines represent the identified skeleton by $\mathtt{FilFinder}$. Red and cyan circles show the positions of the MMCs and second MMCs, respectively. Here, the circle size is constant (not representing core masses). The complete figure set (11 images) is shown after Appendix.
    }
    \label{fig:fil_G24}
\end{figure}

Filamentary structures are identified in almost all clumps (36/39)\footnote{ Clumps in which no filamentary structure is identified are G018.801–00.297, G332.969–00.029, and G340.179–00.242}, some of which contain hub-filament systems. 
The identified skeletons are highlighted as orange lines in Figure~\ref{fig:fil_G24}, for example. In this figure, the convergence point is a hub. 
We have identified 31 hubs in 17 ASHES clumps. We will discuss whether the most massive cores or high-mass cores are preferentially located at such hub positions later (see Section~\ref{sec:mmc}).  
For this work, we focus on the dust continuum emission and the detailed analysis of filaments and hubs is beyond the scope of this paper. 

\section{Discussion}
\label{sec:discussion}
\subsection{Core Mass}
\label{sec:core_mass}
\begin{figure}
    \centering
    \includegraphics[width=8.5cm]{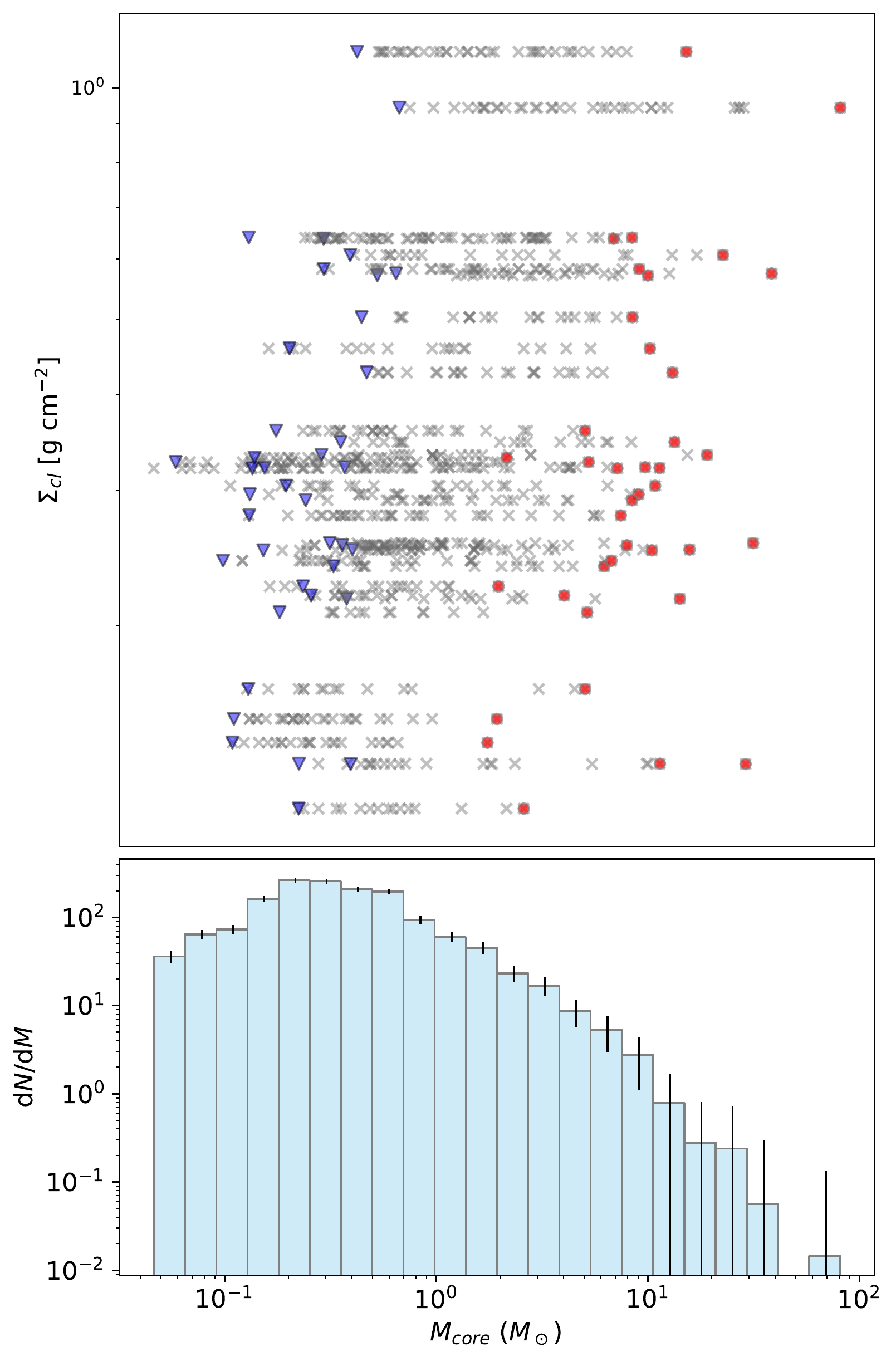}
    \caption{(Top panel) Clump surface density versus core mass. Gray cross symbols correspond to all 839 cores.  Red circles show the most massive cores in each clump. Blue triangles represent mass sensitivity for each clump corresponding to the integrated intensity of 3.5$\sigma$. (Bottom panel) The core mass function of all 893 cores with uniform bin size in log-space. The errorbars represent the statistical error.} 
    \label{fig:MclMcoreCMF}
\end{figure}
The top panel of Figure~\ref{fig:MclMcoreCMF} shows the core mass and the corresponding surface density of the natal clump. Red circles represent the most massive cores in each clump, blue triangles represent mass sensitivity for each clump corresponding to the integrated intensity of 3.5$\sigma$.  
We find a moderate correlation between clump surface density and the maximum core mass with a Spearman’s rank correlation coefficient of $r_s = 0.39$ and $p$–value = 0.01. 
The bottom panel of Figure~\ref{fig:MclMcoreCMF} displays the differential core mass function sharing the horizontal axis of the top panel. 
The core mass distribution peaks at around 0.6\,$M_\odot$ (close to the worst mass sensitivity in the sample), and rapidly decreases toward the high-mass end. 
The detailed analysis for the core mass function (CMF) will be addressed in a dedicated paper under preparation (Morii et al. 2023, in prep.).

\subsection{Most Massive Cores} 
\label{sec:mmc}
As described in Section~\ref{sec:sample}, at least one high-mass star is expected to be formed from each ASHES clump. In this section, we focus on the most massive cores found in each clump (hereafter MMCs), those with $M_\mathrm{core} = M_\mathrm{max}$. 

\subsubsection{Correlation between the Maximum Core Mass and Clump Surface Density}
\begin{figure*}
    \centering
    \includegraphics[width=16cm]{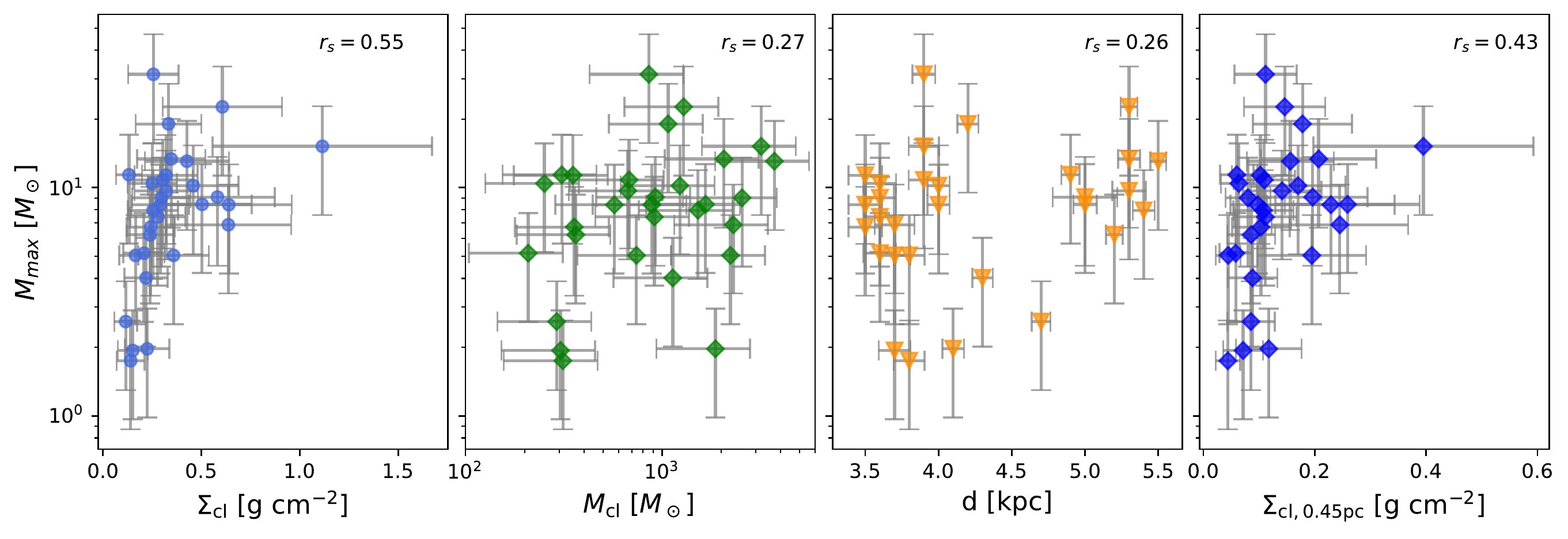}
    \caption{Maximum core mass ($M_\mathrm{max}$) as a function of clump surface density ($\Sigma_\mathrm{cl}$), clump mass ($M_\mathrm{cl}$), distance ($d$), and $M_\mathrm{cl, 0.45 pc}$ for selected 30 clumps. On the top left of each panel, Spearman’s rank correlation coefficients are denoted. }
    \label{fig:mcm}
\end{figure*}

In stellar clusters, higher mass stars are found in higher mass clusters \citep{Larson03}.  
In a similar fashion, it may be expected that higher mass cores should form in higher mass clumps.  
Here, to minimize the effect of having different spatial resolutions, we have limited the sample for this discussion. We have excluded clumps that locate too close ($< 3.5$\,kpc) and too far ($> 5.5$\,kpc), and two more with the worst mass sensitivity ($> 0.45 M_\odot$).  
As a result, the 30 clumps remaining are located between 3.5 and 5.5 kpc and have a mass sensitivity between 0.086 and 0.41 $M_\odot$. 
Figure~\ref{fig:mcm} shows the scatter plots of the maximum core mass ($M_\mathrm{max}$) versus clump surface density, clump mass, distance, and clump surface density within a certain area. As seen in Figure~\ref{fig:mcm}, the clump surface density correlates with $M_\mathrm{max}$, having a Spearman’s rank correlation coefficient $r_s = 0.55$ and $p$–value = 0.0016. 
On the other hand, $M_\mathrm{max}$ and clump mass weakly correlates with a Spearman’s rank correlation coefficient of $r_s = 0.27$ and a $p$–value = 0.15.  
It indicates that the maximum core mass in each clump is not determined by the natal clump mass at least in the very early stages traced in the ASHES survey. 
Considering the relation between the cluster mass (clump mass) and the maximum stellar mass empirically derived by \citet{Larson03}, see Equation~\ref{equ:larson} used earlier, the weak correlation implies the final stellar mass is not determined from the initial core mass. 
The left panel in Figure~\ref{fig:mcm} would rather indicate that more massive cores form in clumps with higher surface density. 
We confirmed that the stronger correlation between the maximum core mass and the clump surface density does not result from the co-dependence on the distance as shown in the right two panels of Figure~\ref{fig:mcm}. We re-calculated the clump surface density within a circle with a radius of 0.45\,pc centered on the mean positions of cores, ($\Sigma_\mathrm{cl, 0.45 pc}$) to reduce the dependence on distance. The circle almost corresponds to the field of view of the closest clump (3.5 kpc). A moderate correlation is present with a Spearman’s rank correlation coefficient of $r_s = 0.43$ and $p$–value = 0.017. 

\subsubsection{The Lack of High-mass Prestellar Cores}
\begin{figure*}
    \centering
    \includegraphics[width=17.5cm]{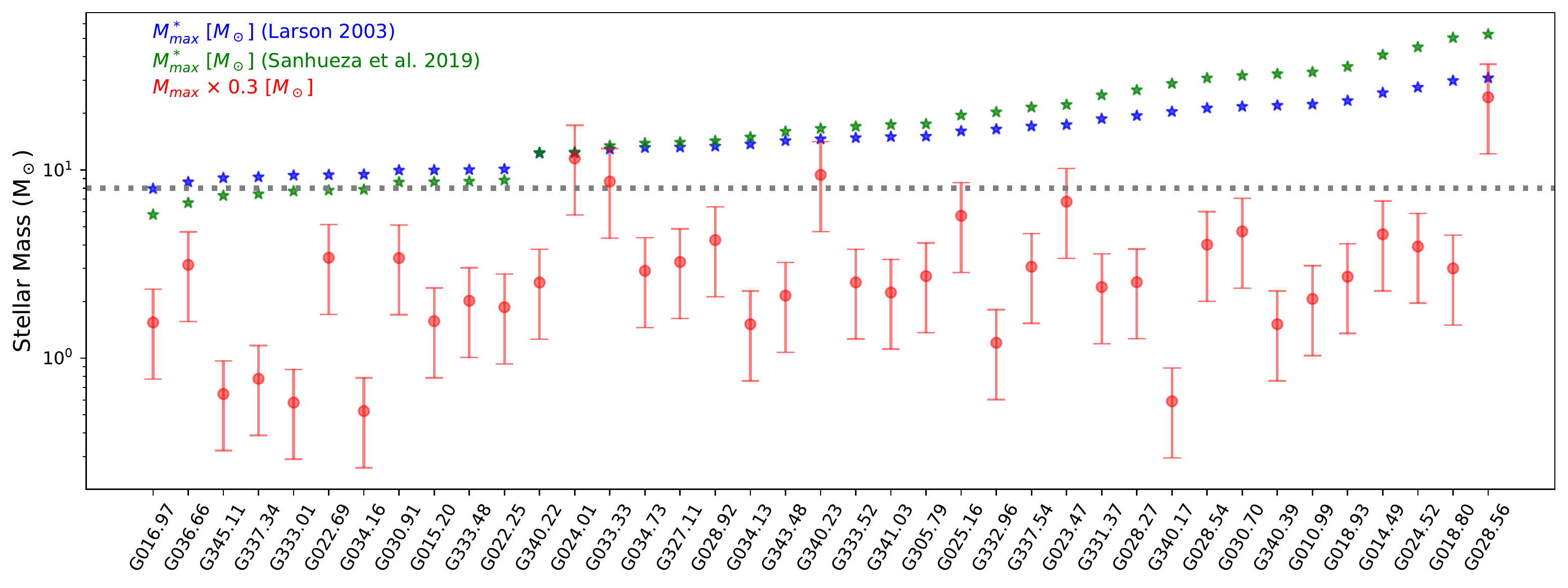}
    \caption{Comparison between stellar mass expected from clump mass and observed maximum core mass in each region. Blue and green star symbols correspond to the maximum stellar mass expected from their clump mass estimated from Equation~\ref{equ:larson} and \ref{equ:sanhueza} assuming $\varepsilon_\mathrm{SFE}=0.3$. Red circles are the maximum core mass in each region estimated in this work, multiplied by 0.3 to indicate the stellar mass with an assumption of SFE (from core to star) of 30\%. Error bar shows the 50\% uncertainty as mentioned in Section~\ref{sec:core-phy}.  The horizontal dotted line represents the stellar mass of 8\,$M_\odot$. The plotting order in the horizontal axis is sorted by the clump mass.}
    \label{fig:M*Mcore}
\end{figure*}
We revealed 839 dense cores, among which about 55\% are low-mass ($<1\,M_\odot$). Although seven high-mass cores ($\gtrsim27\,M_\odot$) are identified thanks to a large sample, more than half of the ASHES  clumps (23/39) host only cores with masses smaller than 10\,$M_\odot$. 
We compared the maximum core mass with the expected maximum stellar mass in Figure~\ref{fig:M*Mcore}. 
Clumps are sorted in order of their masses, and the clump mass (and the expected maximum stellar mass) increases from left to right. 
The expected maximum stellar mass is estimated from the clump mass in two different ways, using Equation~\ref{equ:larson} and \ref{equ:sanhueza} (plotted as blue and green star symbols, respectively). 
We overplotted $M_\mathrm{max}$ multiplied by 0.3 as red circles, assuming a core-to-star formation efficiency (SFE) of 30\%. 

We again find no strong correlation between the maximum core mass and clump mass as mentioned in the previous section. 
Moreover, most red circles (30\% of core mass) show lower masses than blue and green stars (expected stellar masses). The majority of the ASHES clumps (35/39) have currently the most massive cores with insufficient mass to form high-mass stars with the expected mass. 
This plot highlights that most cores in such very early phase have not sufficient mass to form high-mass stars without additional mass feeding, as predicted by clump-fed scenarios, such as competitive accretion, hierarchical collapse, and the inertial-inflow model. 
The most massive cores in only four regions could form a high-mass star and have a mass near to the expected maximum stellar mass, assuming a SFE of 30\% (G024.01, G028.56, G033.33, and G340.23). Even if we assume a higher SFE of 50\%, only two addittional clumps (G023.47 and G025.16) are added. 
We should note that these high-mass cores are associated with outflows traced by CO ($J$=2--1) and also warm line emission such as H$_2$CO and CH$_3$OH ($E_\mathrm{u} > 45$\,K), implying that these high-mass cores present star-formation signatures and are not prestellar (results that will be presented in a forthcoming paper). 

The most massive cores are generally intermediate-mass cores, more massive than the thermal Jeans mass of the host clump (e.g., 1--5\,$M_\odot$).   
As an initial condition, the competitive accretion scenario predicts that clumps fragment into low-mass cores, comparable to the Jeans mass. It does not entirely match our observations in terms of core mass, although some of the maximum core masses may have been initially low-mass but have already grown by accreting gas from the surrounding medium. 
\citet[][]{Contreras18} estimated a core infall rate of 2$\times$10$^{-3}$\,$M_\odot$\,yr$^{-1}$ in one core of the ASHES sample. Assuming that gas accretion with this infall rate continues for $\sim$2.0$\times 10^{4}$\,yr, the mean free-fall time of the 39 most massive cores ($t_{ff} = \sqrt{3\pi / (32G\rho)}$), cores can gain the additional mass of $\sim$40\,$M_\odot$. 
\citet[][]{Sabatini21} estimated a duration of $5 \times 10^4$\,yr for the 70 $\mu$m-dark phase from 110 massive clumps at different evolutionary stages. This time scale, longer than the mean free-fall time for the 39 more massive cores, suggests that cores initially of a Jeans mass have sufficient time to accrete and grow before becoming bright at IR wavelengths. 
In the following subsection, we will investigate whether MMCs are located at the position where efficient mass feeding is expected. 

Using mosaic observations at high sensitivity, we have therefore revealed a large number of low-mass cores  and the lack of high-mass prestellar cores, in dense, massive 70 $\mu$m dark clumps. Considering the large number of cores detected in this study (839 in 39 clumps), following \cite{Sanhueza19}, we can say with more confidence that high-mass prestellar cores do not exist or, if they do, they should form later when environmental conditions are appropriate for their formation.

\subsubsection{Weak Mass Segregation and Strong Density Segregation} 
\label{sec:segregation}
Mass feeding scenarios, such as the competitive accretion model, expect high-mass cores to be formed from low-mass cores located near the bottom of the gravitational potential or (hub-)filaments where cores acquire mass more efficiently. 
The different spatial distribution of massive objects with respect to lower mass objects is called mass segregation \citep[e.g.,][]{Allison09, ParkerGoodwin15}. 
 Segregation caused by two-body relaxation is called dynamical mass segregation \citep[][]{vonHoerner60, Meylan00}, while the segregation found in young clusters, believed to be inherited from the initial fragmentation, is called  primordial mass segregation  \citep[e.g.,][]{Alfaro18}. 
The {\it Herschel} Gould Belt Survey \citep[][]{Andre10} has found mass segregation in some star-forming region such as Aquila, Corona Australis, and W43-MM1 \citep[][]{DibHenning19}, Orion A \citep[][]{Roman-Zuniga19}, Orion B \citep[][]{Parker18, Konyves20}, NGC6334 \citep[][]{Sadaghiani20}, and NGC2264 \citep[][]{Nony21} at $\sim0.01-0.1$\,pc scale. 
\citet{DibHenning19} indicates regions with more active star formation present centrally clustering core distributions and significant mass segregation. 
Our sample is expected to be in the very early phase of high-mass star and cluster formation without IR-bright sources and be the best target to investigate if there is primordial mass segregation in the very early phase. 

Following the pilot survey \citep[][]{Sanhueza19}, we first studied mass segregation ratios (MSRs), $\Lambda_\mathrm{MSR}$ \citep{Allison09} and $\Gamma_\mathrm{MSR}$ \citep{Olczak11}, based on the minimum spanning tree (MST) method developed by \citet[][]{Barrow85}.
MST connects cores (in this case), minimizing the sum of the length and determining a set of straight lines. 
Black line segments on the right panel in Figure~\ref{fig:ashes_cont_1} show an example of the outcome from MST. 
$\Lambda_\mathrm{MSR}$ compares the sum of the edge length of the MST ($l_\mathrm{MST}$) of random cores with that of the same number of massive cores:  
\begin{equation}
    \Lambda_\mathrm{MSR} (N_\mathrm{MST}) = \frac{\langle l^\mathrm{random}_\mathrm{MST} \rangle}{l^\mathrm{massive}_\mathrm{MST}} \pm \frac{\sigma_\mathrm{random}}{l^\mathrm{massive}_\mathrm{MST}},
\label{equ:lambda}
\end{equation}
where $\langle l^\mathrm{random}_\mathrm{MST} \rangle$ is the sum of the edge length of $N_\mathrm{MST}$ random cores averaged by 1000 trial and $l^\mathrm{massive}_\mathrm{MST}$ is that of $N_\mathrm{MST}$ most massive cores. 
$\sigma_\mathrm{random}$ is the standard deviation associated with estimating the average value $\langle l^\mathrm{random}_\mathrm{MST} \rangle$. 
If massive cores are distributed similarly to random cores (i.e., no mass segregation), 
$\Lambda_\mathrm{MSR}$ would be close to unity. 
If massive cores are concentrated, and their distribution is different from the lower-mass cores (mass segregation), $\Lambda_\mathrm{MSR}$ becomes larger than unity. 
In turn, $\Lambda_\mathrm{MSR} <1$ implies massive cores are spread out compared to others (inverse-mass segregation). 
We calculated $\Lambda_\mathrm{MSR} (N_\mathrm{MST})$ varying $N_\mathrm{MST}$ from two to the total number of cores in each clump, $N\mathrm{(core)}$. 
The second method ($\Gamma_\mathrm{MSR}$) uses the geometric mean of the edges ($\gamma_\mathrm{MST}$), not the sum of the edges ($l_\mathrm{MST}$)
\begin{equation}
    \gamma_\mathrm{MST} = \left(\prod_{i=1}^{N\mathrm{(core)}-1} L_i\right)^{1/(N\mathrm{(core)}-1)},
\end{equation}
and defined by
\begin{equation}
    \Gamma_\mathrm{MSR} (N_\mathrm{MST}) = \frac{\gamma^\mathrm{random}_\mathrm{MST}}{\gamma^\mathrm{massive}_\mathrm{MST}} (d\gamma_\mathrm{random})^{\pm 1}, 
\end{equation}
where $d\gamma_\mathrm{random}$ is the geometric standard deviation given by 
\begin{equation}
    d\gamma = \mathrm{exp}\left( \sqrt{\frac{\Sigma^{N\mathrm{(core)}-1}_{i=1} (\mathrm{ln}\,L_i - \mathrm{ln}\, \gamma_\mathrm{MST})^2}{N\mathrm{(core)}-1}} \right),
\end{equation}
where $L_i$ is the $i$-th MST edge. 
The term $\gamma^\mathrm{random}_\mathrm{MST}$ is the geometric mean of the edges for  $N_\mathrm{MST}$ random cores, and $\gamma^\mathrm{massive}_\mathrm{MST}$ is that for the $N_\mathrm{MST}$ massive cores. 
These two different MSRs ($\Lambda_\mathrm{MSR}$ and $\Gamma_\mathrm{MSR}$) are thought to behave in the same way, but \citet{Olczak11} proposed that $\Gamma_\mathrm{MSR}$ is more sensitive to finding weak mass segregation.  

\begin{figure}
\gridline{\fig{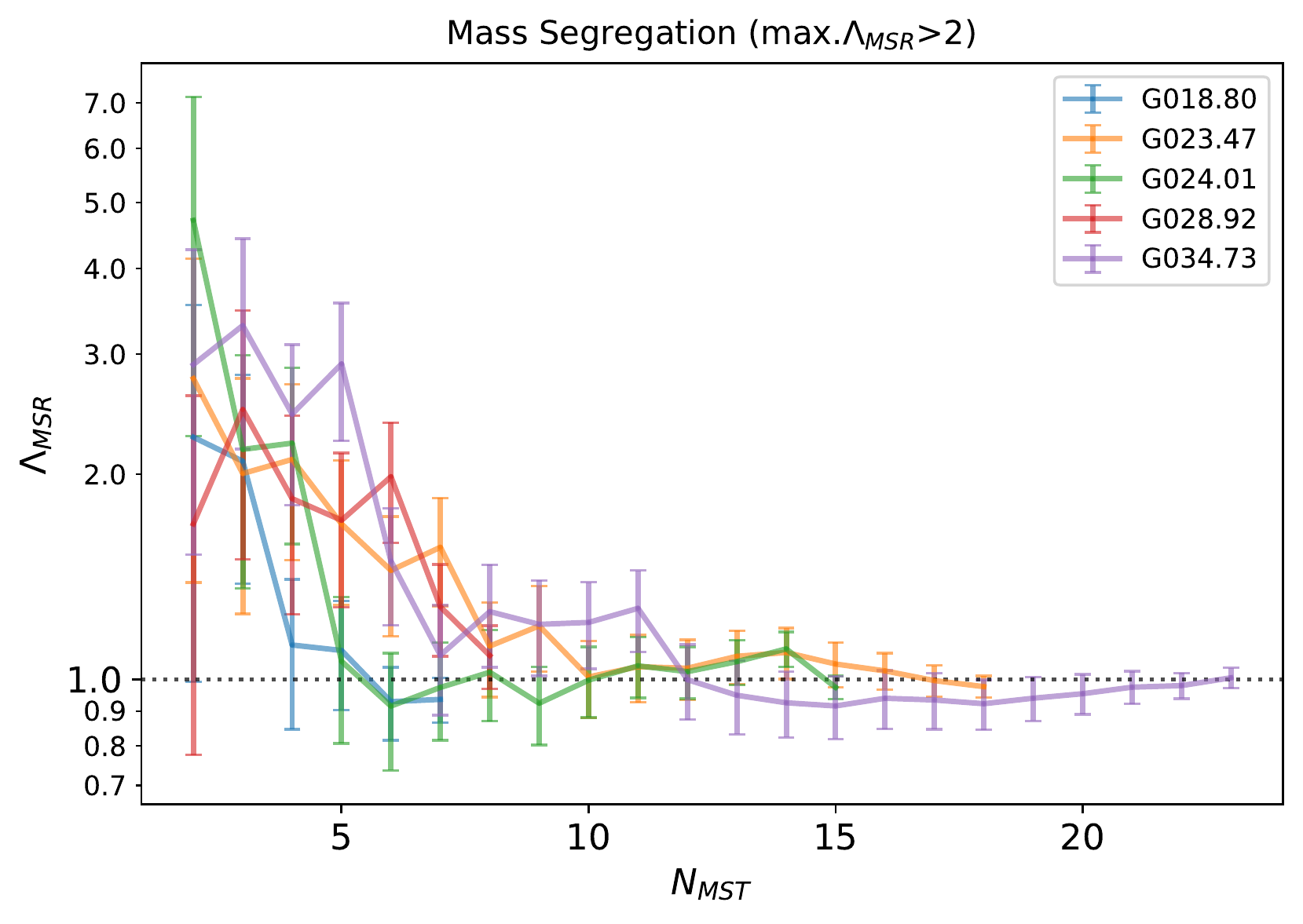}{0.48\textwidth}{}}
\vspace{-0.8cm}
\gridline{\fig{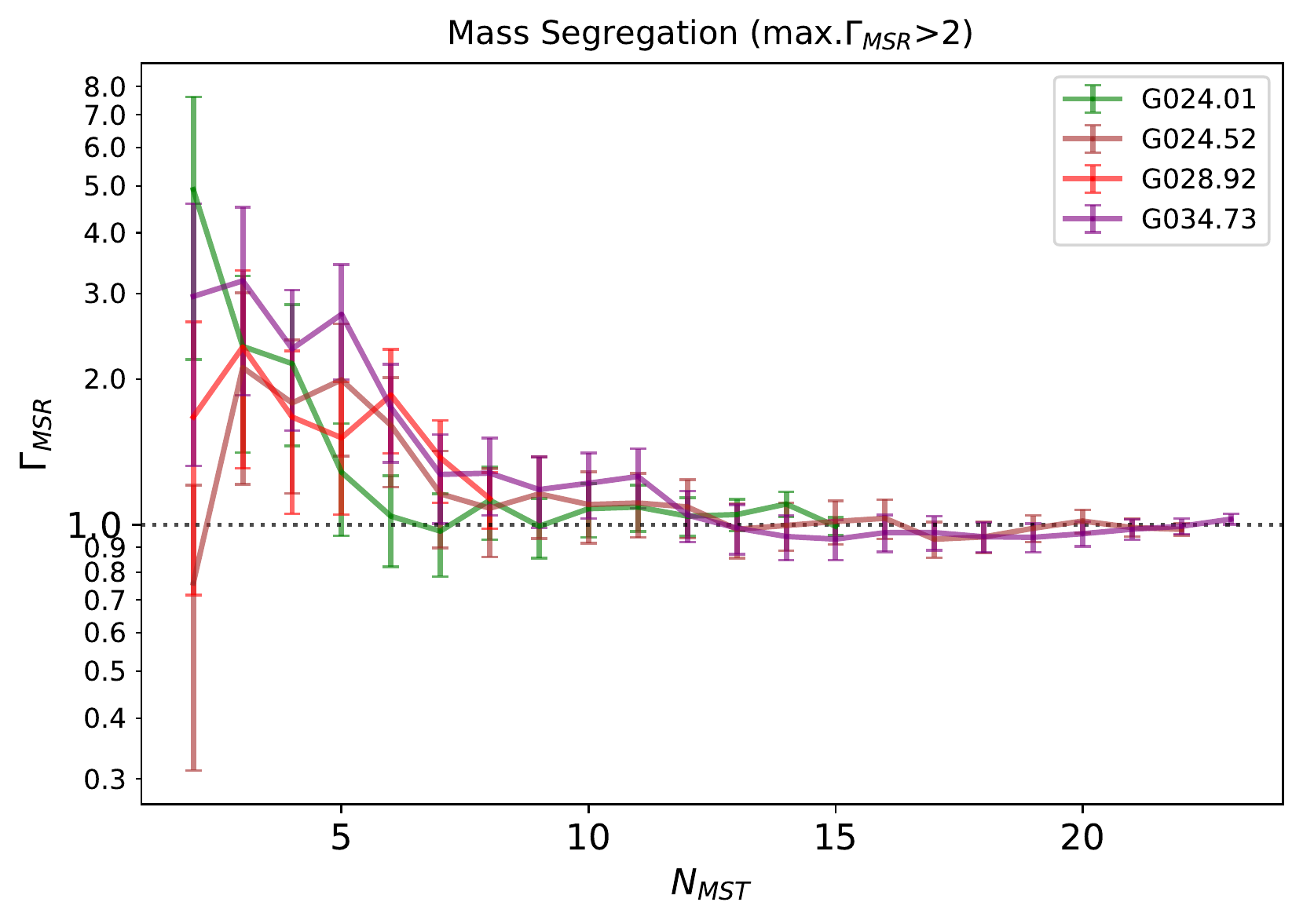}{0.48\textwidth}{}}
\vspace{-0.8cm}
\caption{Mass segregation ratios values ($\Gamma_\mathrm{MSR}$ and $\Lambda_\mathrm{MSR}$) and the number of segments considered ($N_\mathrm{MST}$). Selected regions have relatively higher values i.e., MSR values $\gtrsim2$ or low values (MSR$<$0.5) for $N_\mathrm{MST}=3$.}
\label{fig:segregation}
\end{figure}
Figure~\ref{fig:segregation} shows the mass segregation ratios as $\Gamma_\mathrm{MSR}$ and $\Lambda_\mathrm{MSR}$ with changing $N_\mathrm{MST}$ from 2 to $N\mathrm{(core)}$. 
We display regions with relatively high MSR (max(MSR)$> 2$ at $N_\mathrm{MST}>3$) or low MSR (max(MSR)$< 0.5$ at $N_\mathrm{MST}>3$). 
For the remaining regions, $\Gamma_\mathrm{MSR}$ and $\Lambda_\mathrm{MSR}$ is consistent with unity (i.e., no mass segregation) for $N_\mathrm{MST}>3$. 
The MSRs at $N_\mathrm{MST}=3$ in six clumps is higher than two, but MSRs values gradually decrease for larger $N_\mathrm{MST}$ values. Only G034.73 shows a tentative detection of mass segregation with MSR of $\sim$3 at  $N_\mathrm{MST}=5$. 
Compared with the pilot survey \citep{Sanhueza19}, in which no evidence of mass segregation was found, this larger sample reveals weak detections of mass segregation in G018.80, G023.47, G024.01, G024.52, G028.92, and G034.73. 
We note that most of these clumps host no high-mass cores except for G024.01 ($M_\mathrm{max} = 38.4 M_\odot$), and there is no strong correlation between the maximum core mass and the mass segregation ratio. 
Thus, the weak mass segregation found in the ASHES sample implies that the relatively high-mass cores (e.g., MMCs) form similarly to the lower-mass cores at the initial phase traced by the studied 70 $\mu$m dark IRDCs. 

However, a caveat to keep in mind for this discussion is the small core sample used for estimating mass segregation ratios. Although the previous studies such as \citet[][]{DibHenning19} treat about 100 objects, the sample here contains 40 at most. 

\begin{figure*}
\gridline{\fig{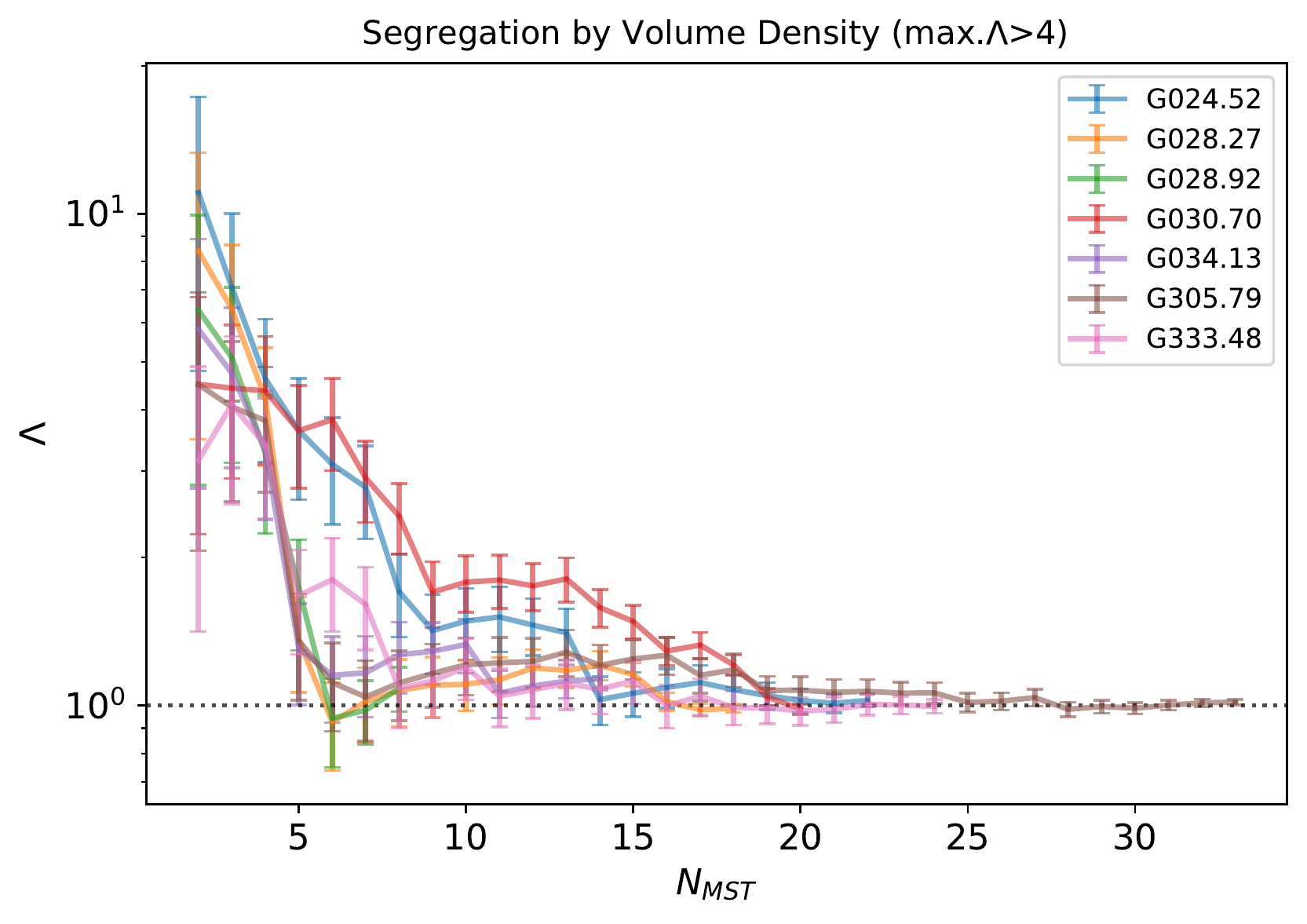}{0.49\textwidth}{}
          \fig{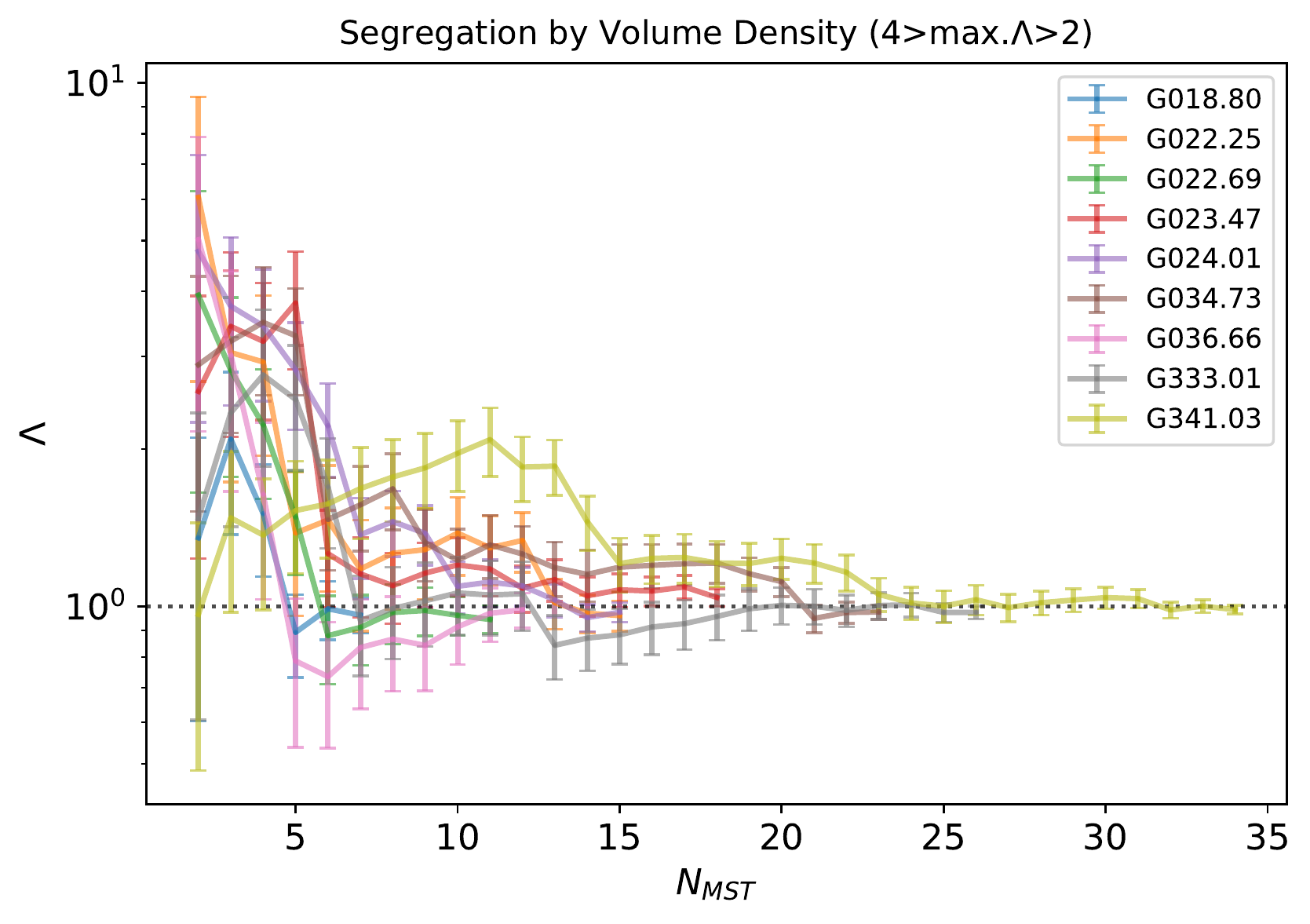}{0.49\textwidth}{}}
\gridline{\fig{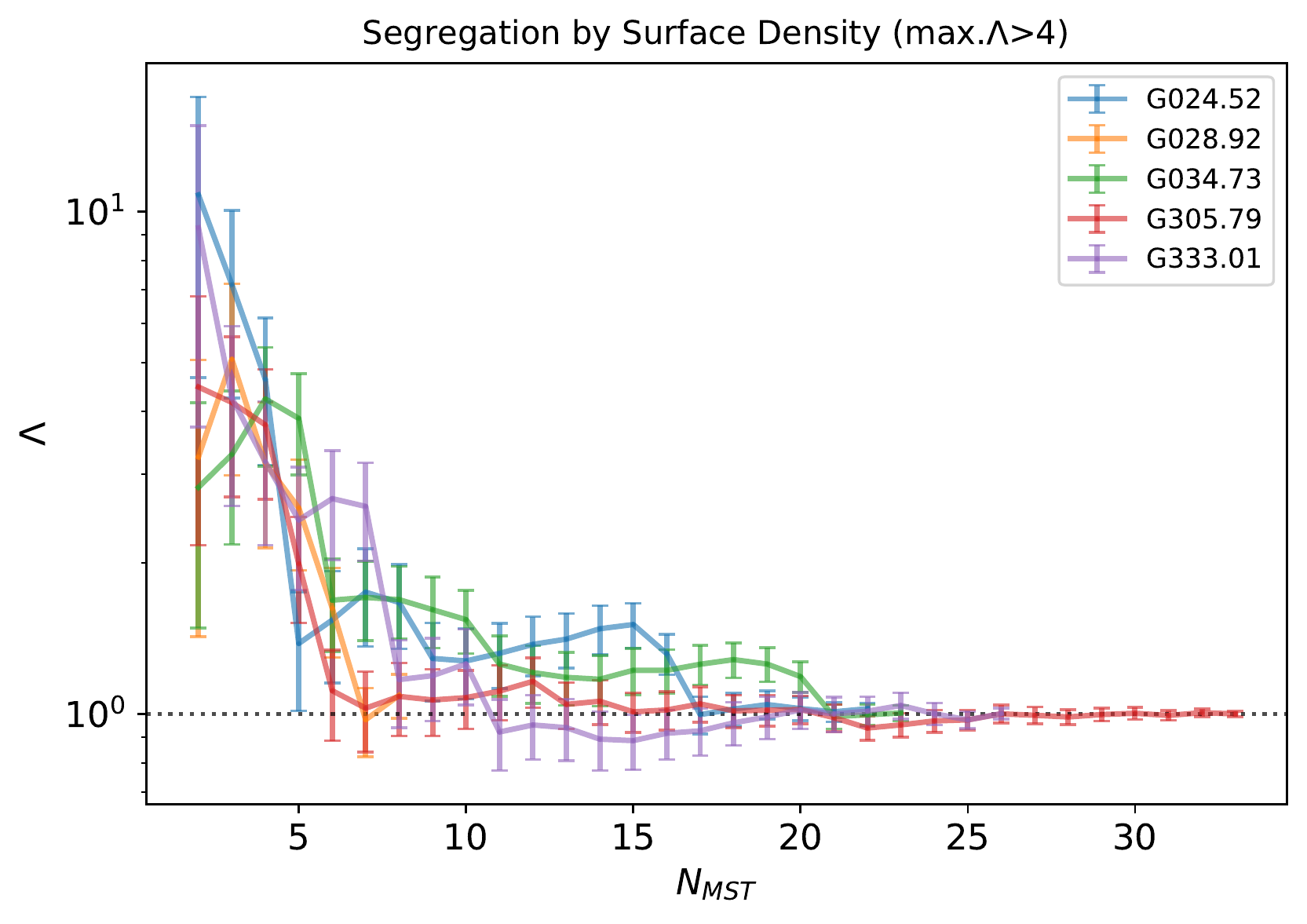}{0.49\textwidth}{}
          \fig{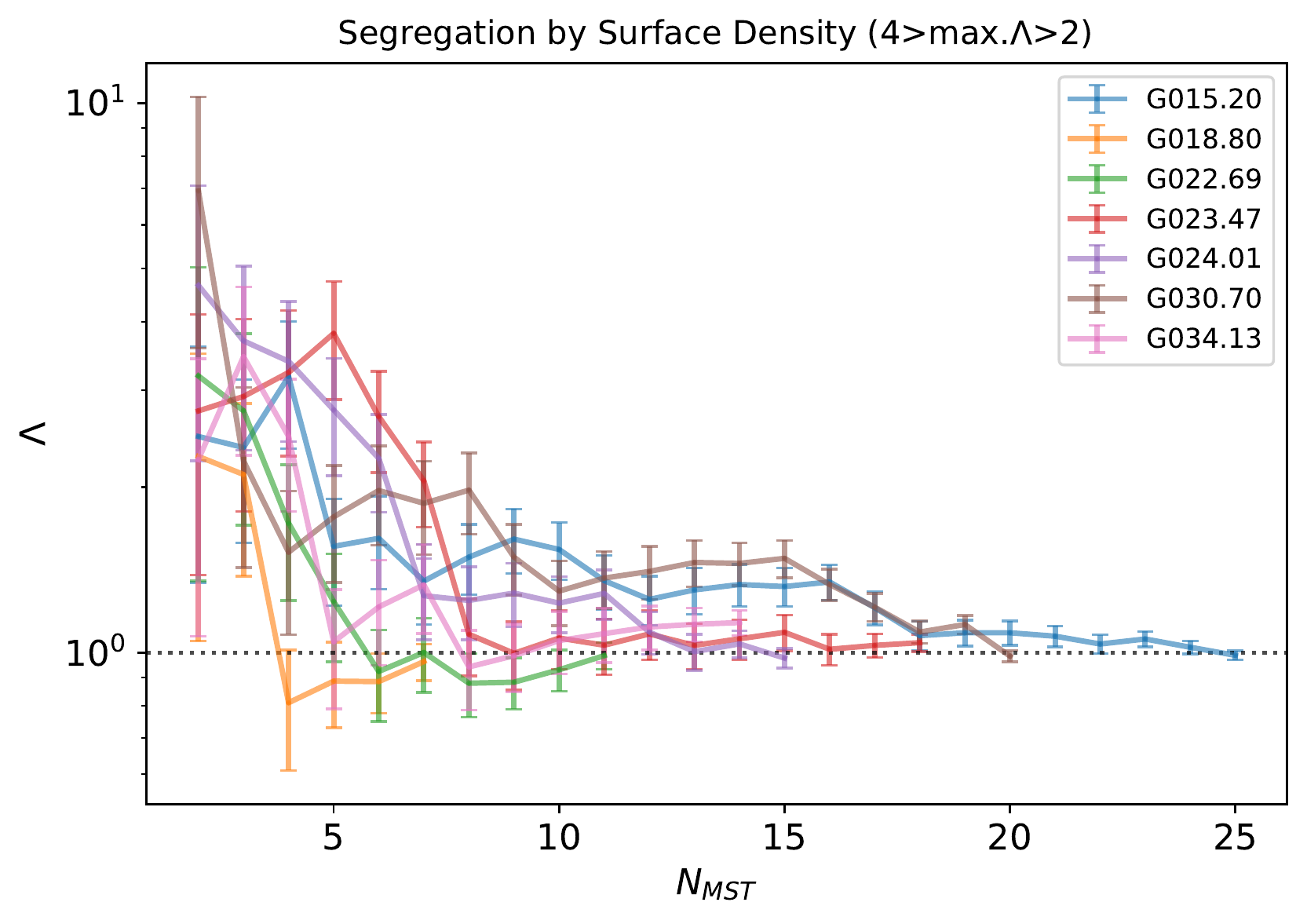}{0.49\textwidth}{}}
\caption{The segregation ratio $\Lambda$ by volume density (top) and surface density (bottom). Selected regions have relatively higher values i.e., $\Lambda>4$ (left) or $4>\Lambda>2$ for (right) for $N_\mathrm{MST}=3$.}
\label{fig:segregation_density}
\end{figure*} 
If we replace $l^\mathrm{massive}_\mathrm{MST}$ in Equation~\ref{equ:lambda} by $l^\mathrm{dense}_\mathrm{MST}$, an edge length of the $N_\mathrm{MST}$ densest cores instead of the most massive cores, we can investigate whether denser cores are distributed differently from relatively less dense cores. We refer to this as  density segregation.  
Figure~\ref{fig:segregation_density} shows the segregation ratio $\Lambda (N_\mathrm{MST})$ calculated by sorting by core volume density, $n(\mathrm{H_2})$ (top panel), and core surface density, $\Sigma$ (bottom panel), instead of core mass. 
Using the same threshold as Figure~\ref{fig:segregation}, we only plot clumps that shows high or low segregation ratio ($\Lambda > 2$ or $< 0.5$, respectively). 
As figures clearly show, contrary to mass segregation, the segregation by density was confirmed in about half of our sample, and their $\Lambda$ values are generally higher than $\Lambda_\mathrm{MSR}$, implying a stronger segregation. 
We note that all six clumps with signs of mass segregation show density segregation as well. By carefully checking core positions, core masses, and densities, we found that mass segregation and density segregation occurs around the same part within these clumps. 
To highlight the density segregation in Figure~\ref{fig:ashes_cont_1}, we colored in orange the cores that are denser than the median of the surface density of all cores in the clump. 

The density segregation may trace a segregation resulting from the different evolutionary stages of cores within a given clump. One may expect that cores become denser as they evolve.  Thus, the density segregation observed may indicate that more evolved cores are segregated. Another possibility is that such denser cores are formed from the fragmentation of denser parts within  clumps. 
We found a strong correlation between the clump surface density and the median of cores' surface density with a Spearman’s rank correlation coefficient of $r_s = 0.53$ and a $p$–value = 4.8$\times$ $10^{-4}$. 
Assuming that this correlations hold even at a smaller scale (i.e., within the clump), a denser region within a non-uniform clump would produce denser cores than a less dense part, resulting in density segregation.  
Therefore, the initial spatial distribution of cores would be dictated by density. 

\citet[][]{Alfaro18} and \citet[][]{Roman-Zuniga19} also reported a spatial segregation by volume density stronger than the segregation by mass in the Pipe Nebula and Orion A. They concluded that density controls the clumpy spatial distribution of prestellar cores at the very early phase. 

\subsubsection{Spatial Distribution of MMCs: Hub-filament System}
\label{sec:hub}
\begin{figure}
    \centering
    \includegraphics[width=9.3cm]{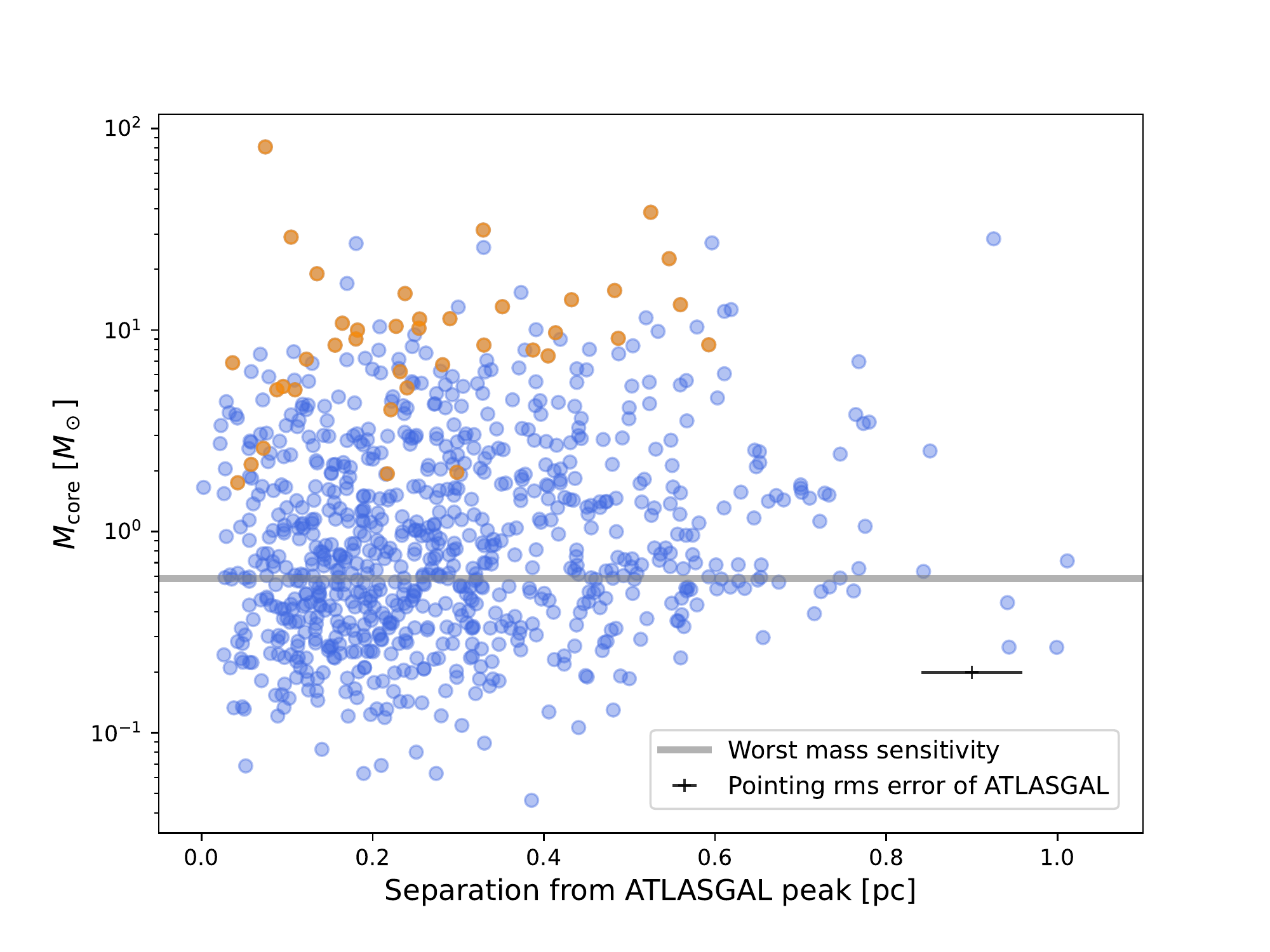}
    \caption{The distribution of core mass versus separation from clump center. The clump center corresponds to the continuum peak obtained from single-dish observations (ATLASGAL survey). Circles represent all 839 cores, and the MMCs are highlighted as oranges. There is no significant difference of separation between blue and orange circles, and it maintains even if the center is replaced with the mean position of cores in each clump.  
    } 
    \label{fig:M_dist}
\end{figure}
In addition to the weak mass segregation, we also found no significant difference in the spatial distribution between MMCs and the lower-mass cores. 
Since we have no information on the gravitational potentials of clumps, we assume that they are around the continuum peak of the single-dish observations. 
Figure~\ref{fig:M_dist} displays the plot of core mass versus separation from the continuum peak of each clump (hereafter clump peak position) obtained by single-dish observations (i.e., ATLASGAL survey).  
The most massive cores in each clump are highlighted as orange, and the other cores are shown in blue circles.  
Above the worst mass sensitivity ($\sim$0.58\,$M_\odot$), 
we cannot reject the null hypothesis at a 5\% level that the separation distributions of MMCs (orange) and the other cores (blue) are identical since the estimated p--value is 0.47 from the KS test. 
It implies that there is no significant difference. 
We note that this lack of difference in spatial distribution is confirmed when the clump peak positions are replaced with the mean position of cores identified by ALMA. 

Alternatively, another preferred location for the most massive cores could be a hub-filament system. 
Our observations resolve 70\,$\mu$m dark massive clumps, revealing cores and filamentary structures within clumps. We revealed prominent hub-filament systems in some clumps, such as G024.524--00.139 and G025.163--00.304, as identified in Section~\ref{sec:filament}. 
This motivates us to study whether there is correlation between the position of the most massive cores and the hub-filament systems. 
Assuming that cores at hub positions can efficiently accumulate gas, more massive cores are expected to form at hub positions. 
We overlaid the MMCs and second MMCs as red and cyan circles, respectively, in Figure~\ref{fig:fil_G24}. 
Most of the MMCs are not located at such hub positions. Indeed, only eight (20\%) MMCs are found at hubs. Even after considering the second most massive core in each clump, only eleven cores in total are located in hubs (11/(39$\times$2)=14\%). 
Considering projection effects, the real percentage would be lower. 
If we focus on high-mass cores ($>27$ \Msun), only a single clump hosts high-mass cores at hub positions among the 17 hub-hosting clumps. 
Our observations imply that the hub-filament systems within massive clumps at very early evolutionary stages are not yet efficiently contributing to core accretion. This is also in agreement with the finding that MMCs are similarly distributed to lower-mass cores at early stages, as we have discussed so far.   
Line emission, such as N$_2$H$^+$, would trace more extended emission, which may reveal more filaments or hub-filament systems. 

However, half of high-mass cores are located within filaments, implying that high-mass cores may form by acquiring gas along filaments, adding further support to previous studies \citep[e.g.,][]{Henshaw14, Peretto14, Williams18, LuX18, Chen19, Sanhueza21, Zhou22}. 
Recently, \cite{Redaelli22} revealed an accretion flow seen in N$_2$H$^+$ ($J$=1--0) in a target of the ASHES sample, G014.492–00.139, and estimated an accretion rate of $2 \times 10^{-4}$ $M_\odot$\,yr$^{-1}$. Assuming that this gas accretion feeds a core for its free-fall time ($\sim$2$\times$10$^4$\,yr)
, the core can acquire an additional mass of 4\,$M_\odot$.
There are additional two MMCs in the whole 39 MMCs 
that can grow into high-mass cores ($\gtrsim27 M_\odot$) with this assumed accretion rate. However, we note that with either a longer accretion time (e.g., $1.5\times t_{ff}$) or higher SFE (e.g., 50\%), accretion through filaments would be substantial and key for the formation of high-mass stars. The study of gas dynamics around the whole population of ASHES cores would make clearer the role of accretion along filaments in the formation of high-mass stars. 

\subsection{Early Phase of High-mass Star Formation}
\label{sec:hmsf}
Using ALMA observations toward 70\,$\mu$m dark massive clumps, we have revealed the properties of hundreds of cores in the very early phases of high-mass star formation. 
The majority of the identified most massive cores have insufficient mass to form high-mass stars following the core accretion scenario. In addition, no high-mass prestellar cores were detected. 
We only found a weak correlation between the maximum core mass and the natal clump mass, in contrast to the correlation between the maximum stellar mass and the cluster's mass (i.e., Larson's relation Equation~\ref{equ:larson}), implying that the initial core mass does not have to be correlated with the final stellar mass \citep[e.g.,][]{Smith09, Pelkonen21}. 
These conditions found in the ASHES sample support clump-fed accretion scenarios, such as competitive accretion, global hierarchical collapse, and the inertial-inflow model. 

In spite of infall rate estimations being so far rare in IRDCs, the few examples available \citep[][]{Contreras18,Chen19,Redaelli22} suggest that cores with masses as those found in ASHES cores can grow significantly in mass in a core free-fall time \citep[as discussed in two ASHES clumps;][]{Contreras18,Redaelli22}. Some ASHES cores have the conditions to become massive as hot cores and form high-mass stars. 
We note that we cannot completely rule out the core accretion picture. However, if high-mass prestellar cores exist, we constrain their formation to later times, once the conditions for their formation may be adequate (e.g., a denser or warmer environment, or under the presence of a more turbulent medium or stronger magnetic field).  

The high-resolution ALMA observations also reveal that 45\% of the ASHES targets (18/39) have developed hub-filament systems. However, our analysis suggests that at the moment, hub-filament systems do not efficiently contribute to the formation of high-mas stars (picture that can change at later evolutionary stages).  
As discussed in Section~\ref{sec:hub}, only one clump  (2.5\%) host high-mass cores and eight (20\%) host their most massive cores at a hub-position. Such low probability implies that the high-mass cores or the most massive cores are not preferentially formed at hub-filament systems, but rather it implies that cores originally formed in hub-systems eventually evolved to become massive at later stages (Liu et al. submited to MNRAS). 

\section{Conclusion}
\label{sec:conclusion}
We have presented the whole ASHES (ALMA Survey of 70 $\mu$m dark High-mass clumps in Early Stages) survey that aims at characterizing the very early phase of high-mass star formation to constrain theoretical  models. 
The sample consists of thirty-nine massive clumps, the central, denser regions of 70 $\mu$m dark IRDCs. 
We have conducted ALMA observations that resolve the whole clumps, mosaicked with ten and three pointing by the 12 and 7 m array, respectively, at a final angular resolution of $\sim$1$\farcs$2. We have characterized the core physical properties and have analysed them in conjunction with the clump properties using the dust continuum emission. We have obtained the following conclusions:

\begin{enumerate}
    \item At $< 10^4$\,au scales, the dust continuum emission shows a diversity of morphologies in the thirty-nine clumps observed, presenting clumpy and filamentary structures, some of which host hub-filament systems. 
    
    \item Using the dendrograms algorithm applied to dust continuum emission, we identified cores from all thirty-nine clumps. The number of the identified cores in each clump ranges from 8 to 39 (median of 20); 839 in total.  

    \item We estimated core masses ranging from 0.05 $M_\odot$ to 81 $M_\odot$, with two orders of magnitude of dynamic range. More than half of cores are low-mass with $M_\mathrm{core}<1\,M_\odot$, and less than 1\% of cores are high-mass ($M_\mathrm{core}>27\,M_\odot$).  
    The identified cores have a size of several times 10$^3$ au and a volume density of 10$^5$--10$^7$\,cm$^{-3}$. About 10\% of cores have surface densities of $>1$\,g\,cm$^{-2}$. 
    
    \item The maximum core mass does not correlate with the clump mass, however, clump surface density is moderately correlated with the maximum core mass. 

    \item Our observations revealed that 35 out of 39 clumps host no high-mass cores that can form high-mass stars at this juncture assuming a star formation efficiency of 30\%. The lack of high-mass prestellar cores implies that high-mass cores do not exist or they form later on in the clump evolution once conditions for their formation are met. 

    \item Using the identified core positions and the minimum spanning tree method, we found weak evidence of mass segregation. Instead, cores are segregated by densities. Besides, there is no sign that the most massive cores (MMCs) are preferentially located near the clump center or at hub-filament systems. These findings indicate that there is no preferred location for high-mass core formation in such an early phase.
\end{enumerate}

\begin{acknowledgments}
K.M is financially supported by Grants-in-Aid for the Japan Society for the Promotion of Science (JSPS) Fellows (KAKENHI Number 22J21529), and also supported by FoPM, WINGS Program, the University of Tokyo. 
PS was partially supported by a Grant-in-Aid for Scientific Research (KAKENHI Number 22H01271) of JSPS. 
HB acknowledges support from the DFG in the Collaborative Research Center SFB 881 - Project-ID 138713538 - “The Milky Way System” (subproject B1). 
G.G. acknowledges support by the ANID BASAL project FB210003. 
Data analysis was in part carried out on the Multi-wavelength Data Analysis System operated by the Astronomy Data Center (ADC), National Astronomical Observatory of Japan. 
This paper uses the following ALMA data: ADS/JAO. ALMA No. 2015.1.01539.S, and 2016.1.01246.S. ALMA is a partnership of ESO (representing its member states), NSF (USA) and NINS (Japan), together with NRC (Canada), $MOST$ and ASIAA (Taiwan), and KASI (Republic of Korea), in cooperation with the Republic of Chile. The Joint ALMA Observatory is operated by ESO, AUI/NRAO, and NAOJ.
Data analysis was in part carried out on the open-use data analysis computer system at the Astronomy Data Center (ADC) of the National Astronomical Observatory of Japan.
\facility {ALMA} 
\software{CASA (v4.5.3, 4.6, 4.7, 5.4, 5.6; \citealt[][]{McMullin07})}
\end{acknowledgments}

\bibliography{reference}
\bibliographystyle{aasjournal}

\appendix
\section{Identification of Filaments}
\label{sec:Appendix_A}
One method to identify filamentary structures is the publicly available filament finding package $\mathtt{FilFinder}$ \citep[][]{Koch15}. 
Following the guideline, we firstly fit a log-normal distribution to the brightness data (continuum emission in our case) to identify the mean ($\mu$) and standard deviation ($\sigma$) of the log-intensity, and flatten the image using an arctangent transform ($I' = I_0 \mathrm{arctan}(I/I_0)$), where the normalization is $I_0 \equiv \mathrm{exp}(\mu + 2\sigma)$. 
This aims to suppress significantly brighter objects than filamentary structures such as dense cores. 
Next, we smooth the flattened image with a Gaussian to minimize variations within the filamentary structures. We create a mask of filamentary structure using the smoothed image. 
The width of the element used for the adaptive threshold mask is set by a parameter of $\mathtt{adapt\,thresh}$. 
In this step, pixels, where the intensity is much greater than the medium of the neighborhood, are extracted. Additionally, the global threshold mask is combined as a final mask, which exclude pixels below the noise level in the image. 
The parameters used for making masks are $\mathtt{smooth\,size}$ of 1.5 times beam-width, $\mathtt{adapt\,thresh}$ of 2.5 times beam-width, $\mathtt{glob\,thresh}$ of 2.5$\sigma$, and $\mathtt{size\,thresh}$ of 10 times beam-area, where $\sigma$ is the RMS noise level of each continuum image (input data), beam-width is the geometrical mean of the beam size (Table~\ref{tab:obs-ashes}), and beam-area is the total pixel number in the synthesized beam. They are selected by checking outputs. 

Structures within the mask are reduced to a skeleton using a medial axis transform. 
The algorithm drives the shortest path between each pair of endpoints and calculates the positions of the medial axis, a single-pixel width, for a skeleton.
Thus, filamentary structures are identified as skeletons. 
The geometrical cleaning is applied to remove isolated nodes and very short components. 
We impose a lower length limit of 10 times beam-width for each skeleton, $\mathtt{nbeam lengths}$ in $\mathtt{FilFinder}$ method and, and 5 times beam-width for each branch ($\mathtt{branch nbeam lengths}$). 
The pruning is tried 300 times in total. 
We adopted the same parameters for all regions. The extracted skeletons are drawn in Figure~\ref{fig:fil_G24} as orange lines. 
It should be noted that output skeletons depend on the setting parameters, but the main conclusion is insensitive to them. 


\begin{figure*}
    \gridline{\fig{G010.99_IR.pdf}{0.97\textwidth}{}}\vspace{-1em}
    \gridline{\fig{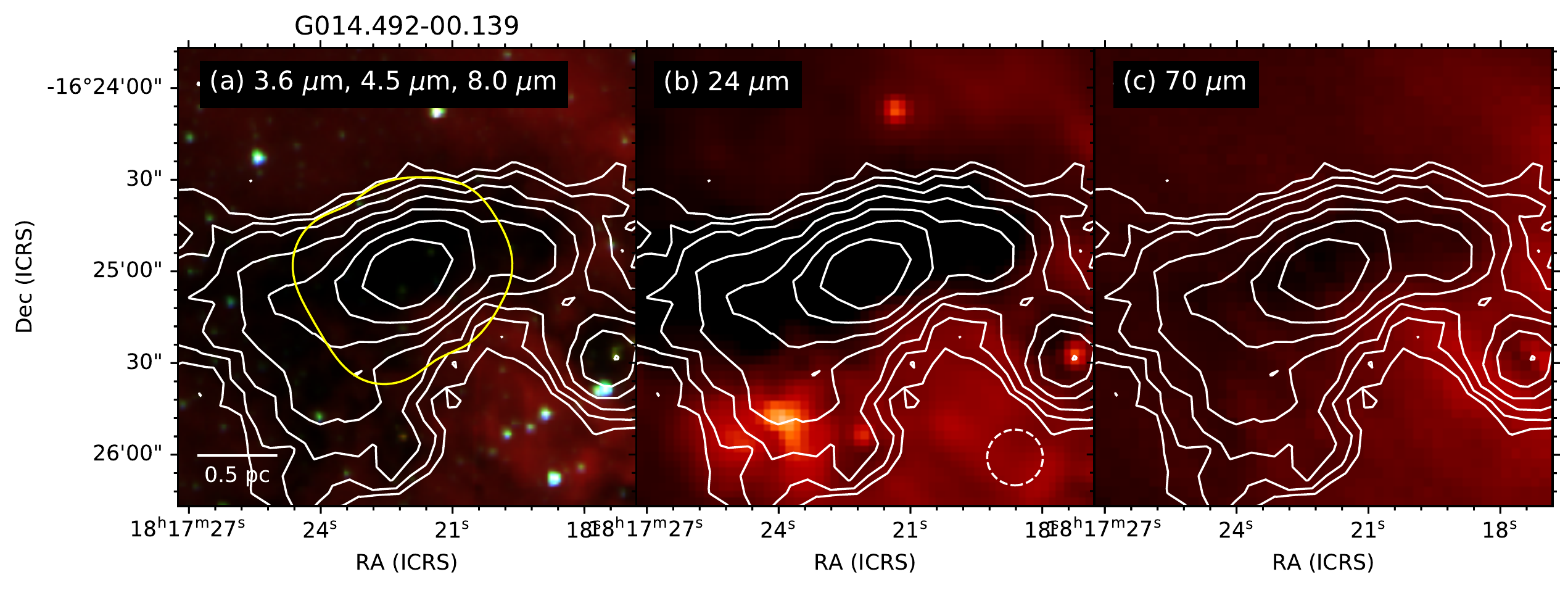}{0.97\textwidth}{}}\vspace{-1em}
    \gridline{\fig{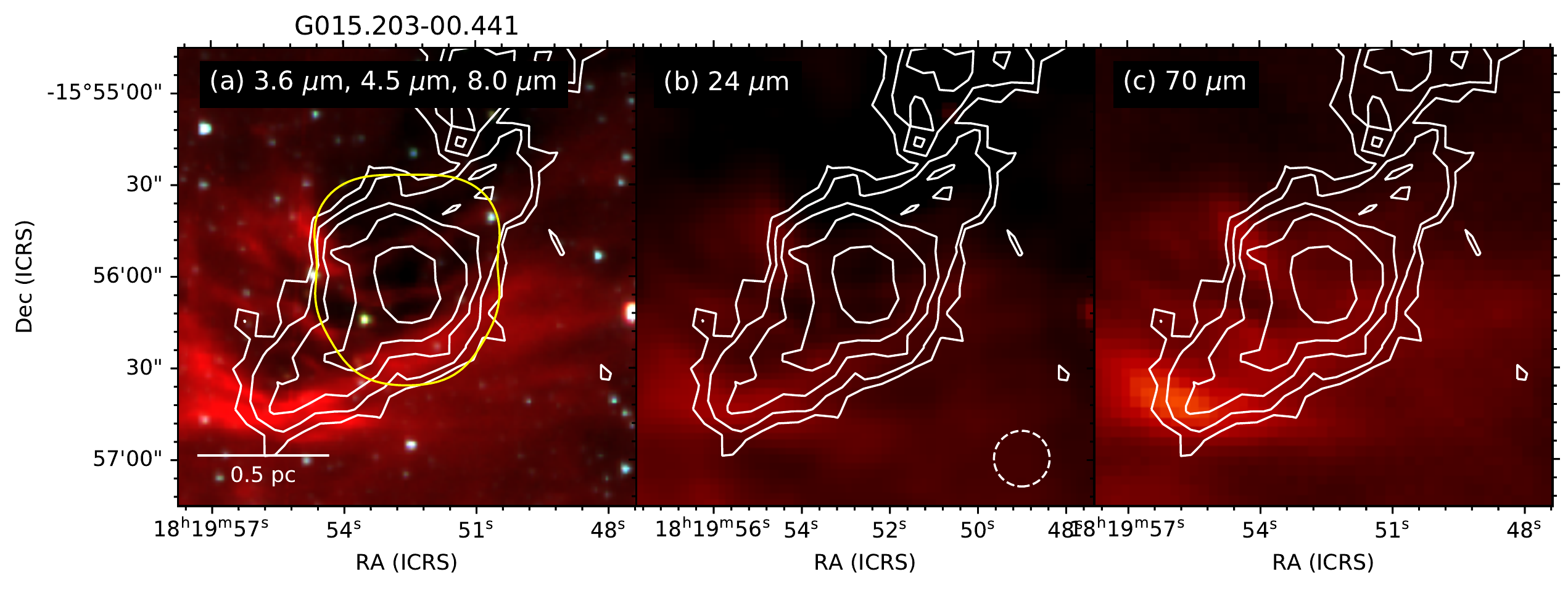}{0.97\textwidth}{}}\vspace{-1em}
    \caption{Same as Figure~\ref{fig:IR_G10} except for contour levels for the 870 $\mu$m dust continuum emission, which are (3, 5, 7, 9, 11, 15, 19, 23, and 27)$\times \sigma$ with $\sigma=83$ mJy\,beam$^{-1}$ for G014.492--00.139 and (3, 5, 7, 9, 11, 14 and 16)$\times \sigma$ with $\sigma=80$ mJy\,beam$^{-1}$ for G015.203--00.441.}
    \label{fig:Appendix_IR_1}
\end{figure*}

\begin{figure*}
\gridline{\fig{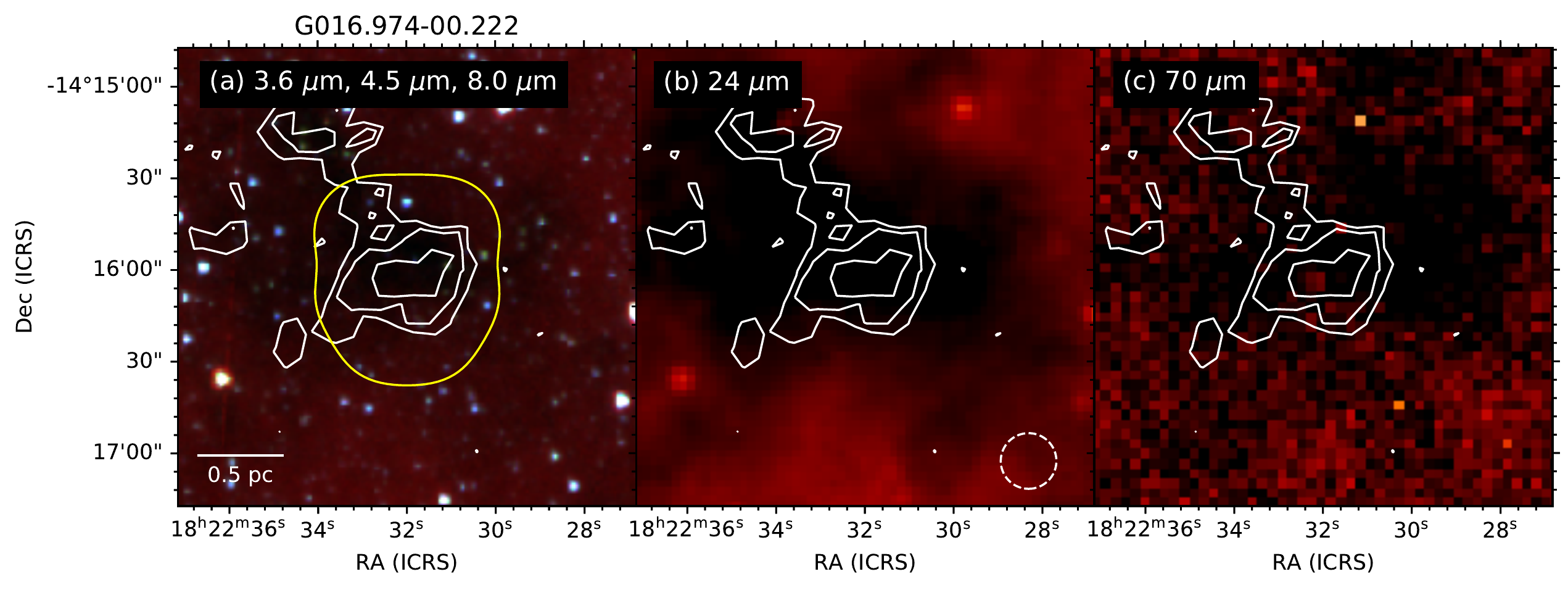}{0.97\textwidth}{}}\vspace{-1em}
\gridline{\fig{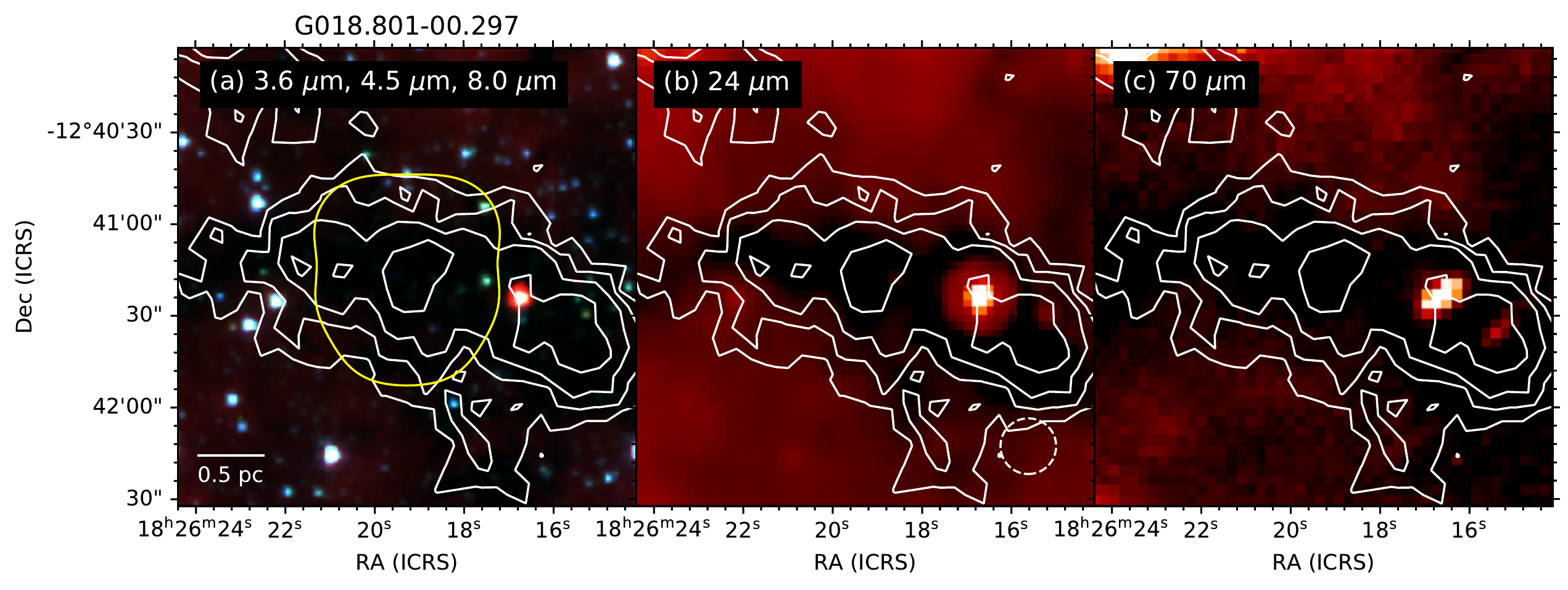}{0.97\textwidth}{}}\vspace{-1em}
\gridline{\fig{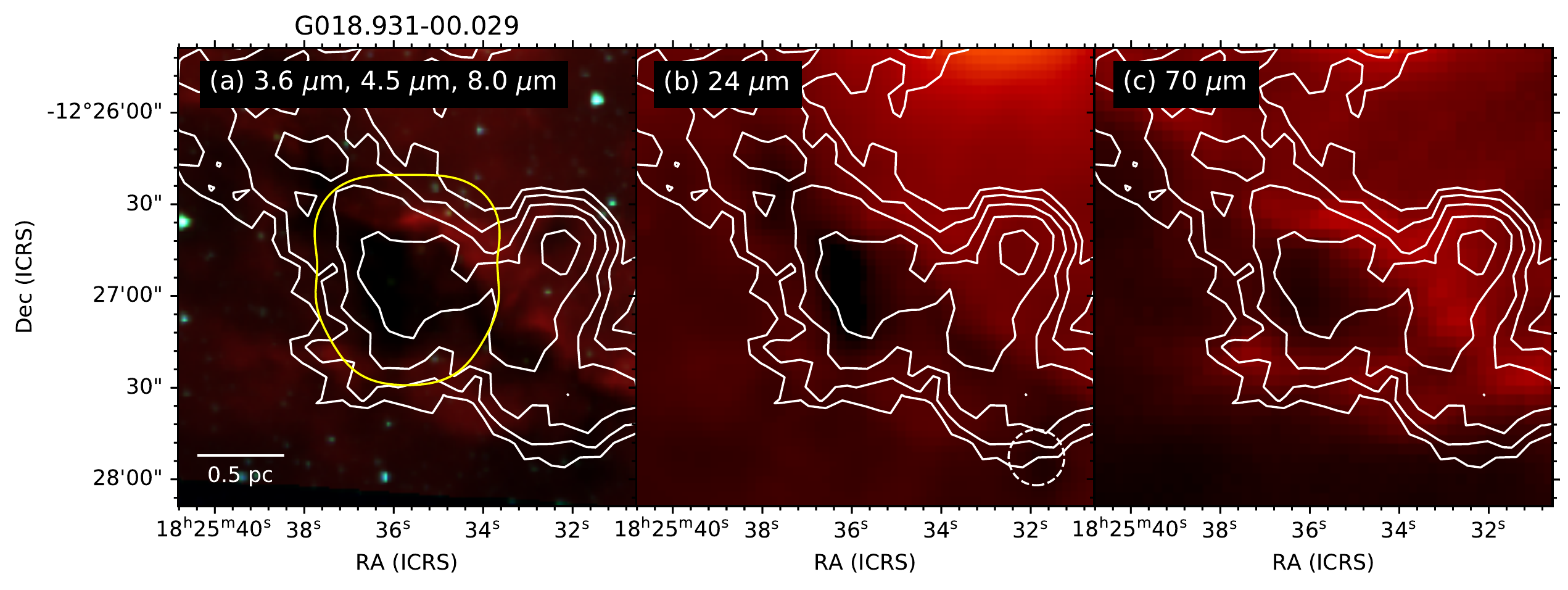}{0.97\textwidth}{}}\vspace{-1em}
\caption{Same as Figure~\ref{fig:IR_G10} except for contour levels for the 870 $\mu$m dust continuum emission, which are (3, 4, 5, 6 and 7)$\times \sigma$ with $\sigma=56$ mJy\,beam$^{-1}$ for G016.974--00.222, (3, 5, 7, 9, 11, 13, 15 and 20)$\times \sigma$ with $\sigma=77$ mJy\,beam$^{-1}$ for G018.801--00.297, and (3, 5, 7, 9, 10 and 13)$\times \sigma$ with $\sigma=77$ mJy\,beam$^{-1}$ for G018.931--00.029.}
\end{figure*}

\begin{figure*}
\gridline{\fig{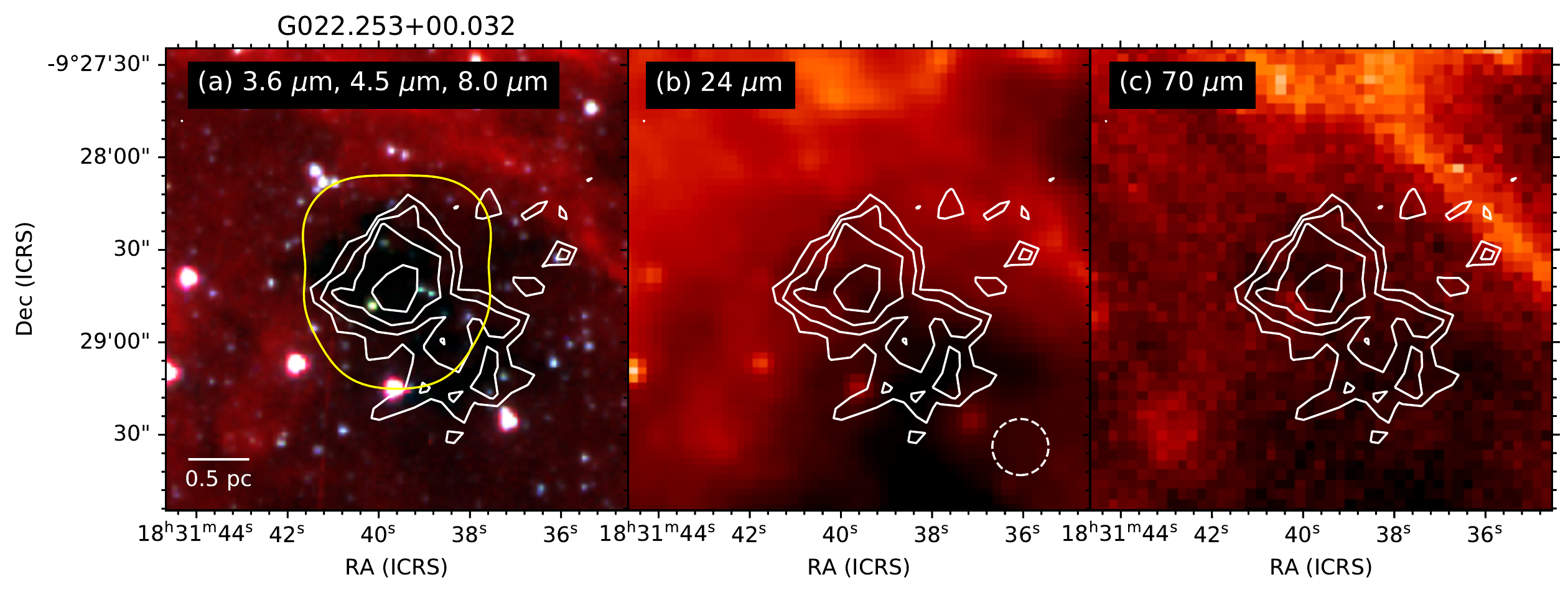}{0.97\textwidth}{}}\vspace{-2em}
\gridline{\fig{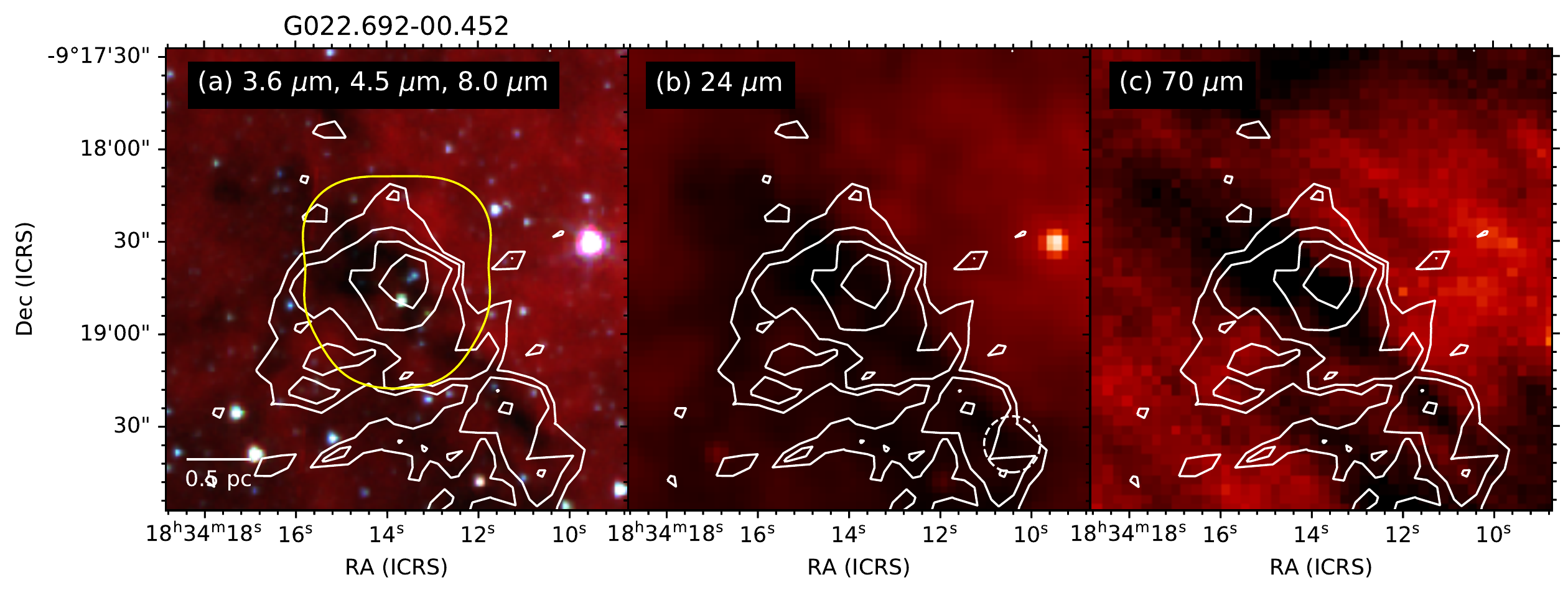}{0.97\textwidth}{}}\vspace{-2em}
\gridline{\fig{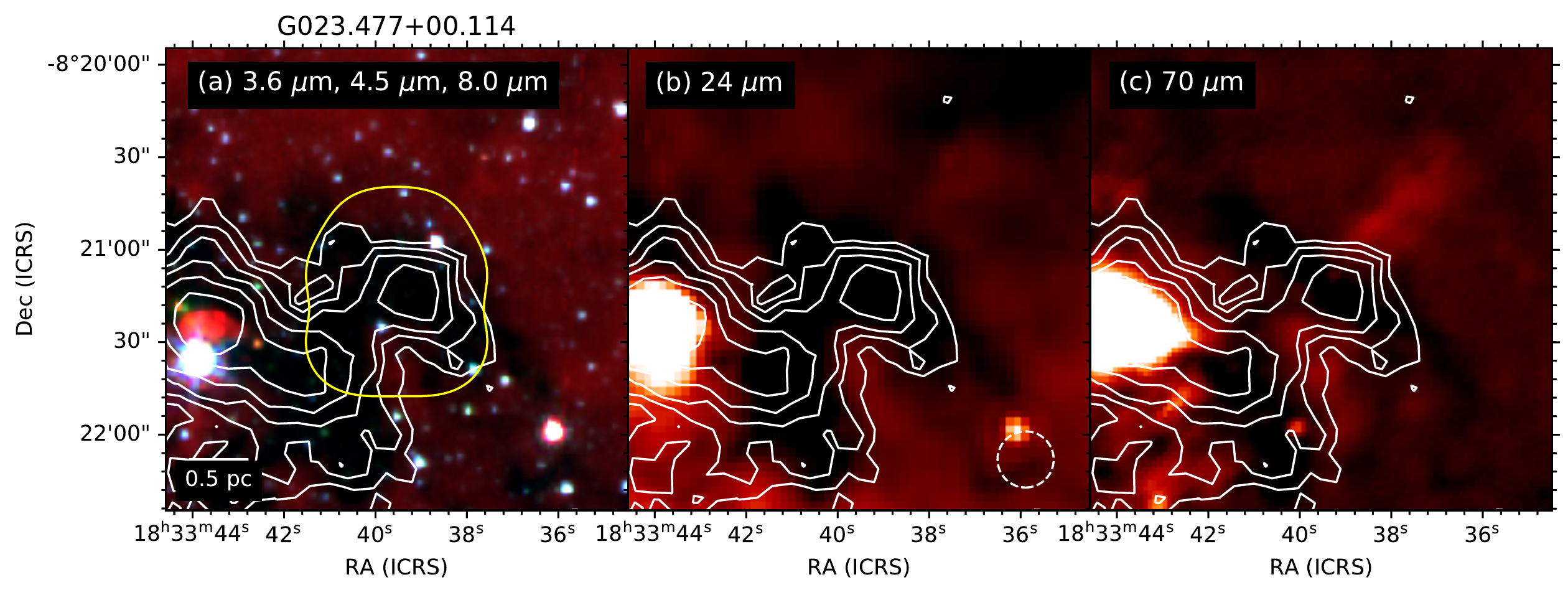}{0.97\textwidth}{}}\vspace{-2em}
\caption{Same as Figure~\ref{fig:IR_G10} except for contour levels for the 870 $\mu$m dust continuum emission, which are (3, 5, 7, 9, 11 and 13)$\times \sigma$ with $\sigma=53$ mJy\,beam$^{-1}$ for G022.253+00.032, with $\sigma=61$ mJy\,beam$^{-1}$ for G022.692--00.452, and $\sigma=86$ mJy\,beam$^{-1}$ for G023.477+00.114.}
\vspace{-2pt}
\end{figure*}

\begin{figure*}
\gridline{\fig{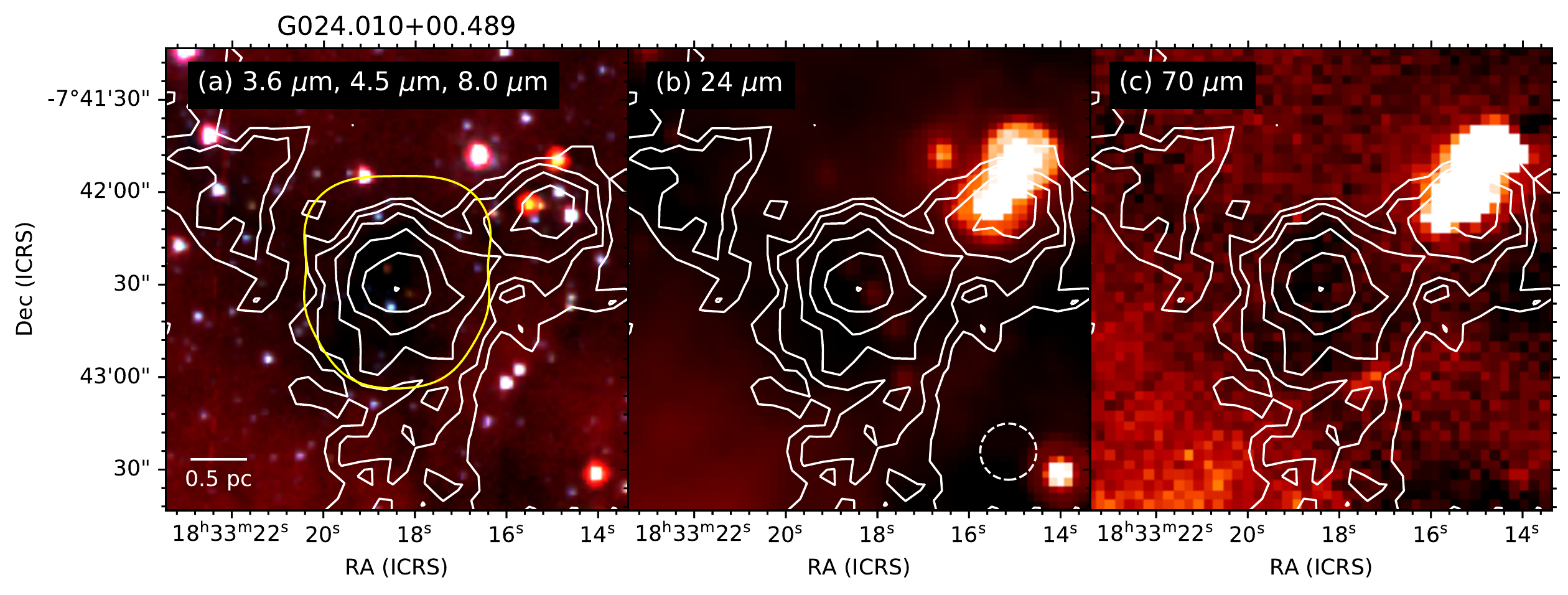}{0.97\textwidth}{}}\vspace{-2em}
\gridline{\fig{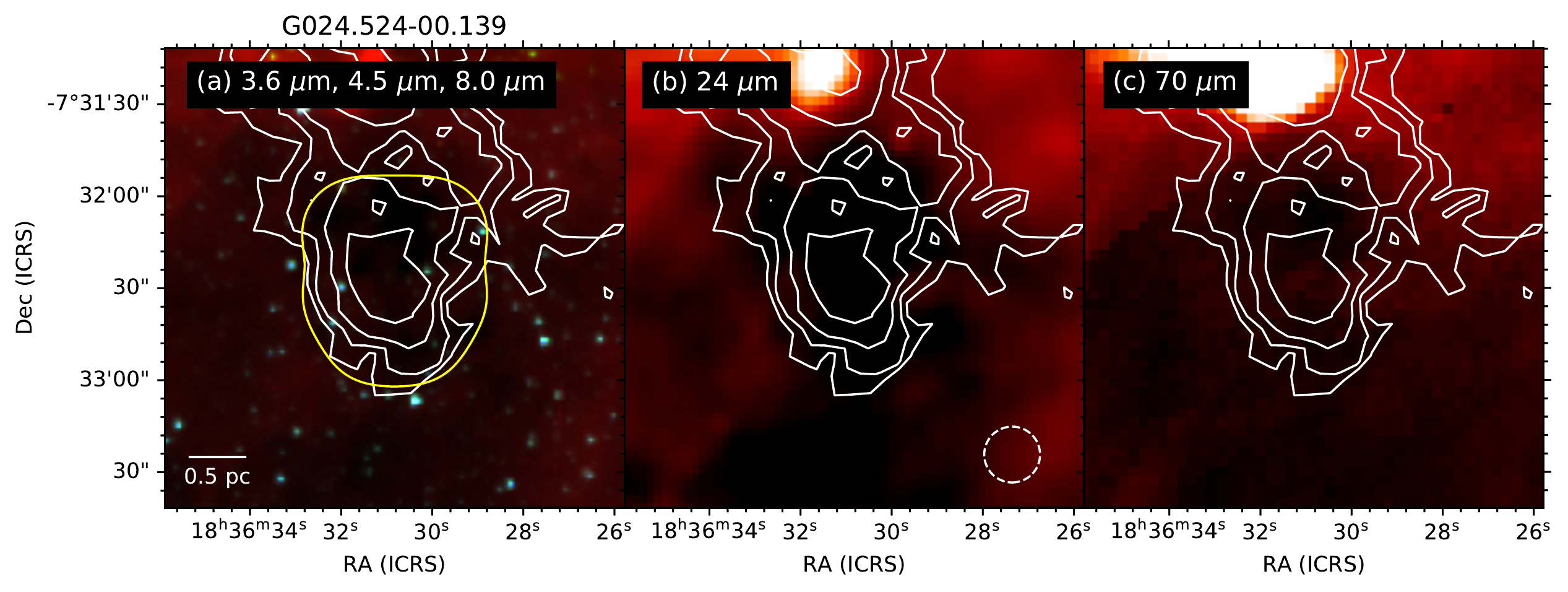}{0.97\textwidth}{}}\vspace{-2em}
\gridline{\fig{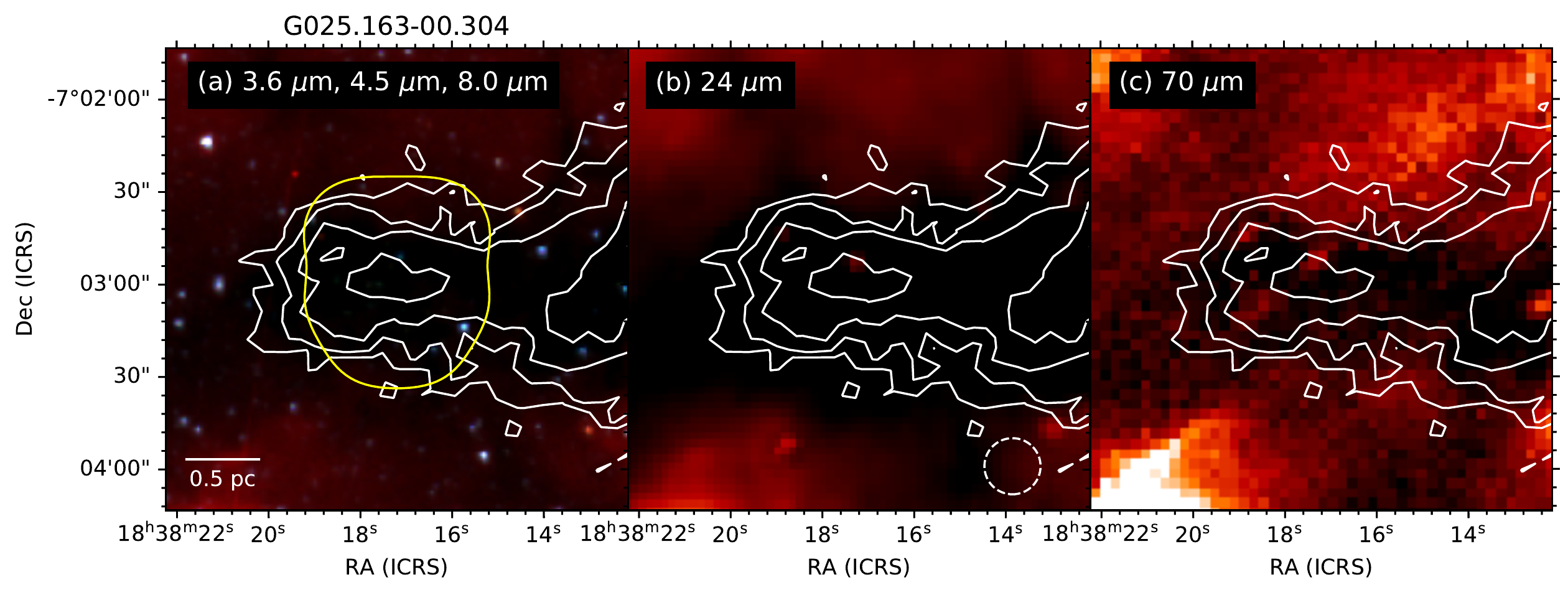}{0.97\textwidth}{}}\vspace{-2em}
\caption{Same as Figure~\ref{fig:IR_G10} except for contour levels for the 870 $\mu$m dust continuum emission, which are (3, 5, 7, 9, 11 and 13)$\times \sigma$ with $\sigma=56$ mJy\,beam$^{-1}$ for for G024.010+00.489, $\sigma=72$ mJy\,beam$^{-1}$ for G024.524--00.139, and $\sigma=92$ mJy\,beam$^{-1}$ for G025.163--00.304.}
\end{figure*}

\begin{figure*}
\gridline{\fig{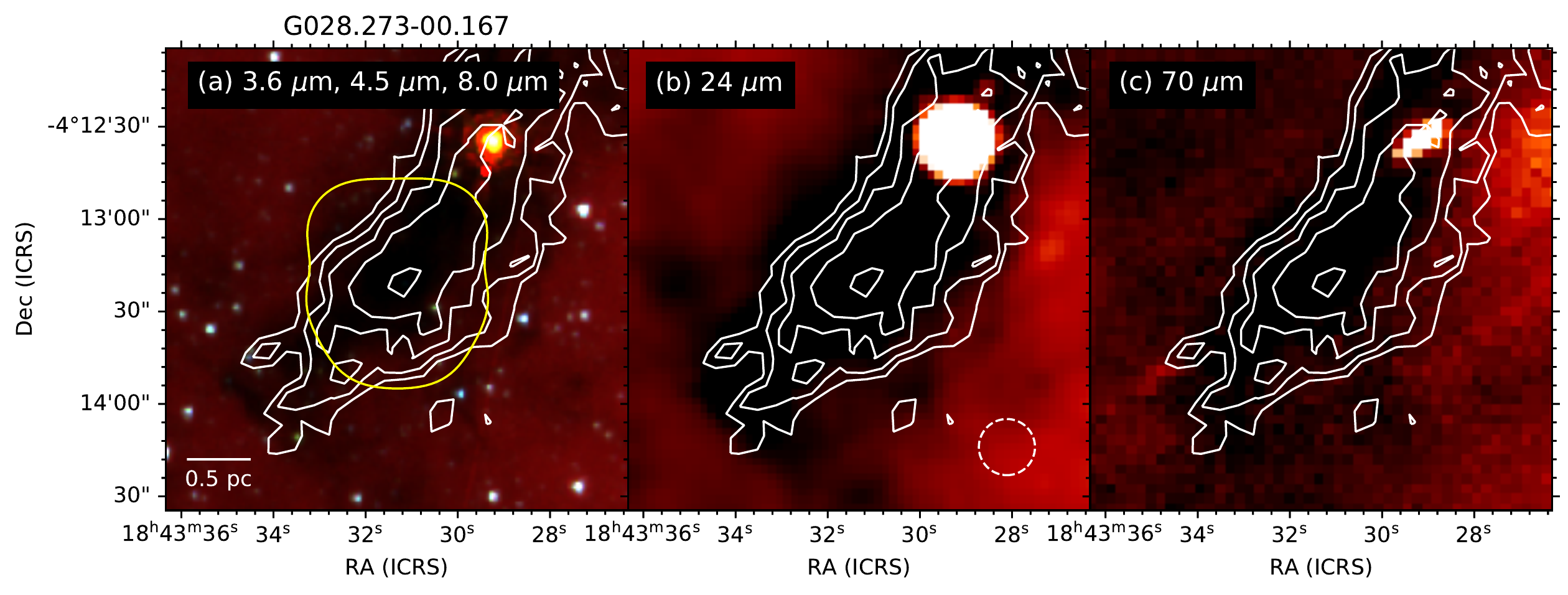}{0.97\textwidth}{}}\vspace{-1em}
\gridline{\fig{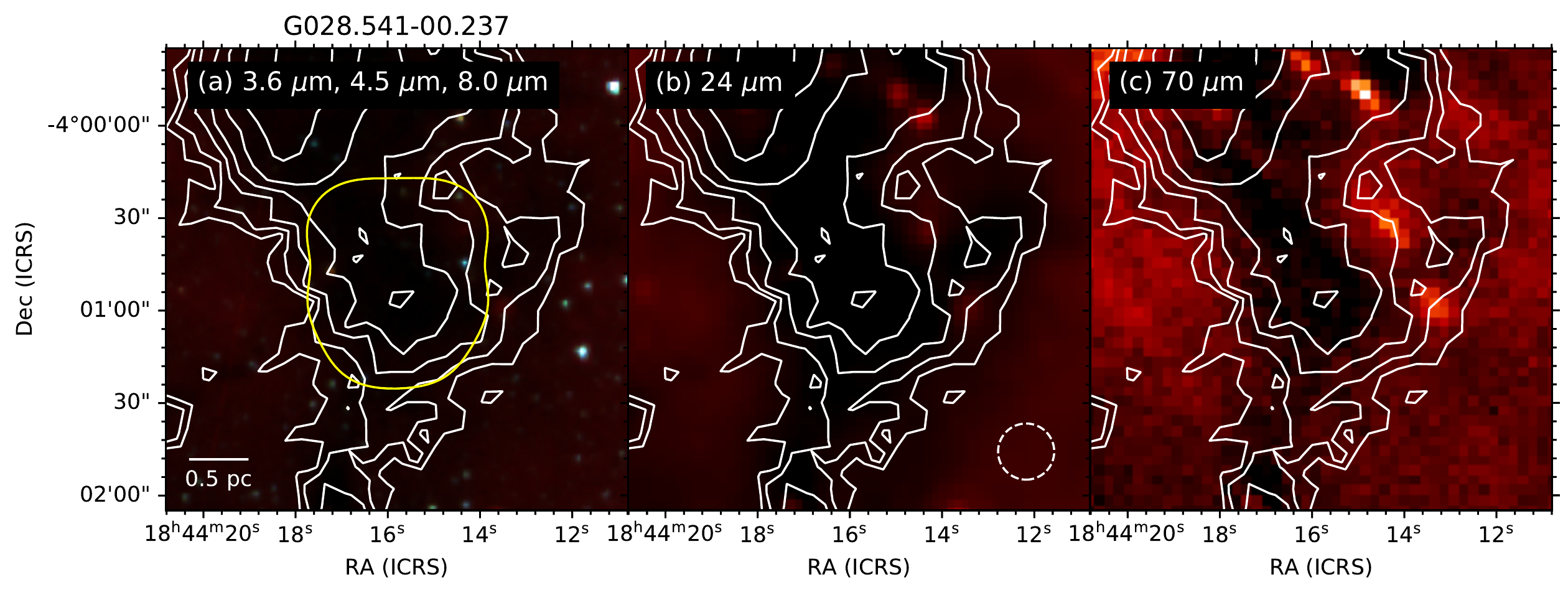}{0.97\textwidth}{}}\vspace{-1em}
\gridline{\fig{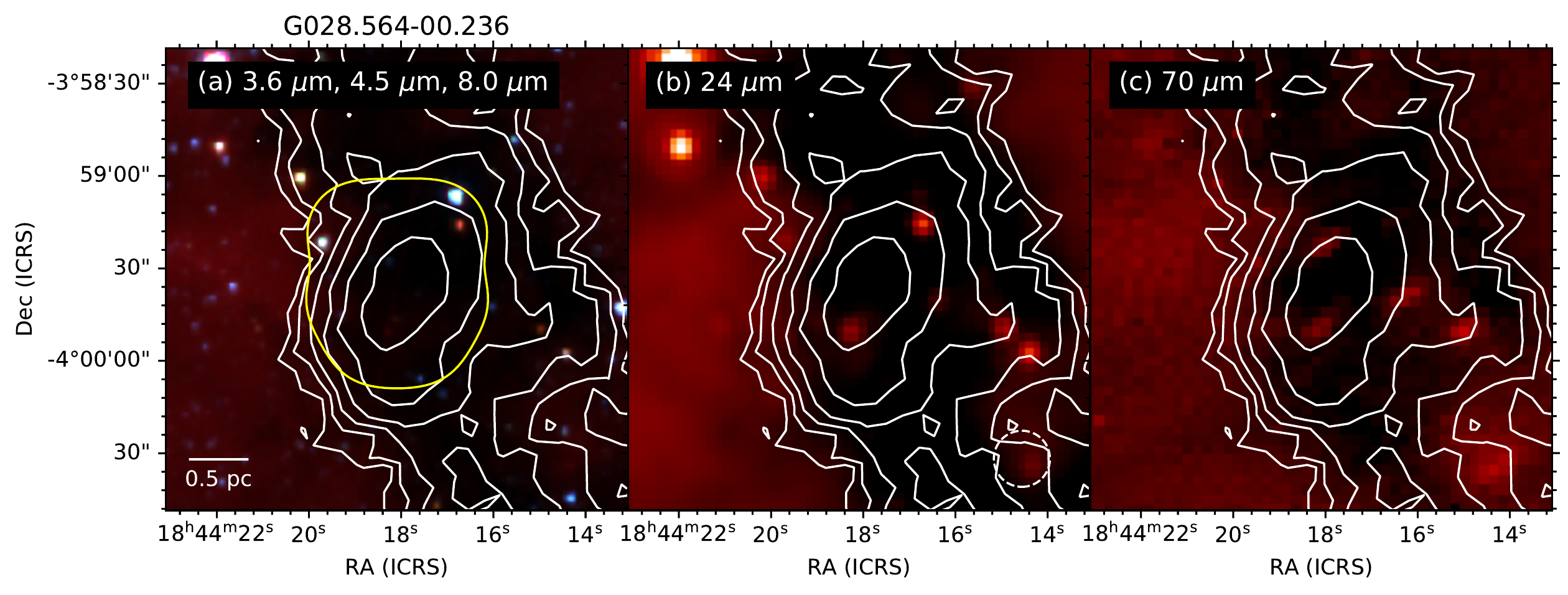}{0.97\textwidth}{}}\vspace{-1em}
\caption{Same as Figure~\ref{fig:IR_G10} except for contour levels for the 870 $\mu$m dust continuum emission, which are (3, 5, 7, 9, 10 and 12)$\times \sigma$ with $\sigma=76$ mJy\,beam$^{-1}$ for G028.273--00.167 and (3, 5, 7, 9, 12, 15, 18 and 21)$\times \sigma$ with $\sigma=88$ mJy\,beam$^{-1}$ for G028.541--00.237 and $\sigma=111$ mJy\,beam$^{-1}$ for G028.564--00.236.}
\vspace{-5pt}
\end{figure*}

\begin{figure*}
\gridline{\fig{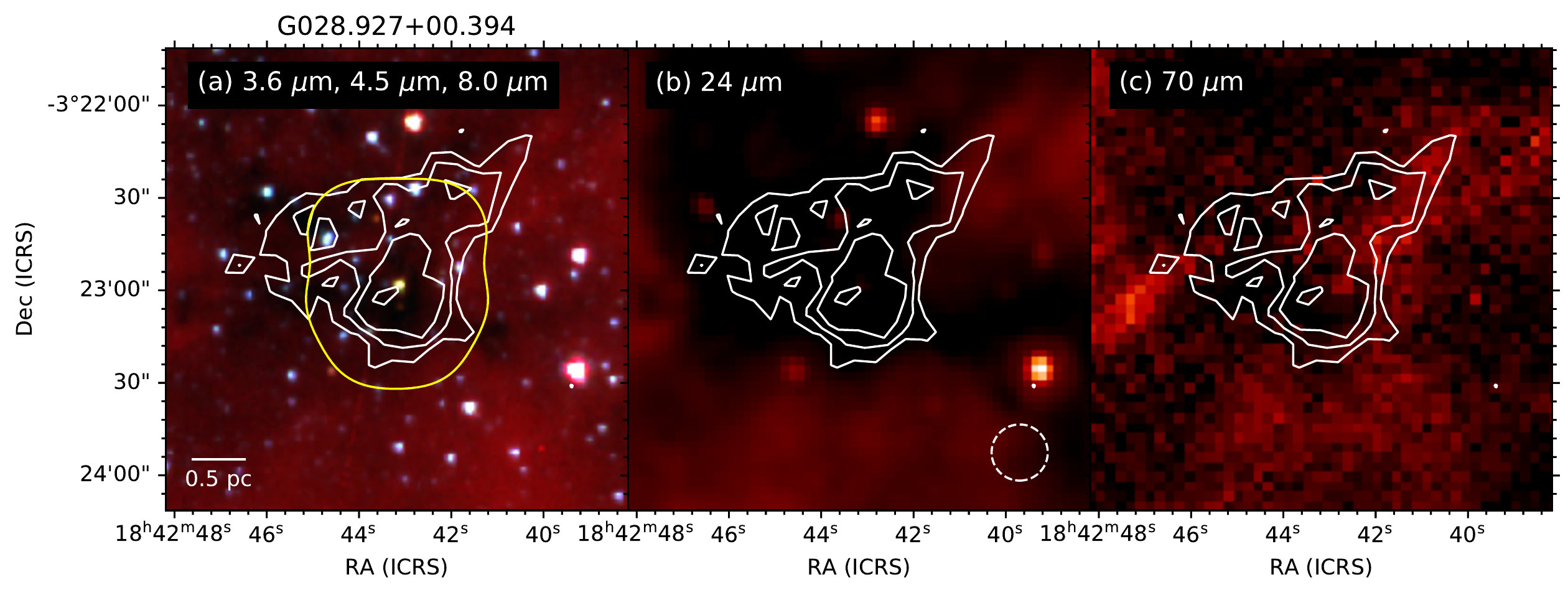}{0.97\textwidth}{}}\vspace{-1em}
\gridline{\fig{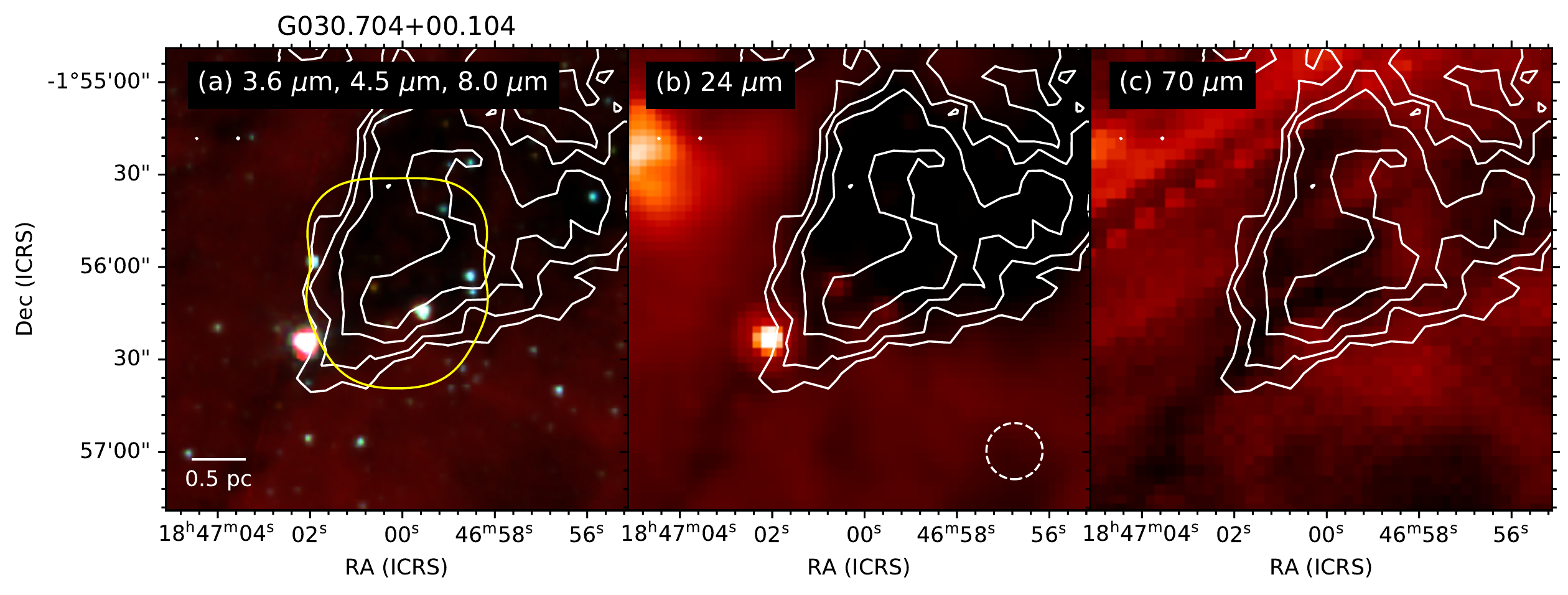}{0.97\textwidth}{}}\vspace{-1em}
\gridline{\fig{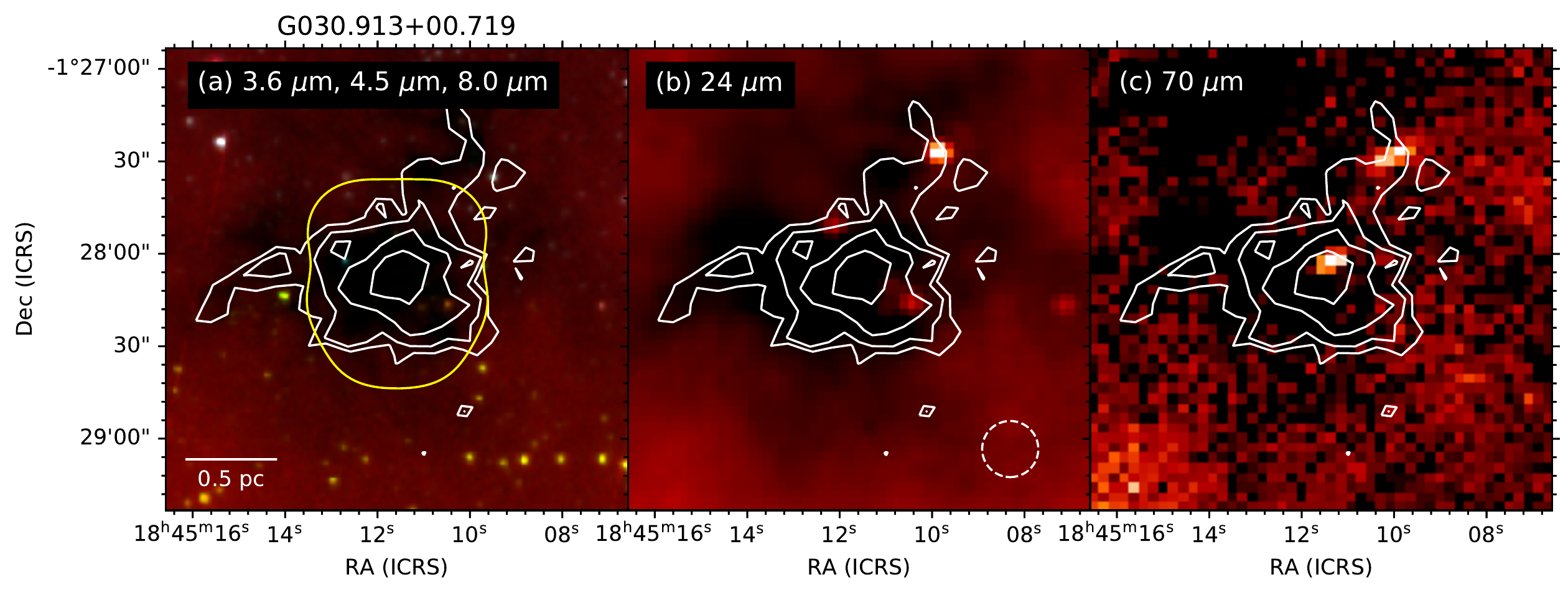}{0.97\textwidth}{}}\vspace{-1em}
\caption{Same as Figure~\ref{fig:IR_G10} except for contour levels for the 870 $\mu$m dust continuum emission, which are (3, 5, 7, 8 and 9)$\times \sigma$ with $\sigma=65$ mJy\,beam$^{-1}$ for G028.927+00.394 and (3, 5, 7, 8, 9 and 10)$\times \sigma$ with $\sigma=75$ mJy\,beam$^{-1}$ for G030.704+00.104 and $\sigma=63$ mJy\,beam$^{-1}$ for G030.913+00.7519.}
\vspace{-5pt}
\end{figure*}

\begin{figure*}
\gridline{\fig{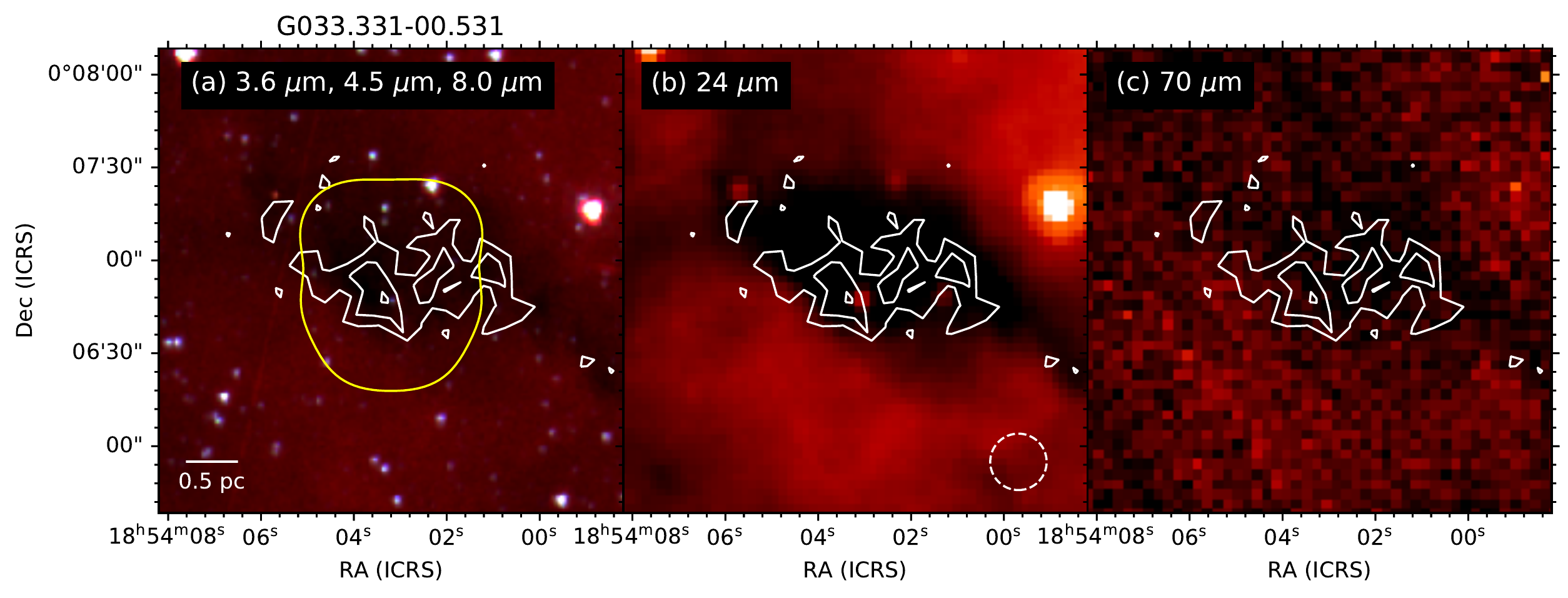}{0.97\textwidth}{}}\vspace{-2em}
\gridline{\fig{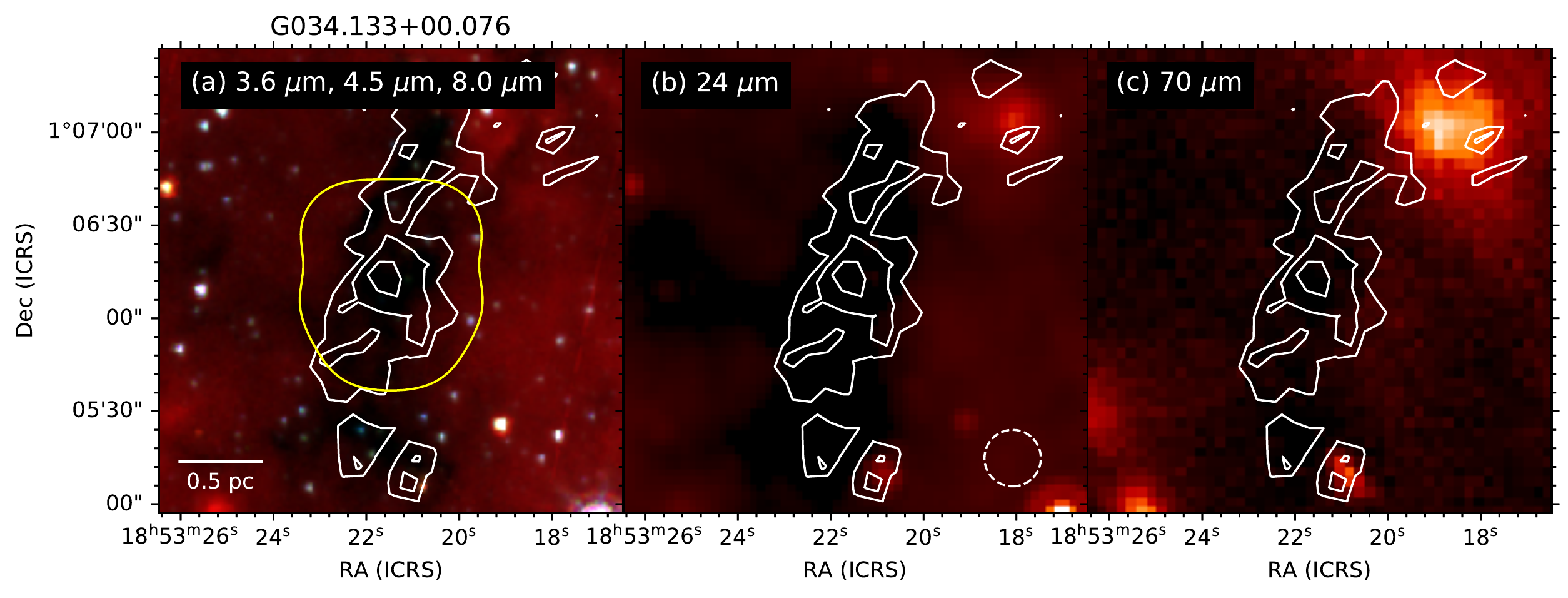}{0.97\textwidth}{}}\vspace{-2em}
\gridline{\fig{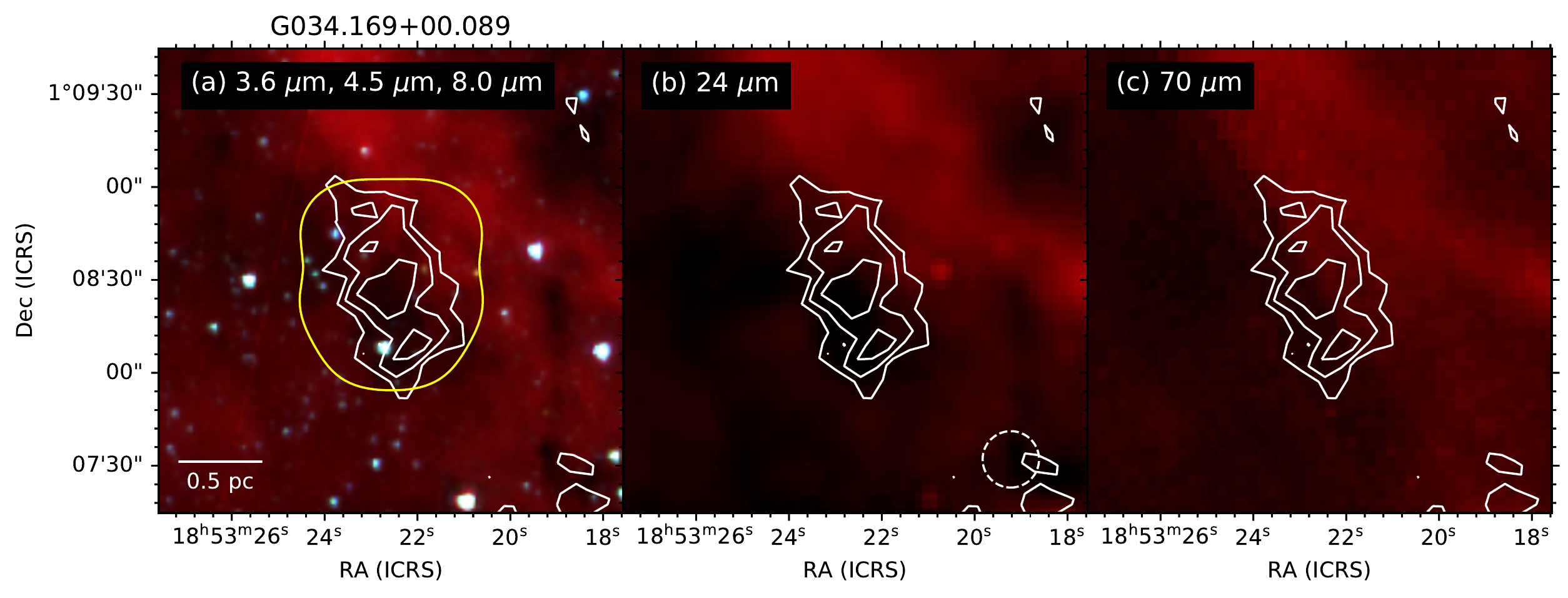}{0.97\textwidth}{}}\vspace{-2em}
\caption{Same as Figure~\ref{fig:IR_G10} except for contour levels for the 870 $\mu$m dust continuum emission, which are (2, 3, 4 and 5)$\times \sigma$ with $\sigma=87$ mJy\,beam$^{-1}$ for G033.331--00.531, (2, 3, 4, 5 and 6)$\times \sigma$ with $\sigma=68$ mJy\,beam$^{-1}$ for G034.133+00.076, and (2, 4, 6 and 8)$\times \sigma$ with $\sigma=65$ mJy\,beam$^{-1}$ for G034.169+00.08.}
\vspace{-7pt}
\end{figure*}

\begin{figure*}
\gridline{\fig{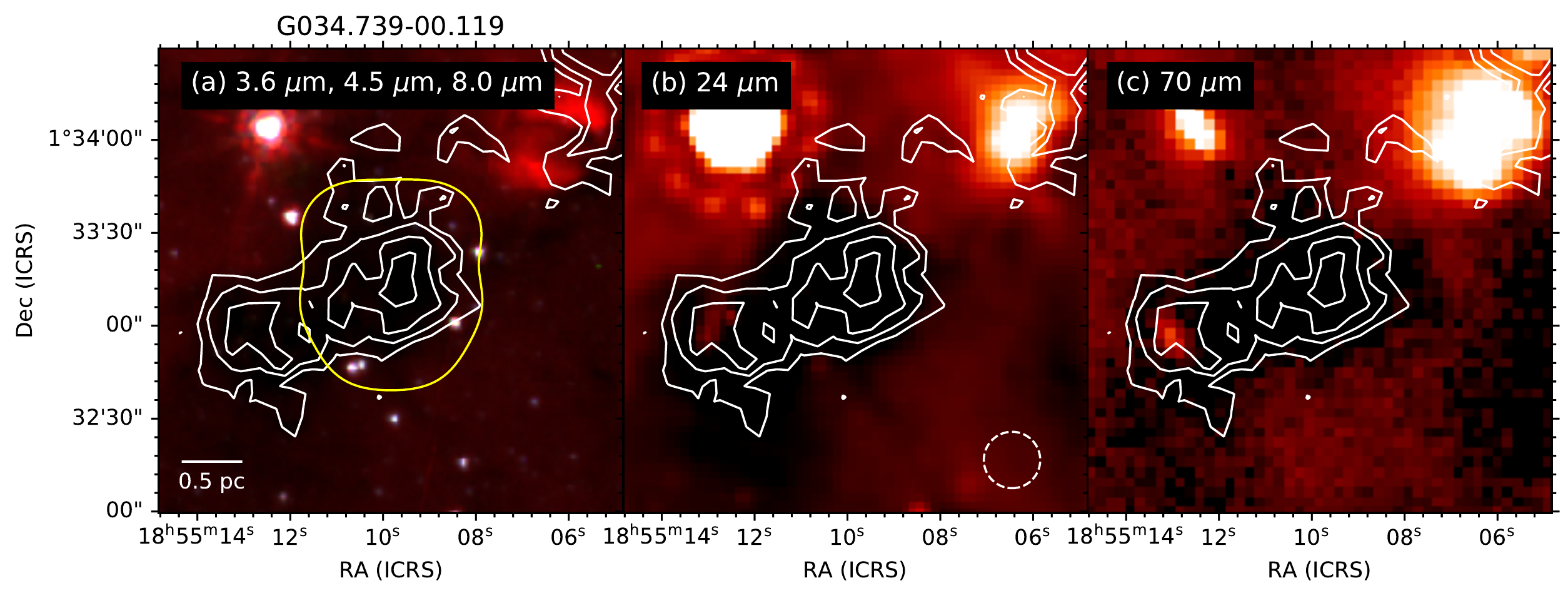}{0.97\textwidth}{}}\vspace{-1em}
\gridline{\fig{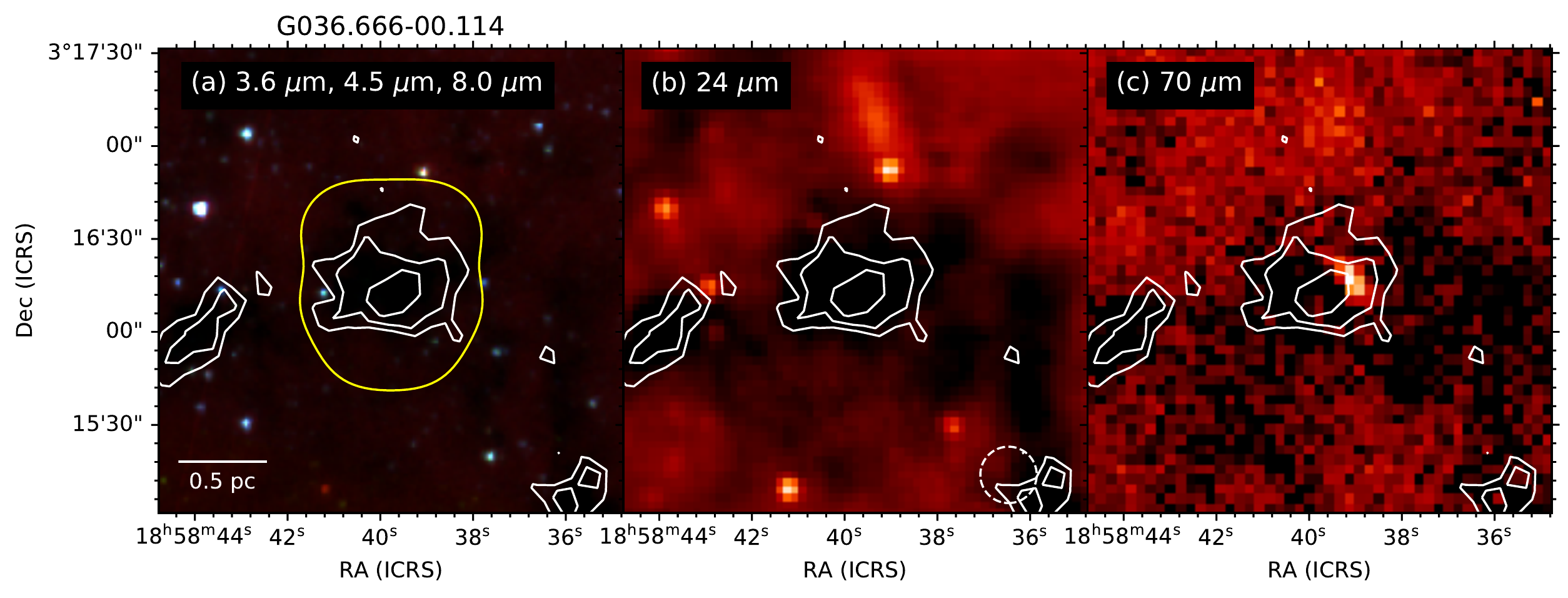}{0.97\textwidth}{}}\vspace{-1em}
\gridline{\fig{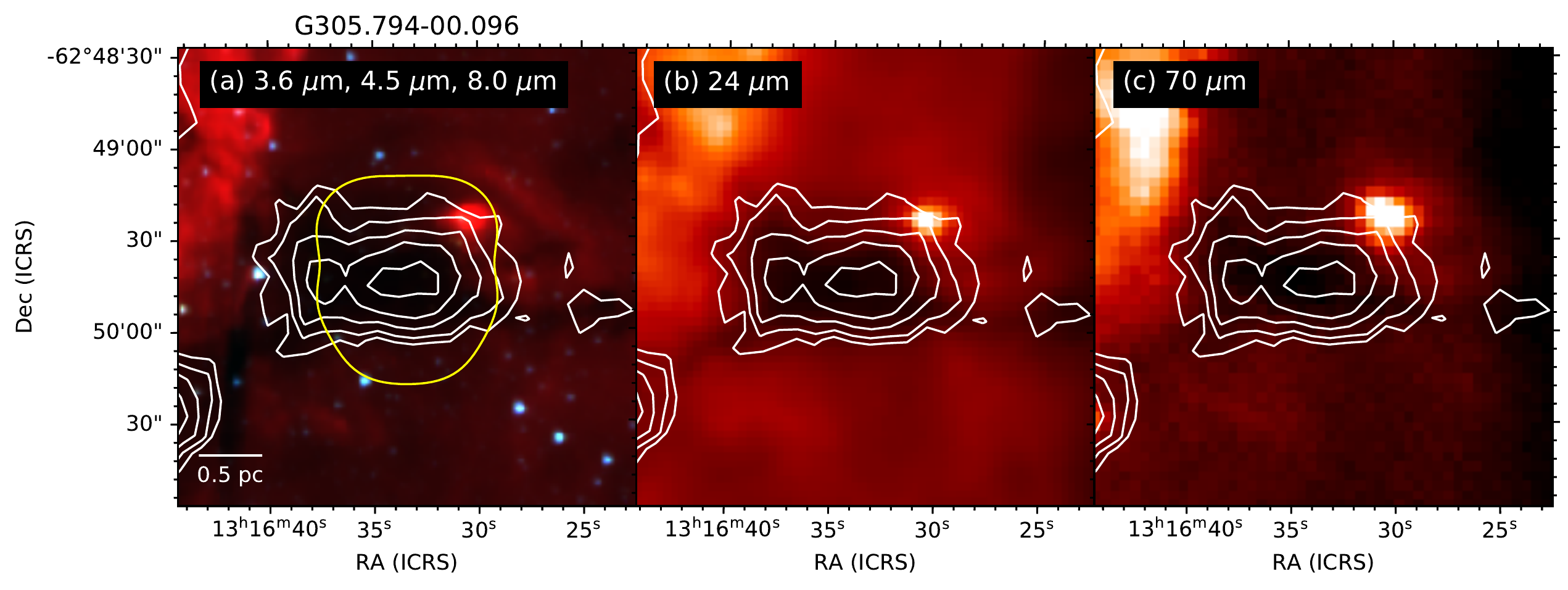}{0.97\textwidth}{}}\vspace{-1em}
\caption{Same as Figure~\ref{fig:IR_G10} except for contour levels for the 870 $\mu$m dust continuum emission, which are (3, 5, 7, 8, 9 and 11)$\times \sigma$ with $\sigma=76$ mJy\,beam$^{-1}$ for G034.739--00.119, (3, 4, 5 and 6)$\times \sigma$ with $\sigma=75$ mJy\,beam$^{-1}$ for G036.666--00.114, and (3, 5, 7, 9, 11 and 13)$\times \sigma$ with $\sigma=107$ mJy\,beam$^{-1}$ for G305.794--00.096.}
\vspace{-5pt}
\end{figure*}

\begin{figure*}
\gridline{\fig{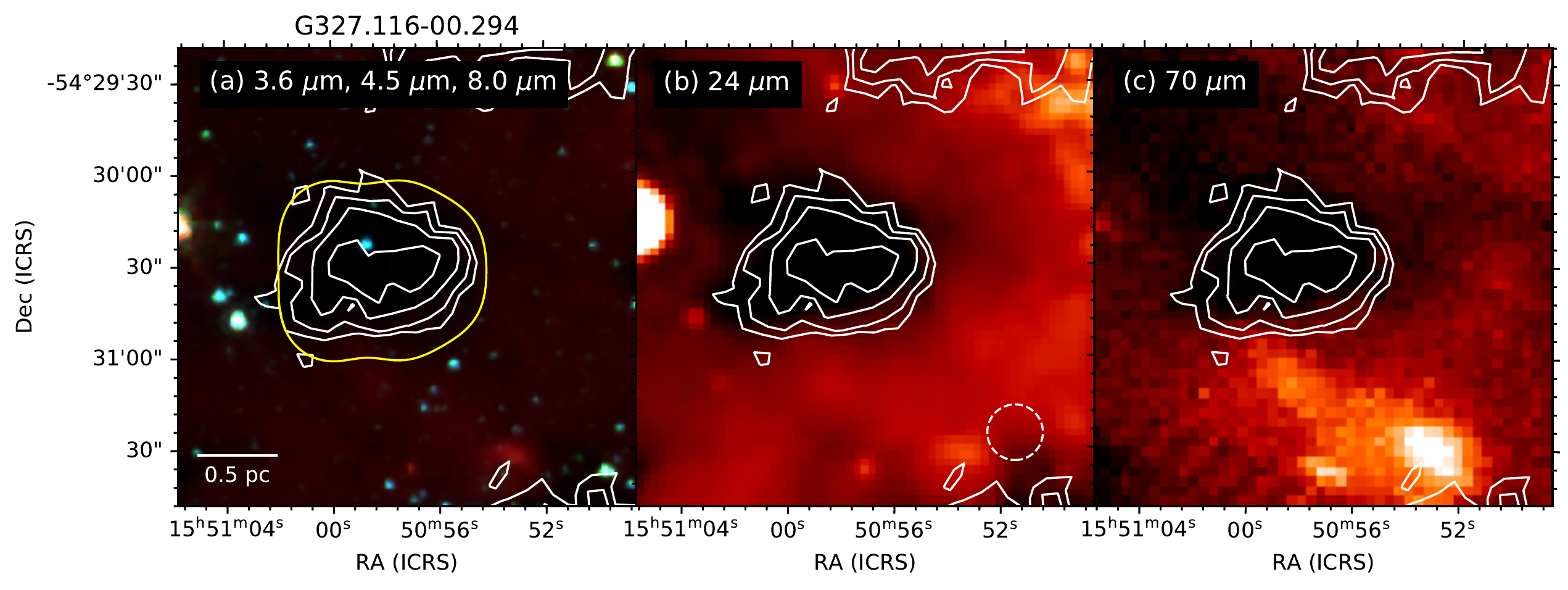}{0.97\textwidth}{}}\vspace{-1em}
\gridline{\fig{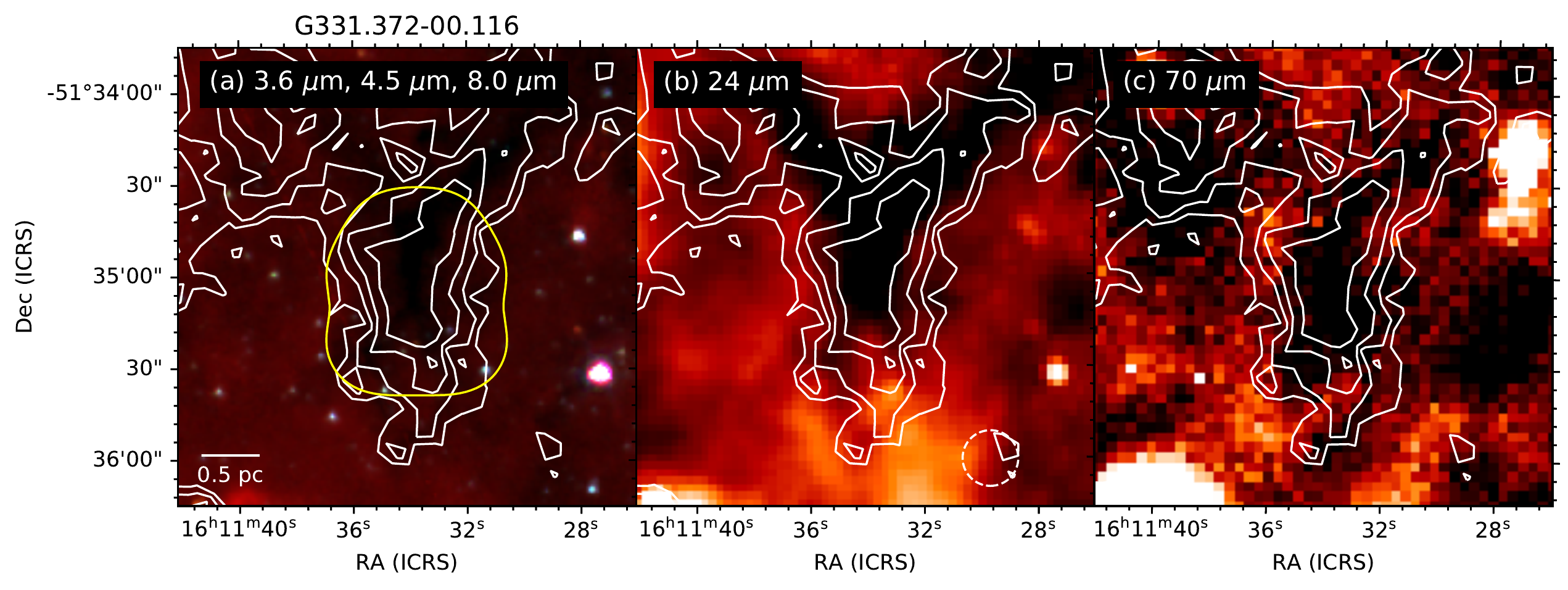}{0.97\textwidth}{}}\vspace{-1em}
\gridline{\fig{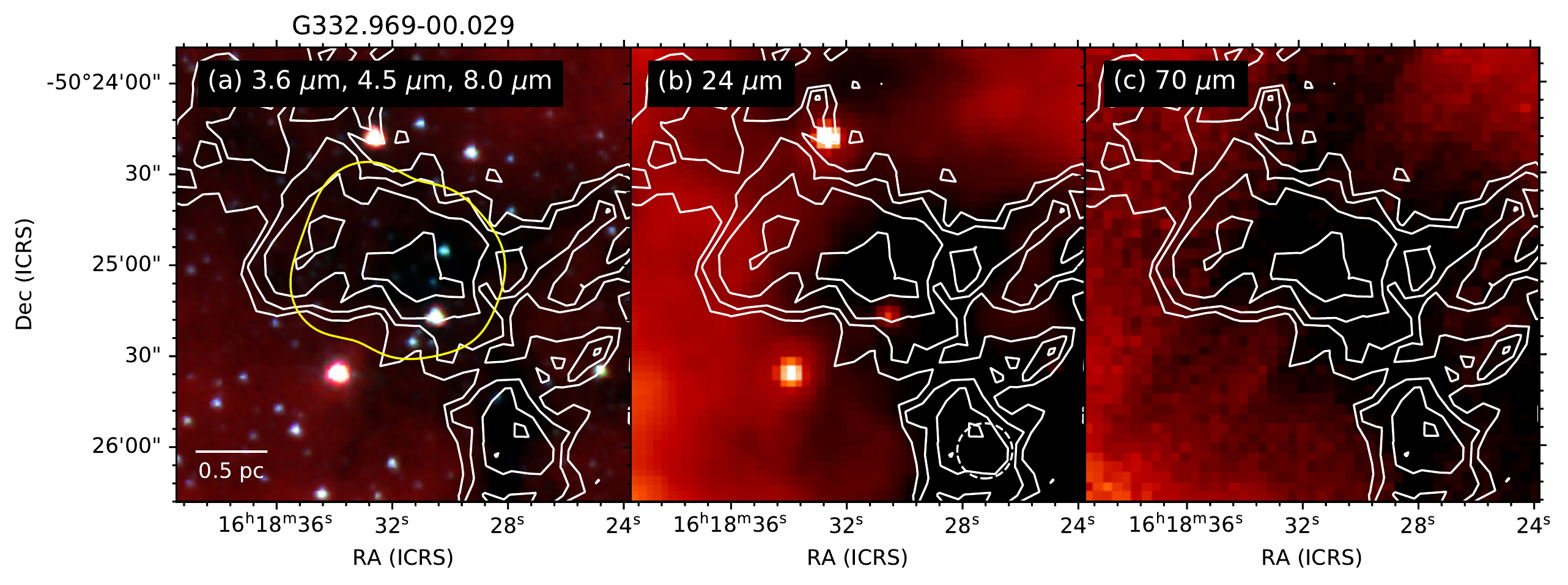}{0.97\textwidth}{}}\vspace{-1em}
\caption{Same as Figure~\ref{fig:IR_G10} except for contour levels for the 870 $\mu$m dust continuum emission, which are (3, 5, 7, 8, 9, 11, 15 and 19)$\times \sigma$ with $\sigma=71$ mJy\,beam$^{-1}$ for G327.116--00.294 and (3, 5, 7, 8, 9, 11, 15 and 19)$\times \sigma$ with $\sigma=56$ mJy\,beam$^{-1}$ for G331.372--00.116 and $\sigma=47$ mJy\,beam$^{-1}$ for G332.969--00.029.}
\end{figure*}

\begin{figure*}
\gridline{\fig{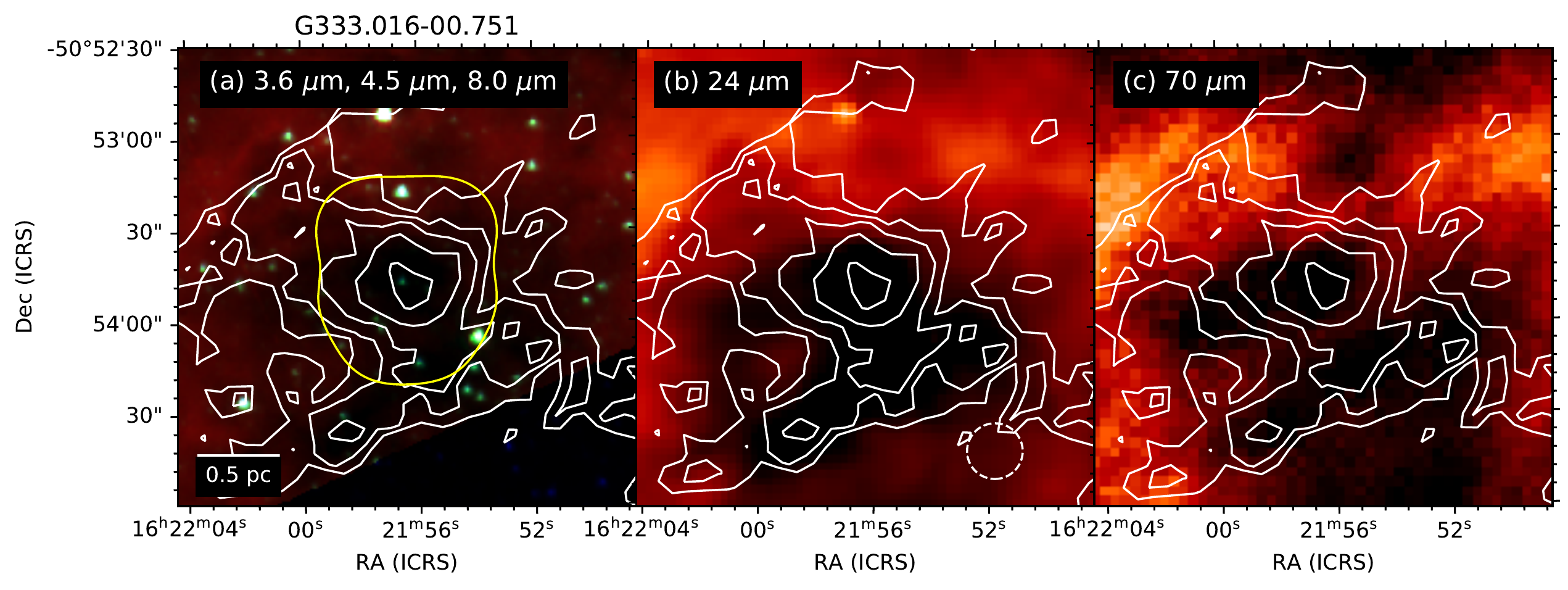}{0.97\textwidth}{}}\vspace{-1em}
\gridline{\fig{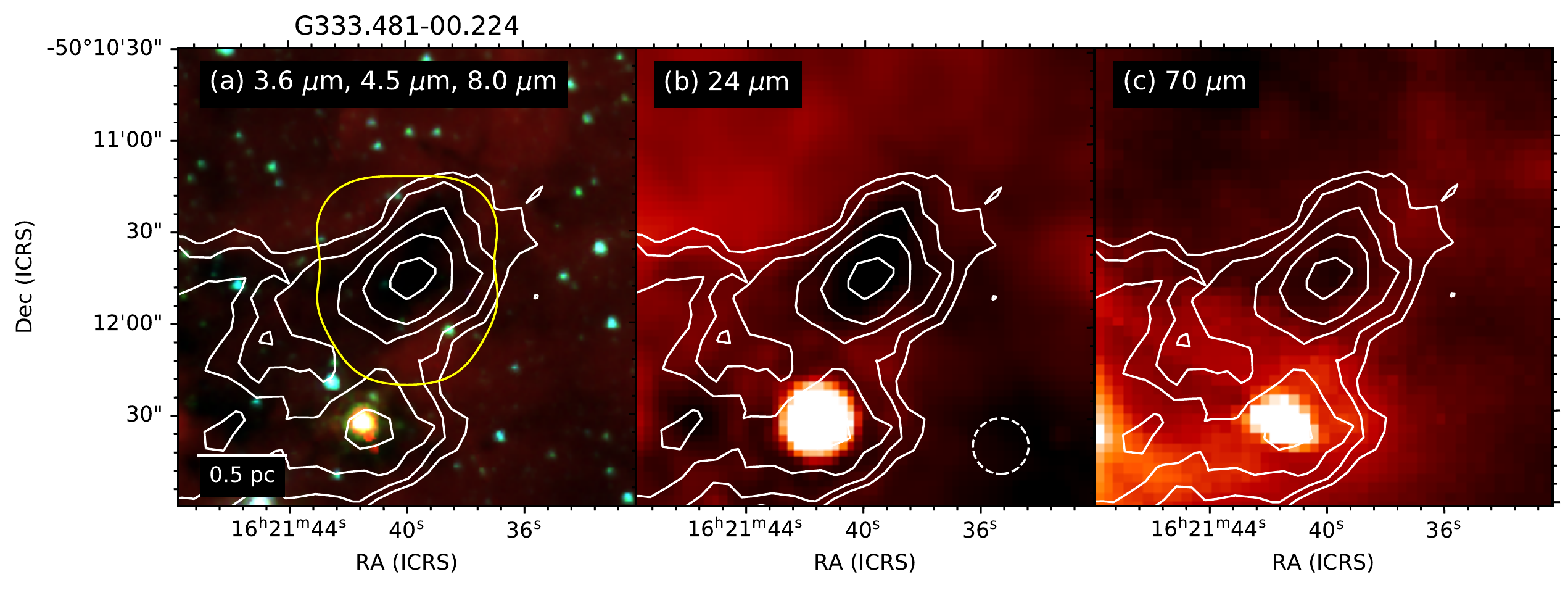}{0.97\textwidth}{}}\vspace{-1em}
\gridline{\fig{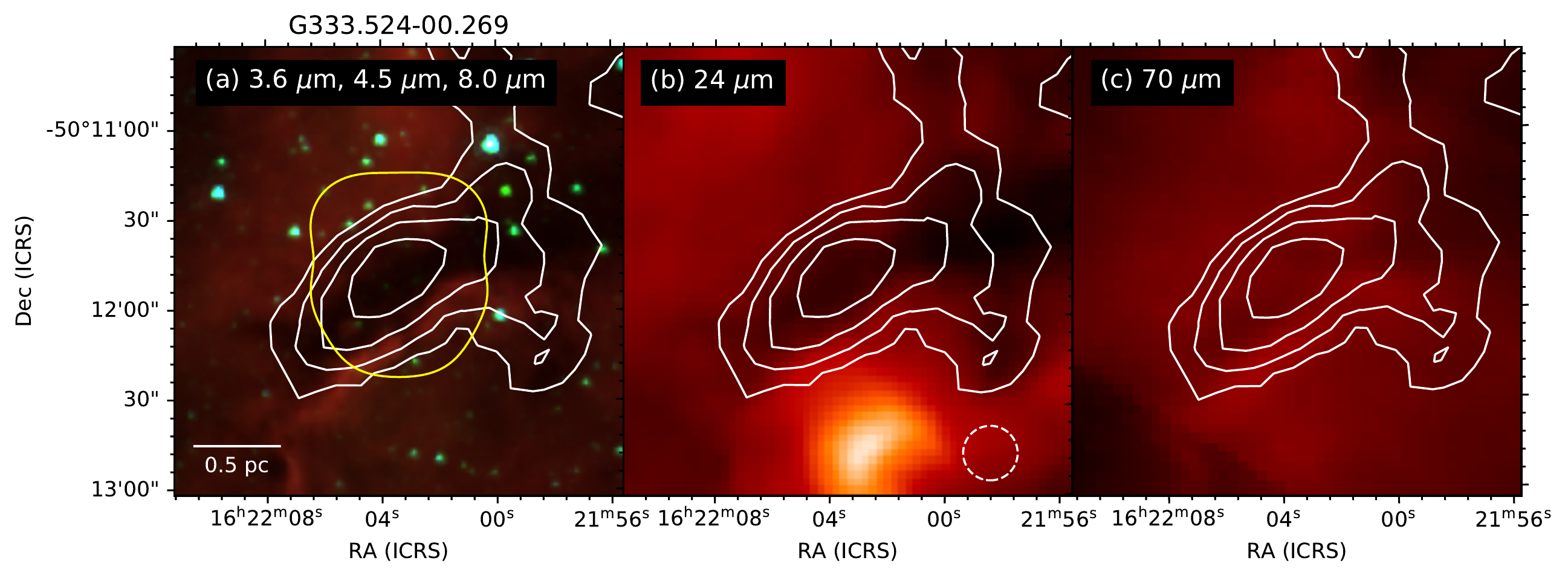}{0.97\textwidth}{}}\vspace{-1em}
\caption{Same as Figure~\ref{fig:IR_G10} except for contour levels for the 870 $\mu$m dust continuum emission, which are (3, 5, 7, 9, 11 and 13)$\times \sigma$ with $\sigma=57$ mJy\,beam$^{-1}$ for G333.016--00.7551, $\sigma=102$ mJy\,beam$^{-1}$ for G333.481--00.224, and $\sigma=268$ mJy\,beam$^{-1}$ for G333.524--00.269.}
\end{figure*}

\begin{figure*}
\gridline{\fig{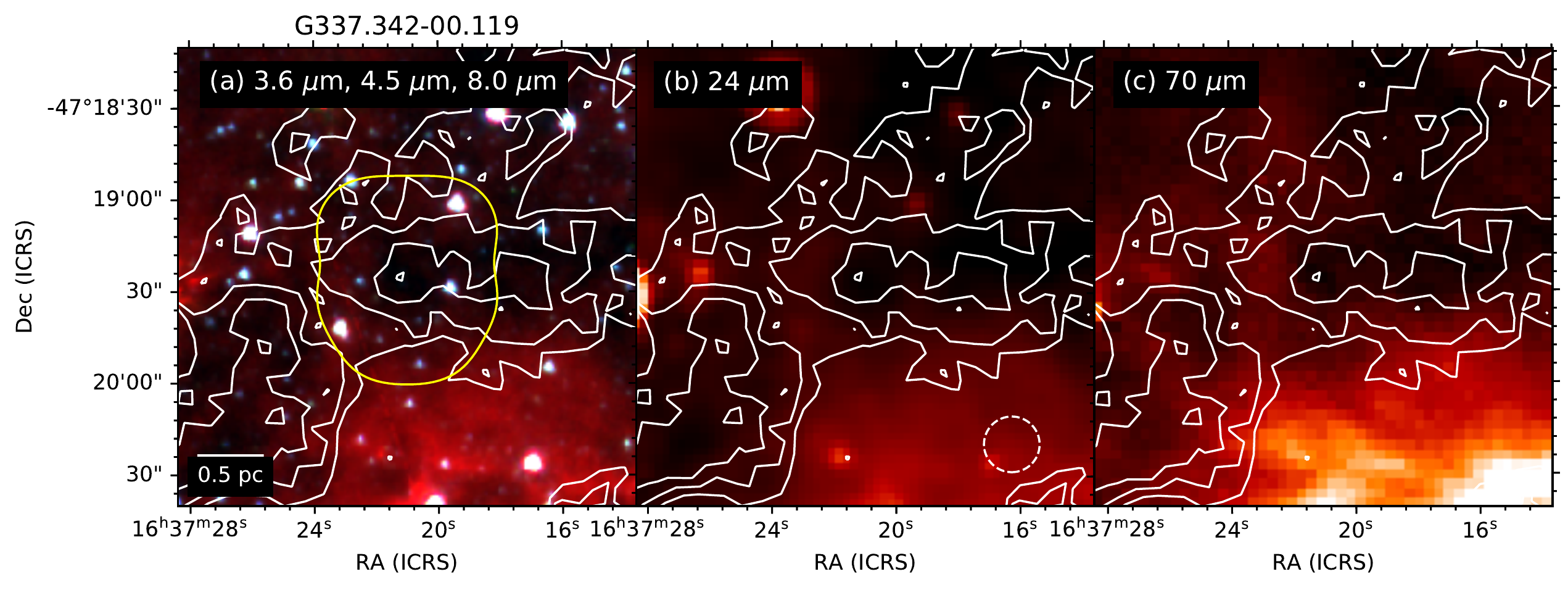}{0.97\textwidth}{}}\vspace{-1em}
\gridline{\fig{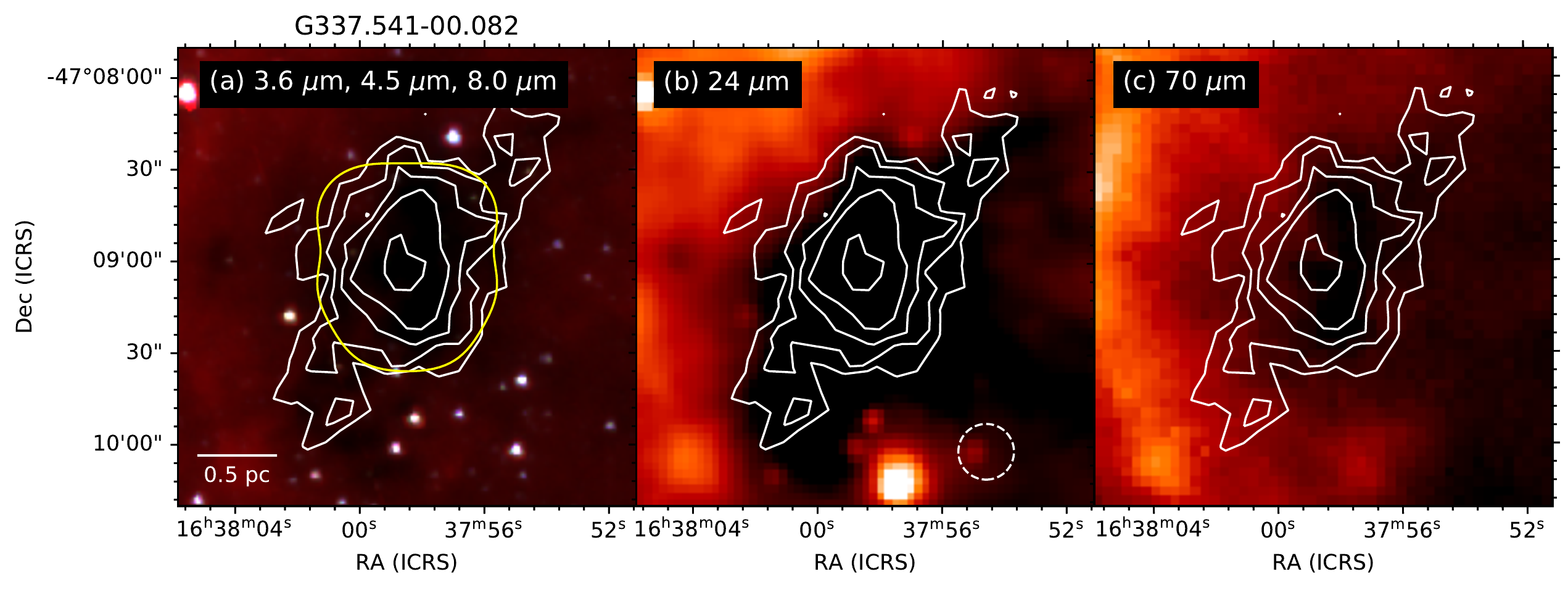}{0.97\textwidth}{}}\vspace{-1em}
\gridline{\fig{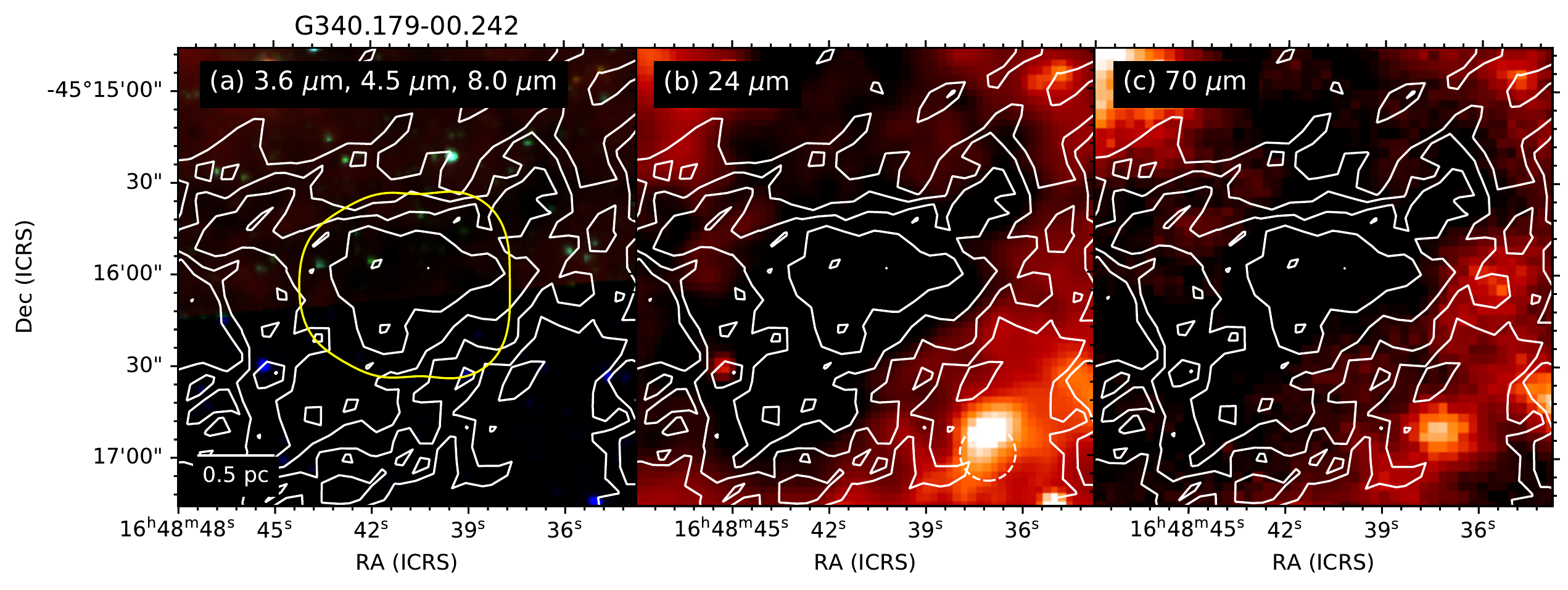}{0.97\textwidth}{}}\vspace{-1em}
\caption{Same as Figure~\ref{fig:IR_G10} except for contour levels for the 870 $\mu$m dust continuum emission, which are (3, 4, 5, 6 and 7)$\times \sigma$ with $\sigma=70$ mJy\,beam$^{-1}$ for G337.342--00.119 and (3, 5, 7, 9, 11, 15 and 19)$\times \sigma$ with $\sigma=66$ mJy\,beam$^{-1}$ for G337.541--00.082 and with $\sigma=57$ mJy\,beam$^{-1}$ for G340.179--00.242.}
\end{figure*}

\begin{figure*}
\gridline{\fig{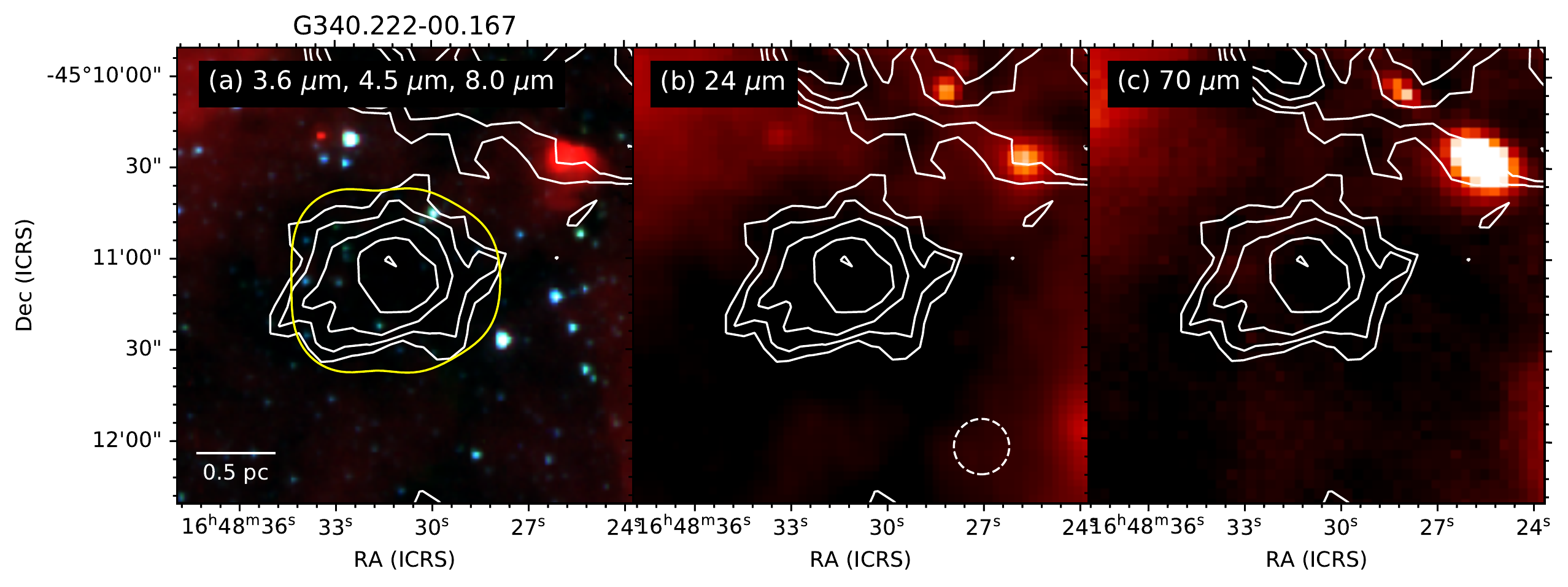}{0.97\textwidth}{}}\vspace{-1em}
\gridline{\fig{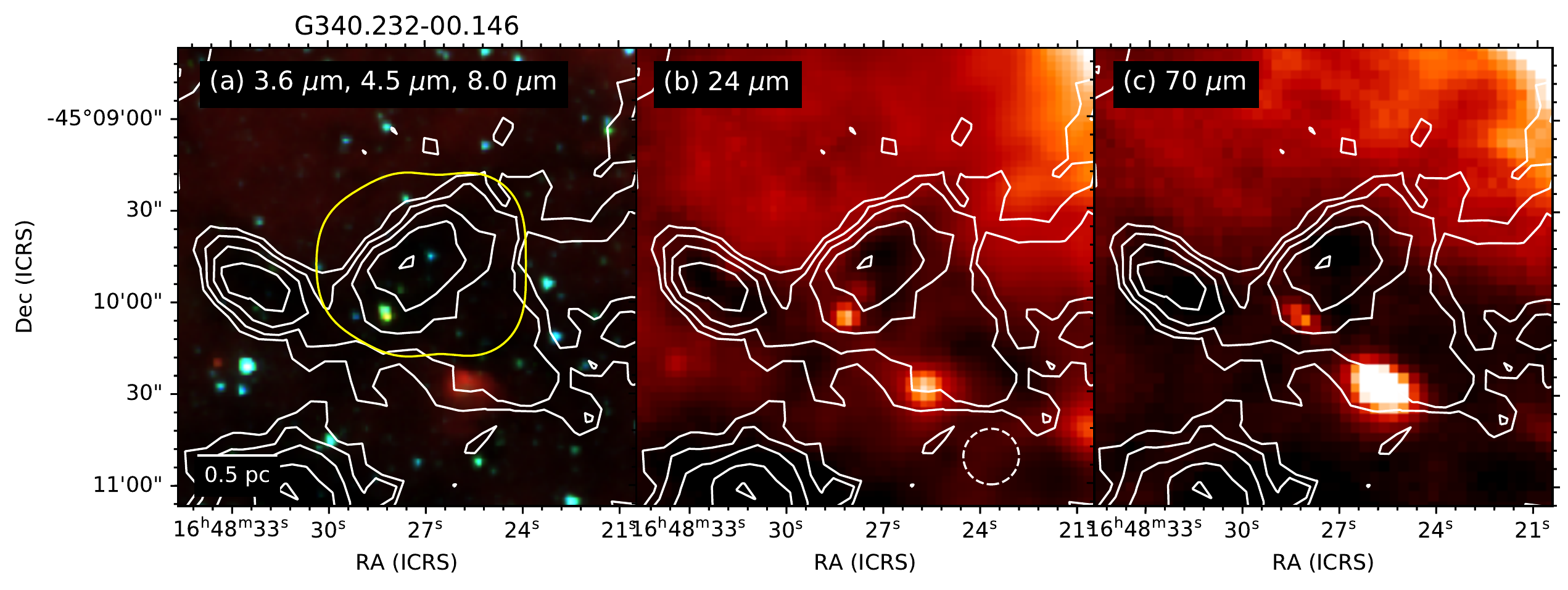}{0.97\textwidth}{}}\vspace{-1em}
\gridline{\fig{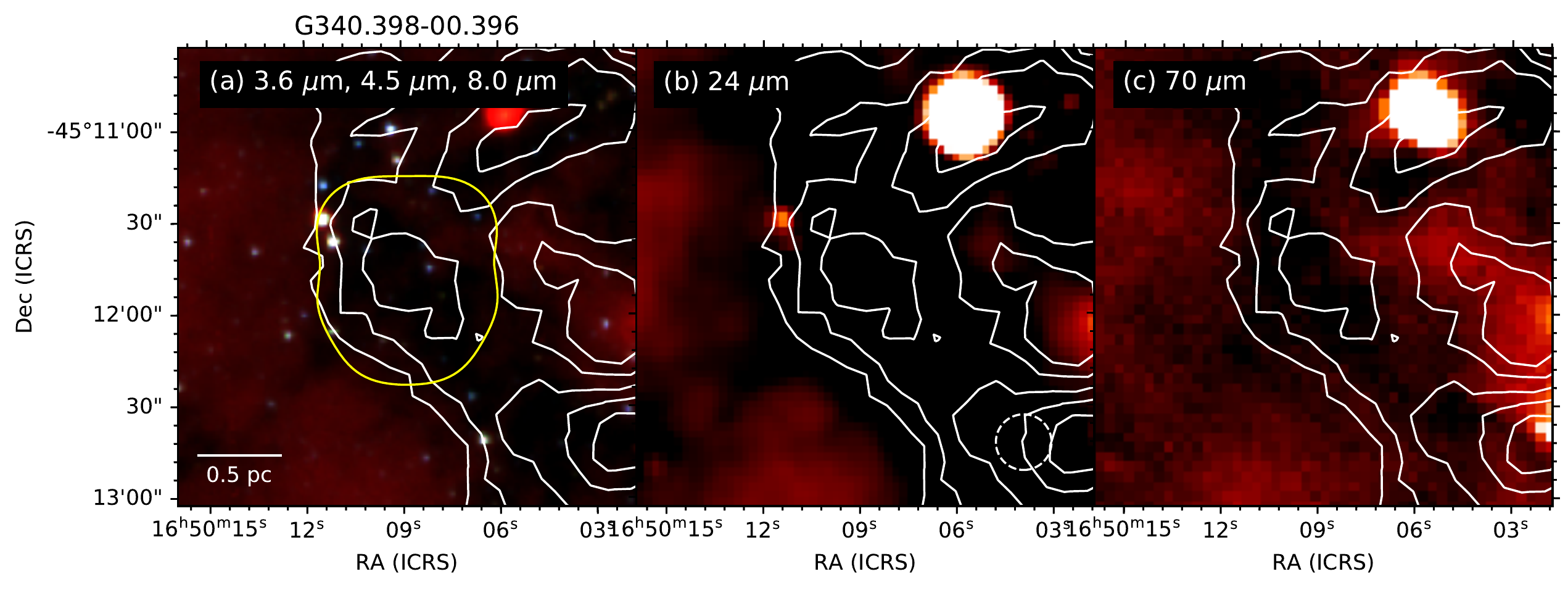}{0.97\textwidth}{}}\vspace{-1em}
\caption{Same as Figure~\ref{fig:IR_G10} except for contour levels for the 870 $\mu$m dust continuum emission, which are (3, 5, 7, 9, 11, 15 and 19)$\times \sigma$ with $\sigma=66$ mJy\,beam$^{-1}$ for G340.222--00.167 and $\sigma=65$ mJy\,beam$^{-1}$ for G340.232--00.146, and (3, 4, 5, 6 and 7)$\times \sigma$ with $\sigma=139$ mJy\,beam$^{-1}$ for G340.398--00.396.}
\end{figure*}

\begin{figure*}
\gridline{\fig{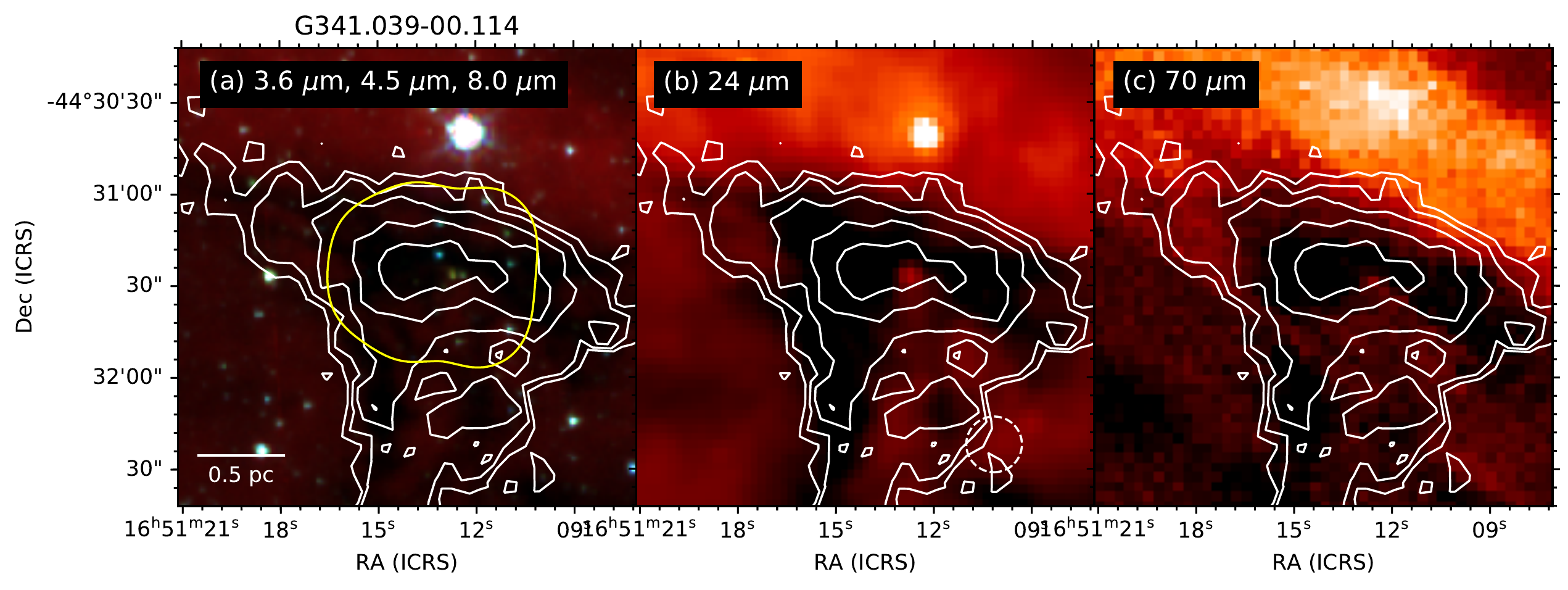}{0.97\textwidth}{}}\vspace{-1em}
\gridline{\fig{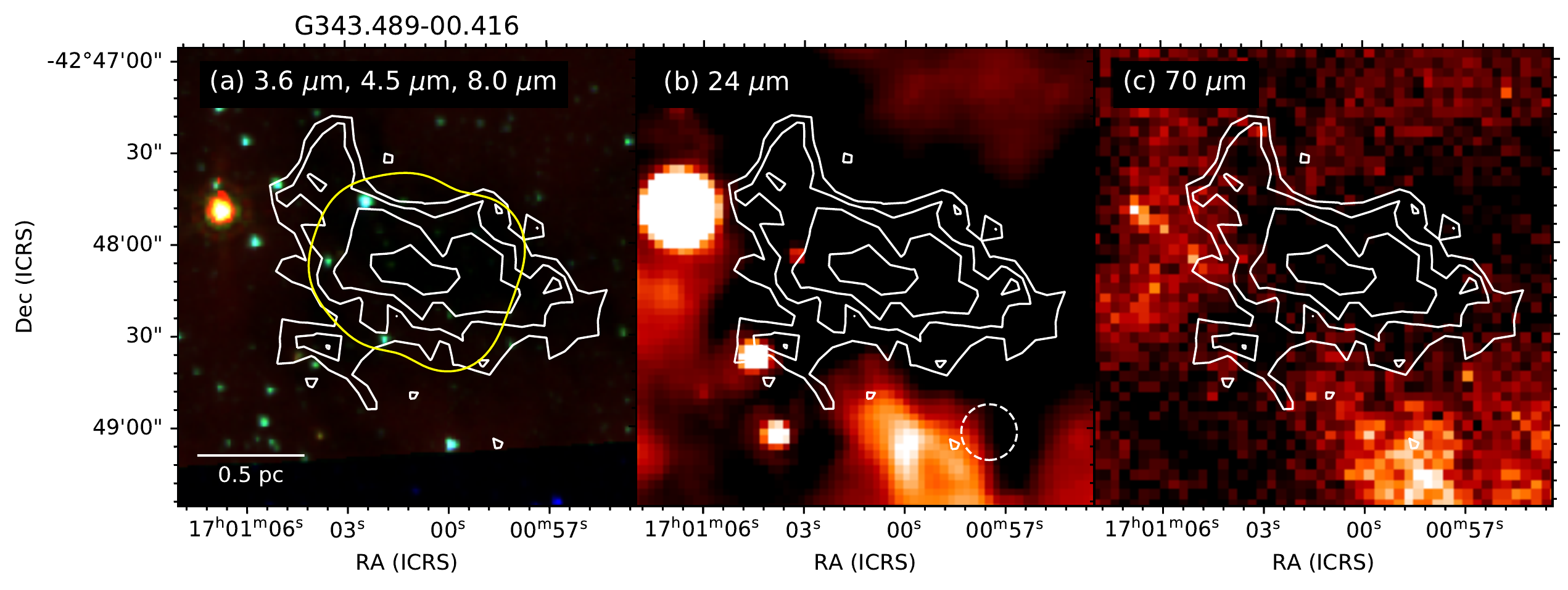}{0.97\textwidth}{}}\vspace{-1em}
\gridline{\fig{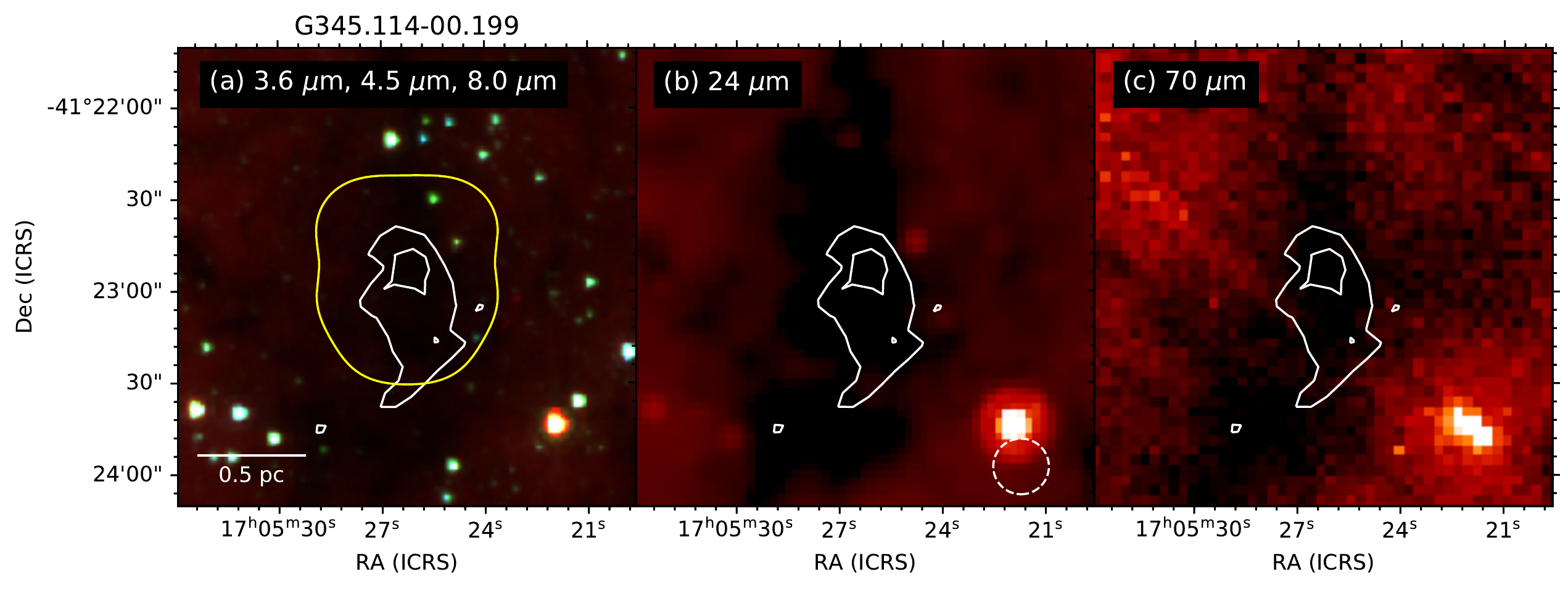}{0.97\textwidth}{}}\vspace{-1em}
\caption{Same as Figure~\ref{fig:IR_G10} except for contour levels for the 870 $\mu$m dust continuum emission, which are (3, 5, 7, 11, 15 and 19)$\times \sigma$ with$\sigma=52$ mJy\,beam$^{-1}$ for G341.039--00.114, (3, 5, 7, 11, 15 and 19)$\times \sigma$ with $\sigma=54$ mJy\,beam$^{-1}$ for G343.489--00.416, and (2, 3, 4, 5 and 6)$\times \sigma$ with $\sigma=141$ mJy\,beam$^{-1}$ for G345.114--00.199.}
\label{fig:Appendix_IR_last}
\end{figure*}


\begin{figure*}
\gridline{\fig{G010.99_12m7m_25105.pdf}{0.75\textwidth}{}} \vspace{-3em}
\gridline{\fig{G014.49_12m7m_25105.pdf}{0.75\textwidth}{}} \vspace{-3em}
\gridline{\fig{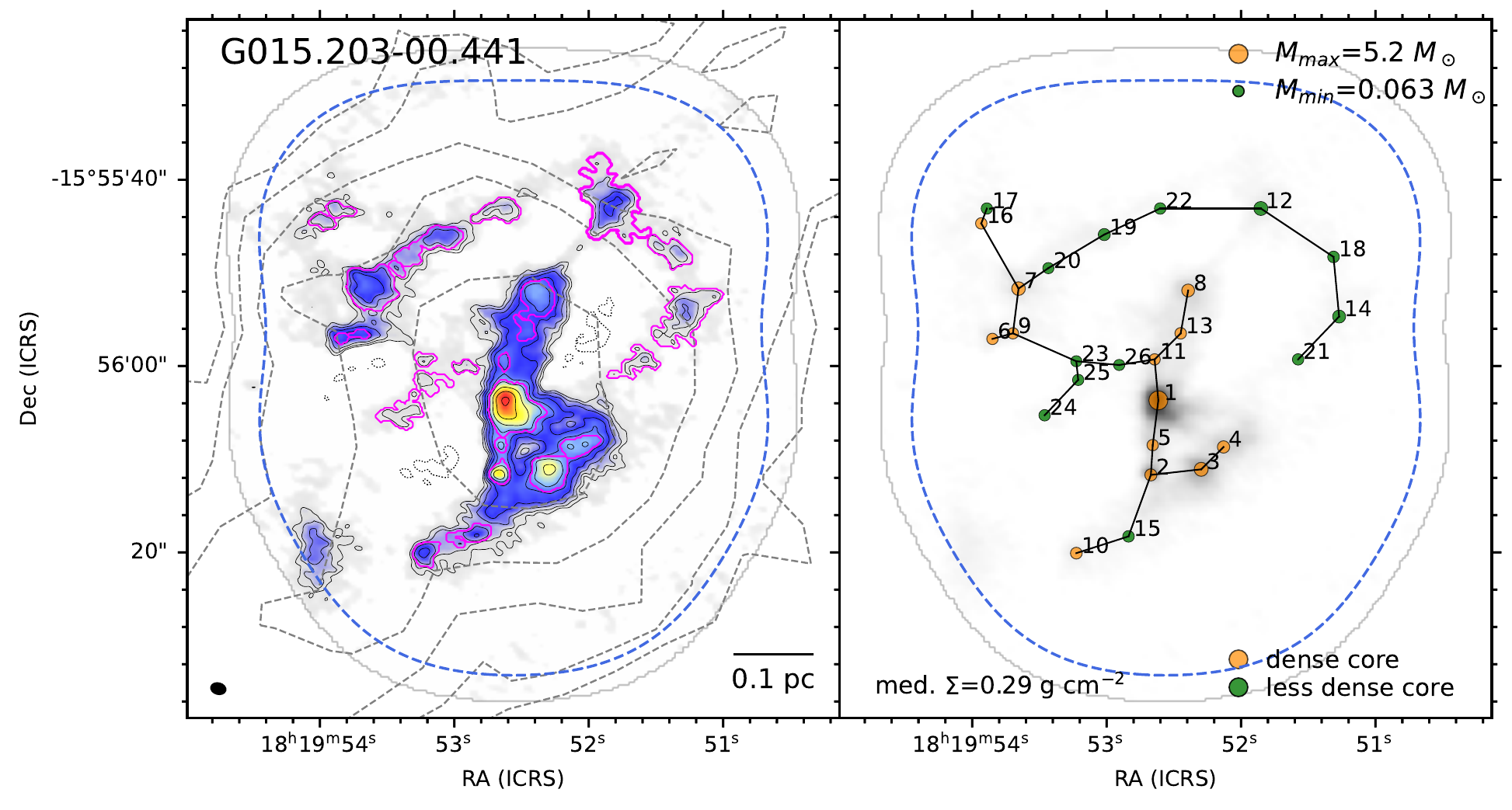}{0.75\textwidth}{}}\vspace{-3em}
\caption{Same as Figure~\ref{fig:ashes_cont_1}.}
\label{fig:Appendix_cont_1}
\end{figure*}

\begin{figure*}
    \gridline{\fig{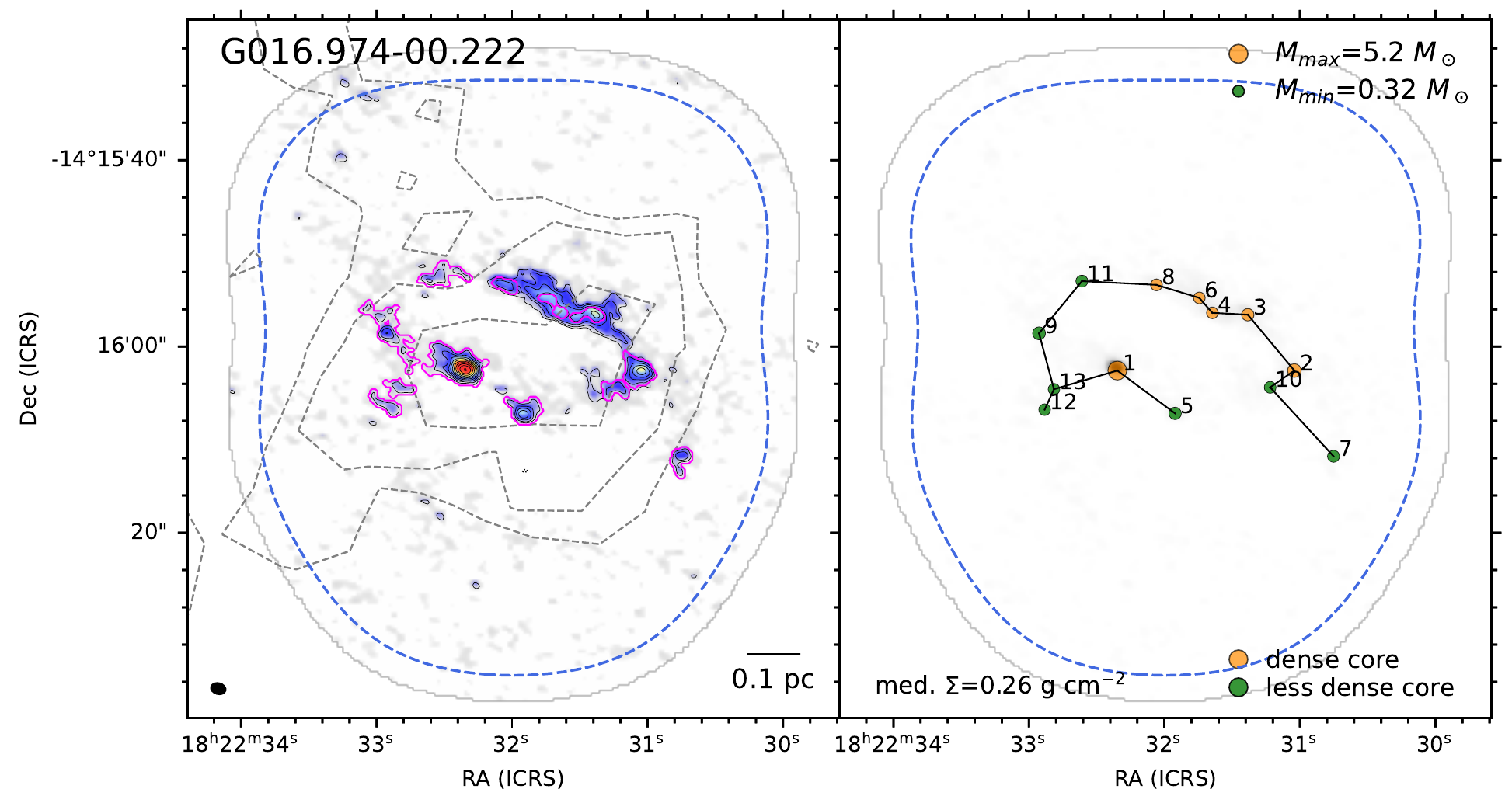}{0.75\textwidth}{}}\vspace{-3em}
    \gridline{\fig{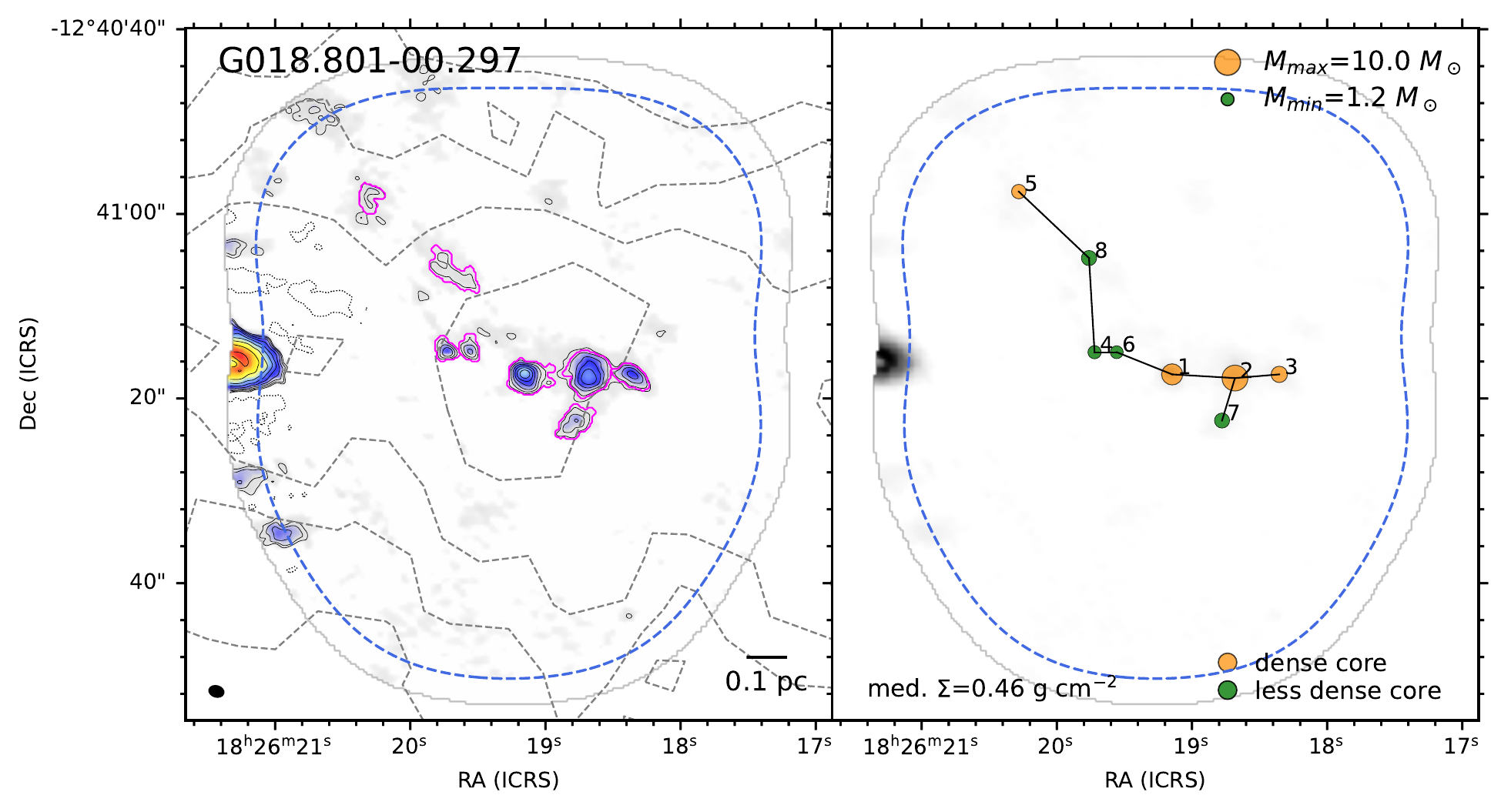}{0.75\textwidth}{}}\vspace{-3em}
    \gridline{\fig{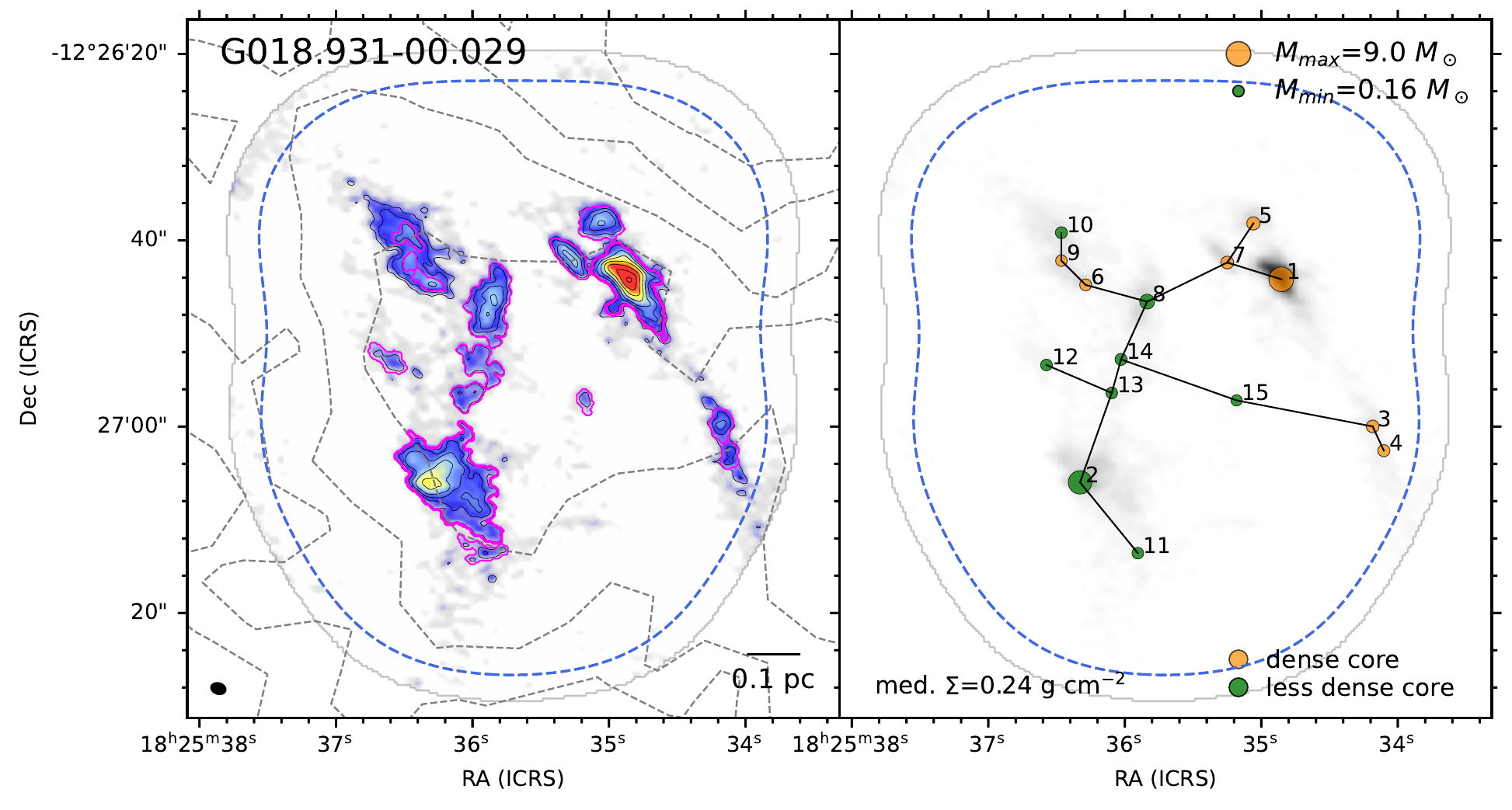}{0.75\textwidth}{}}\vspace{-3em}
    \caption{Same as Figure~\ref{fig:ashes_cont_1}.}
\end{figure*}

\begin{figure*}
    \gridline{\fig{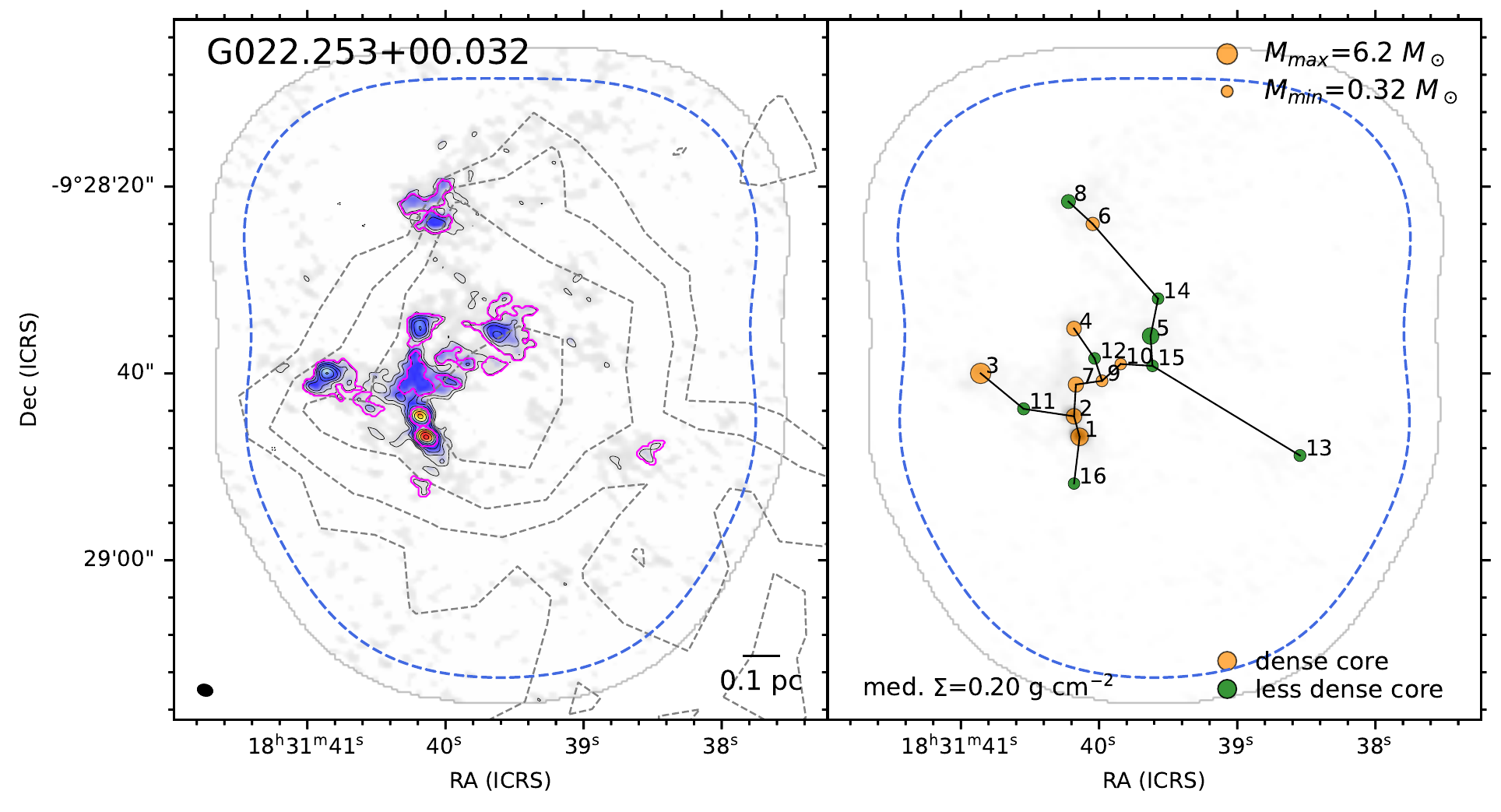}{0.75\textwidth}{}}\vspace{-3em}
    \gridline{\fig{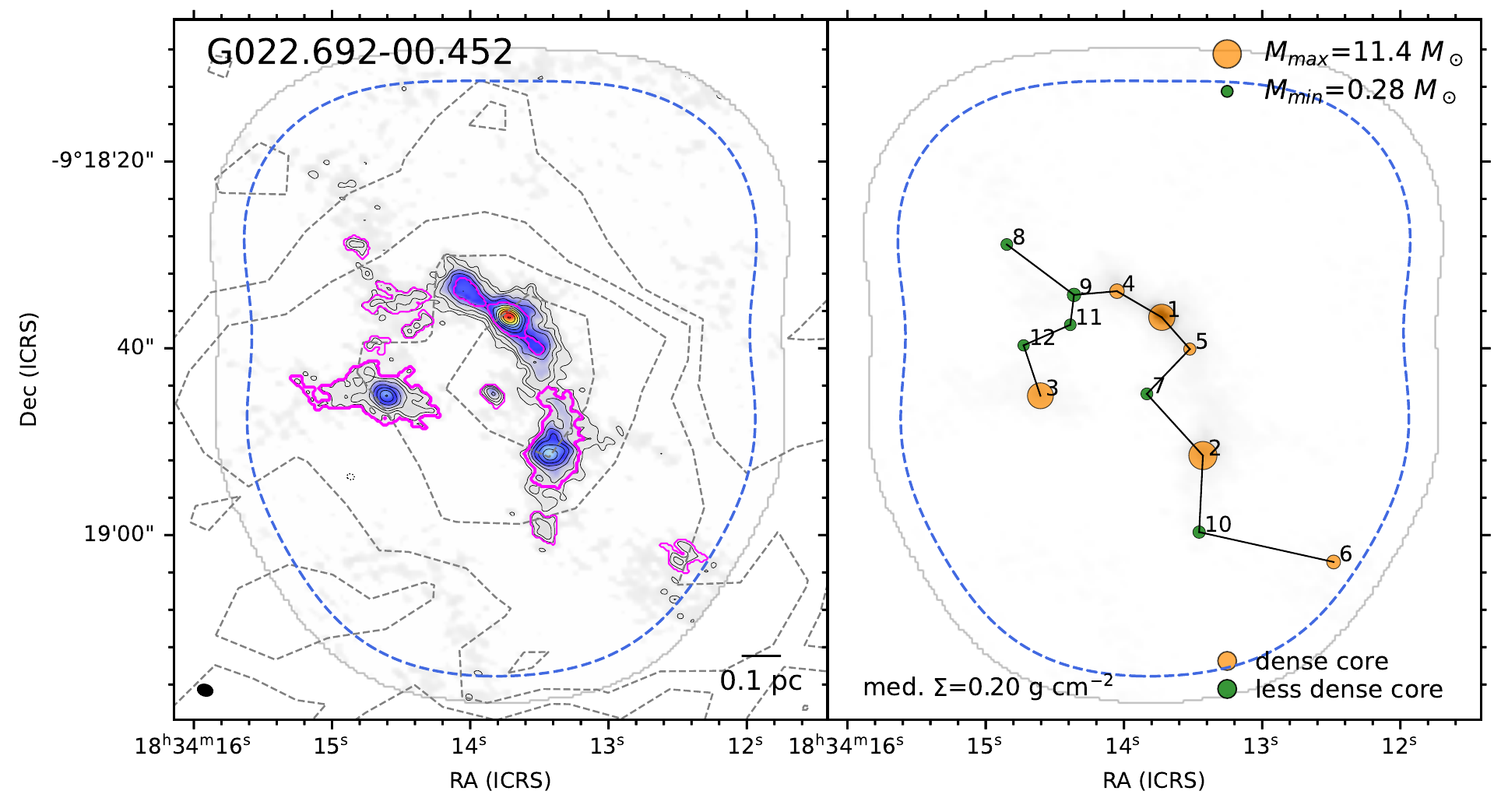}{0.75\textwidth}{}}\vspace{-3em}
    \gridline{\fig{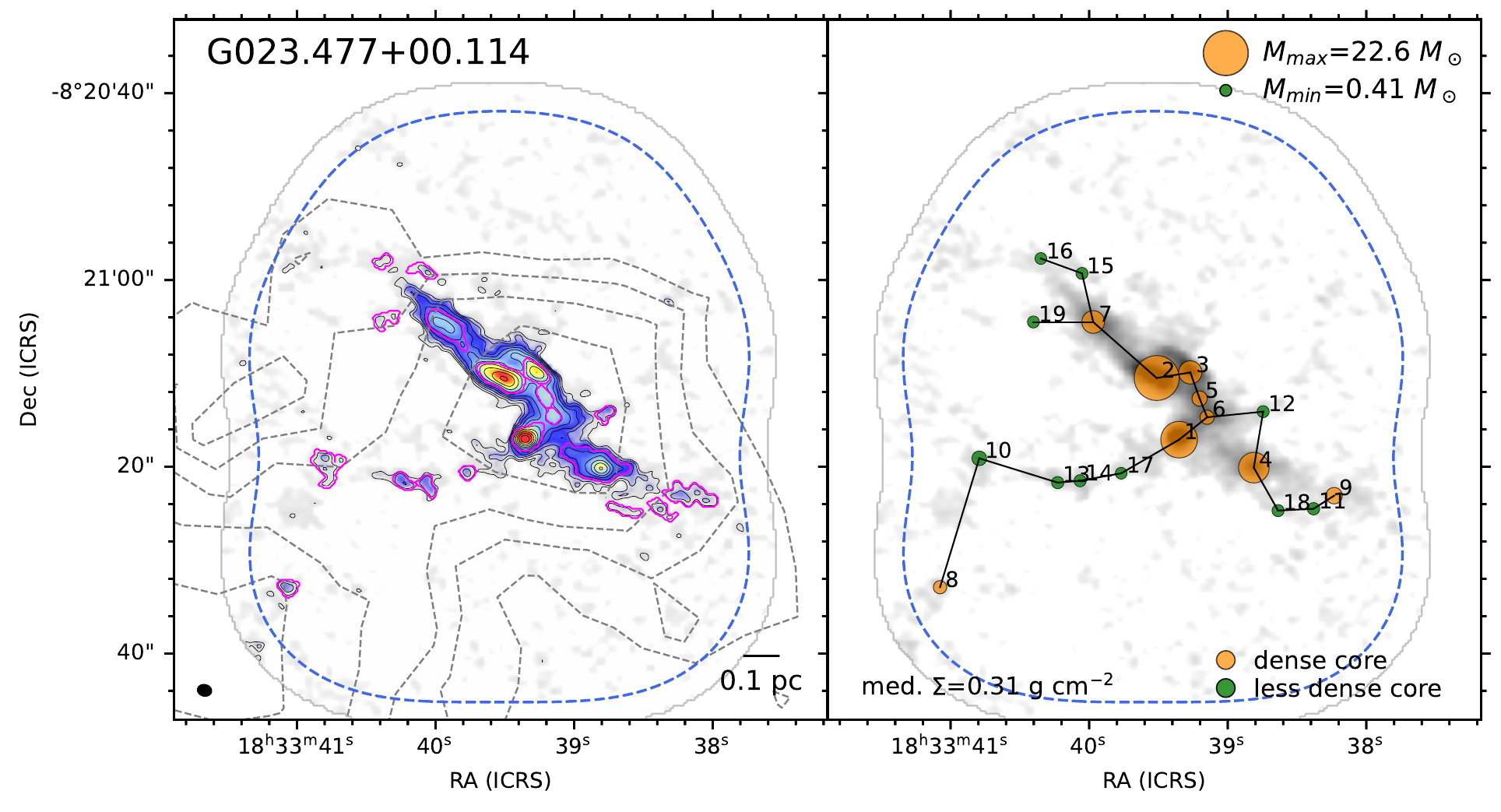}{0.75\textwidth}{}}\vspace{-3em}
    \caption{Same as Figure~\ref{fig:ashes_cont_1}.}
    \vspace{-5pt}
\end{figure*}

\begin{figure*}
    \gridline{\fig{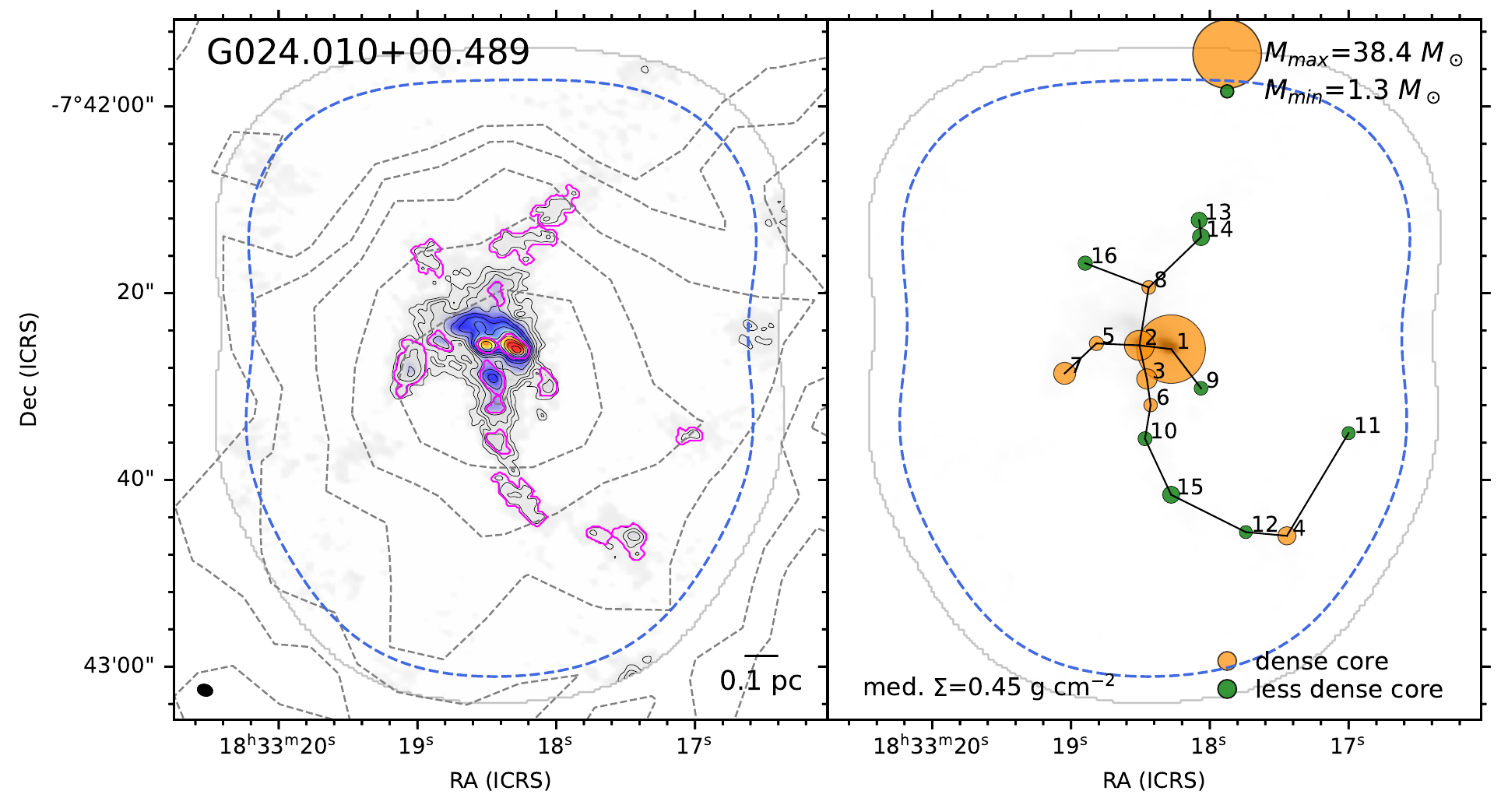}{0.75\textwidth}{}}\vspace{-3em}
    \gridline{\fig{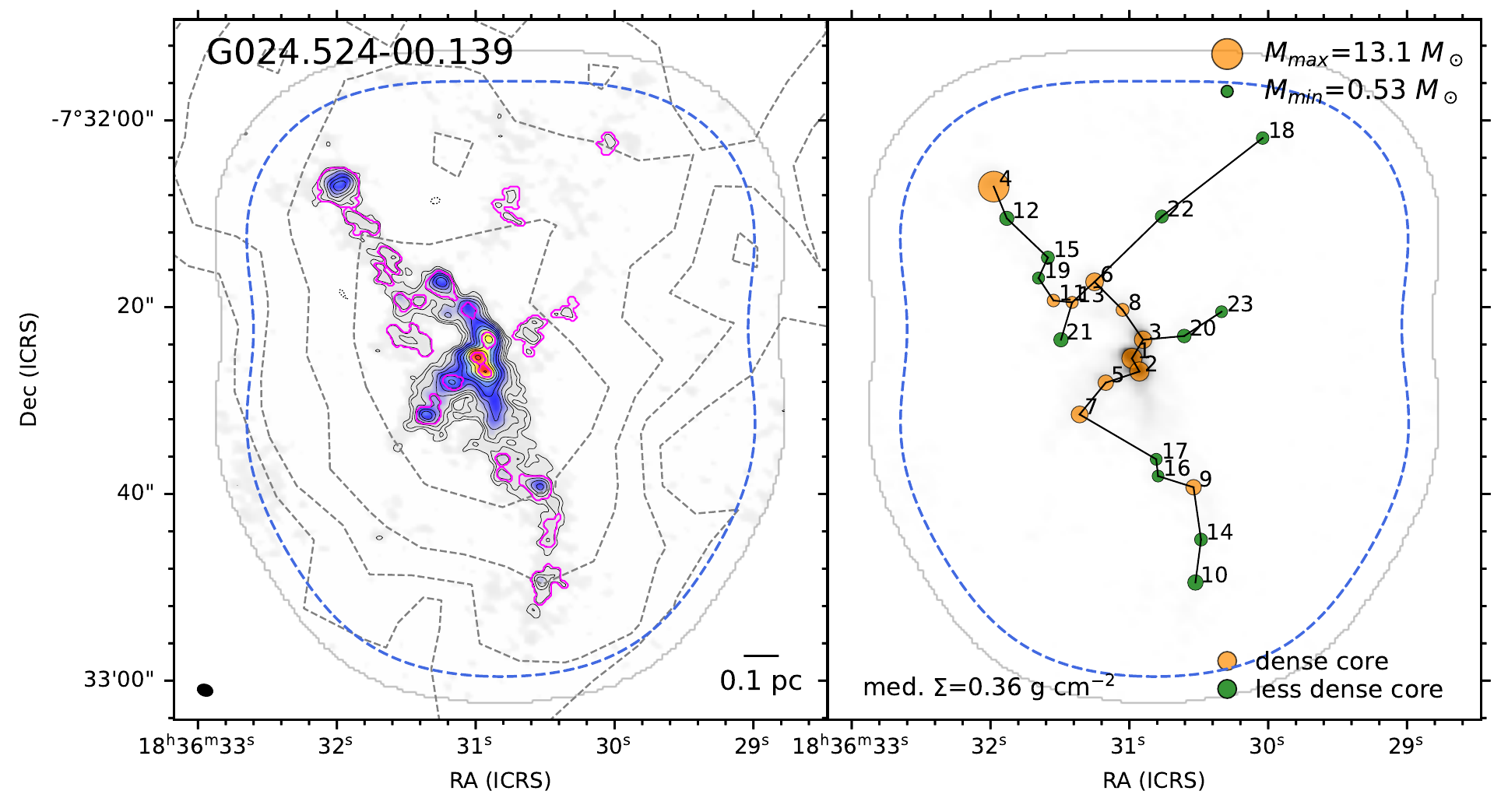}{0.75\textwidth}{}}\vspace{-3em}
    \gridline{\fig{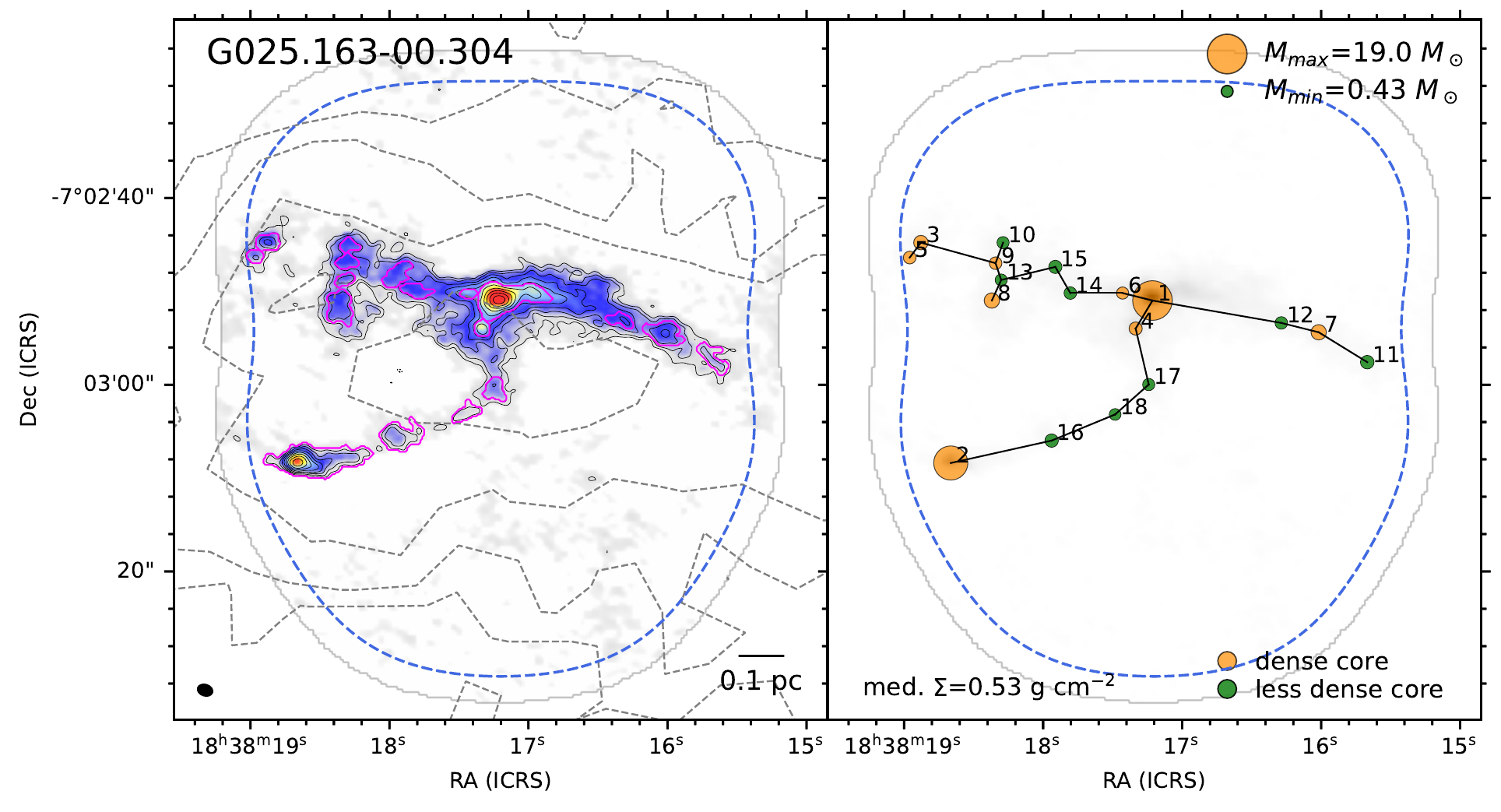}{0.75\textwidth}{}}\vspace{-3em}
    \caption{Same as Figure~\ref{fig:ashes_cont_1}.}
    \vspace{-5pt}
\end{figure*}

\begin{figure*}
    \gridline{\fig{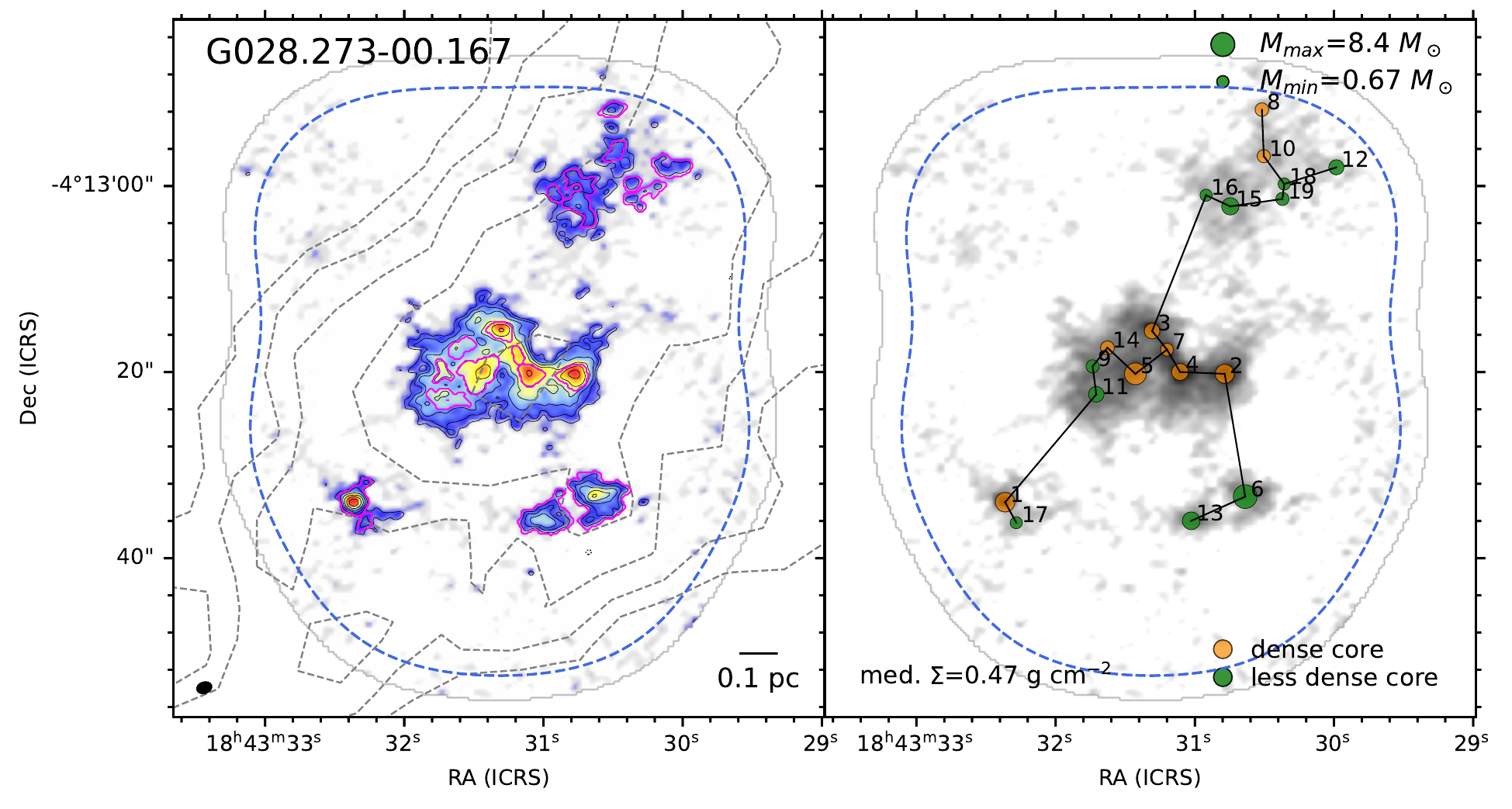}{0.75\textwidth}{}}\vspace{-3em}
    \gridline{\fig{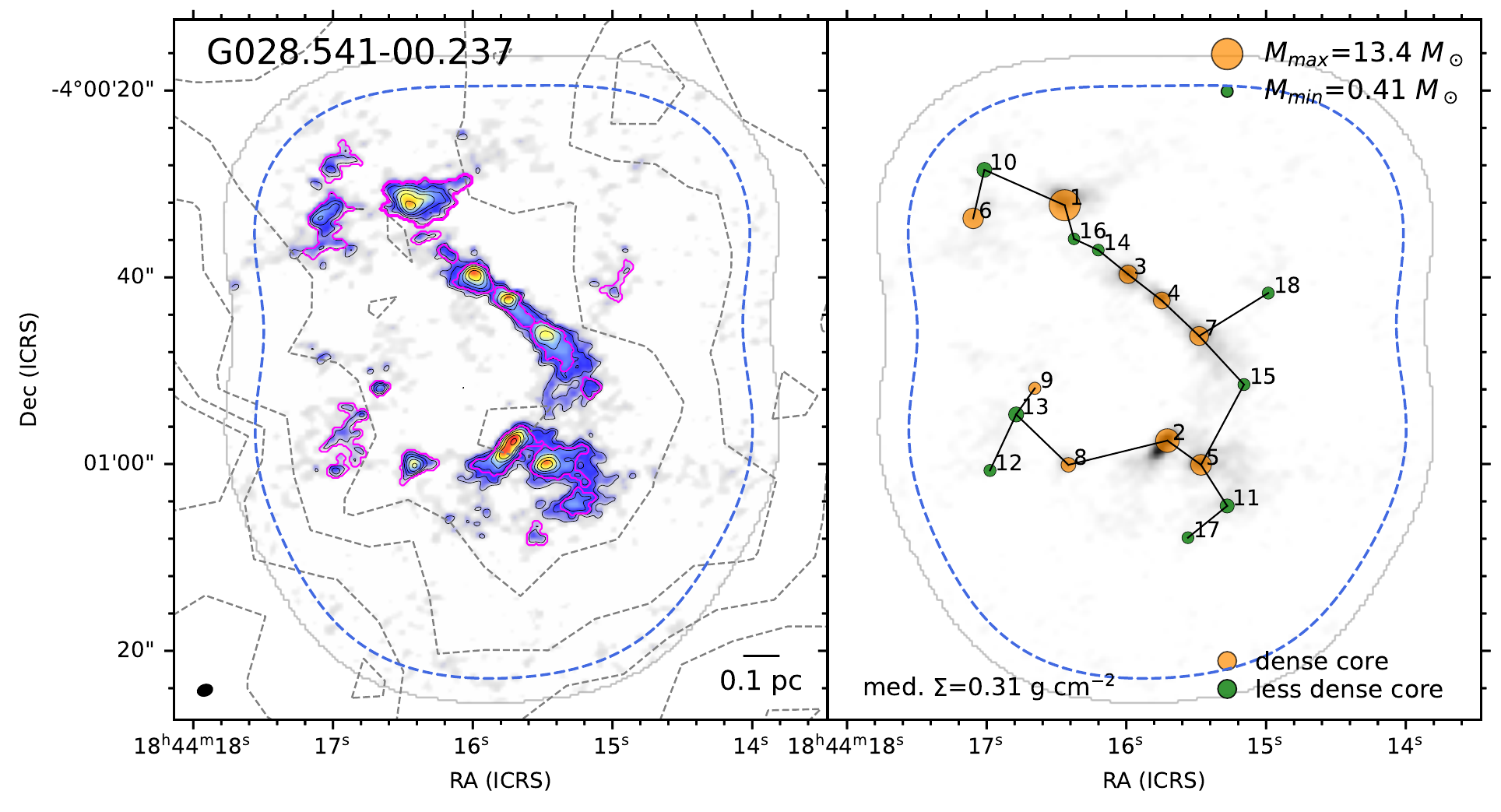}{0.75\textwidth}{}}\vspace{-3em}
    \gridline{\fig{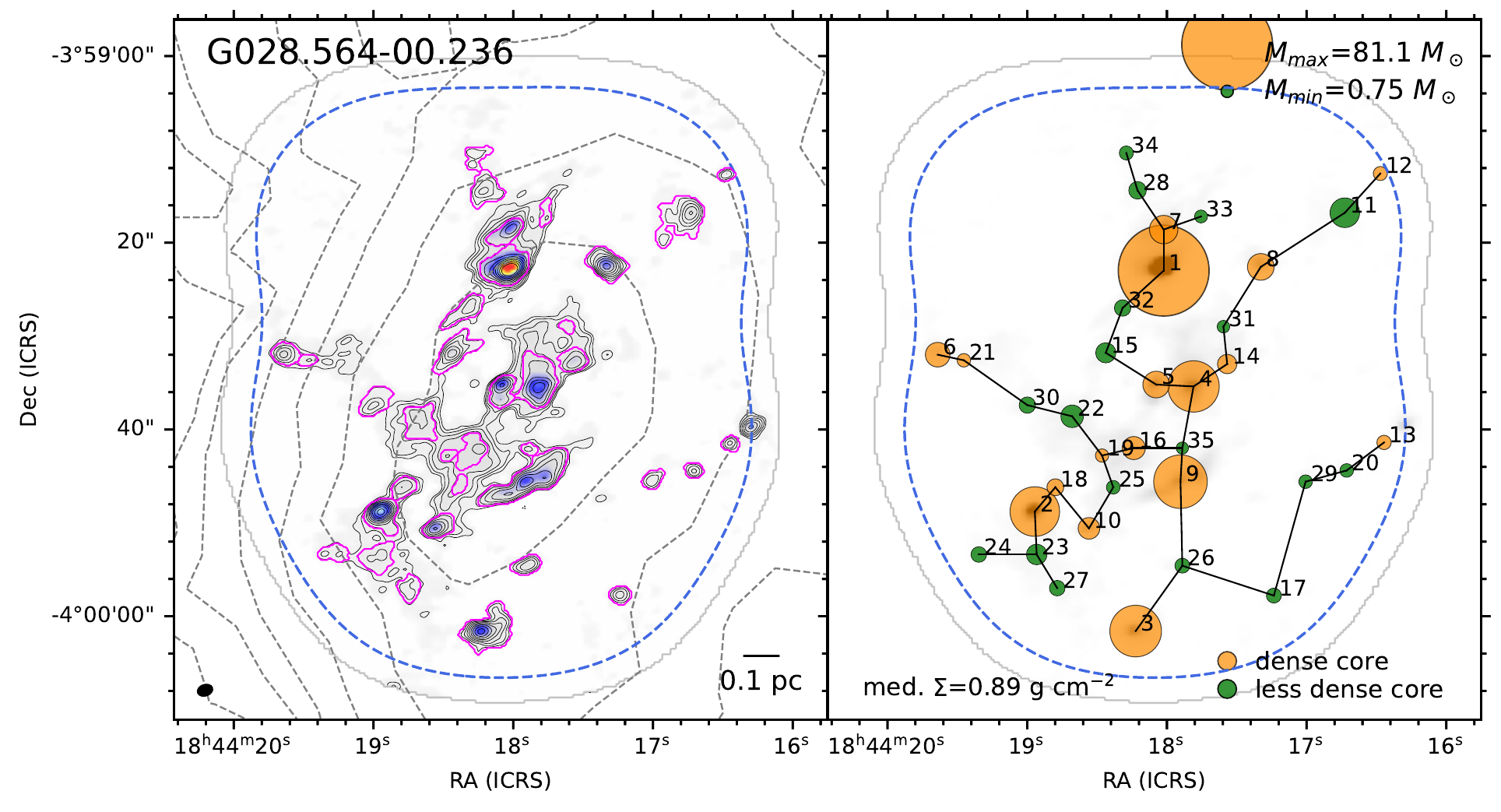}{0.75\textwidth}{}}\vspace{-3em}
    \caption{Same as Figure~\ref{fig:ashes_cont_1}.}
    \vspace{-5pt}
\end{figure*}

\begin{figure*}
    \gridline{\fig{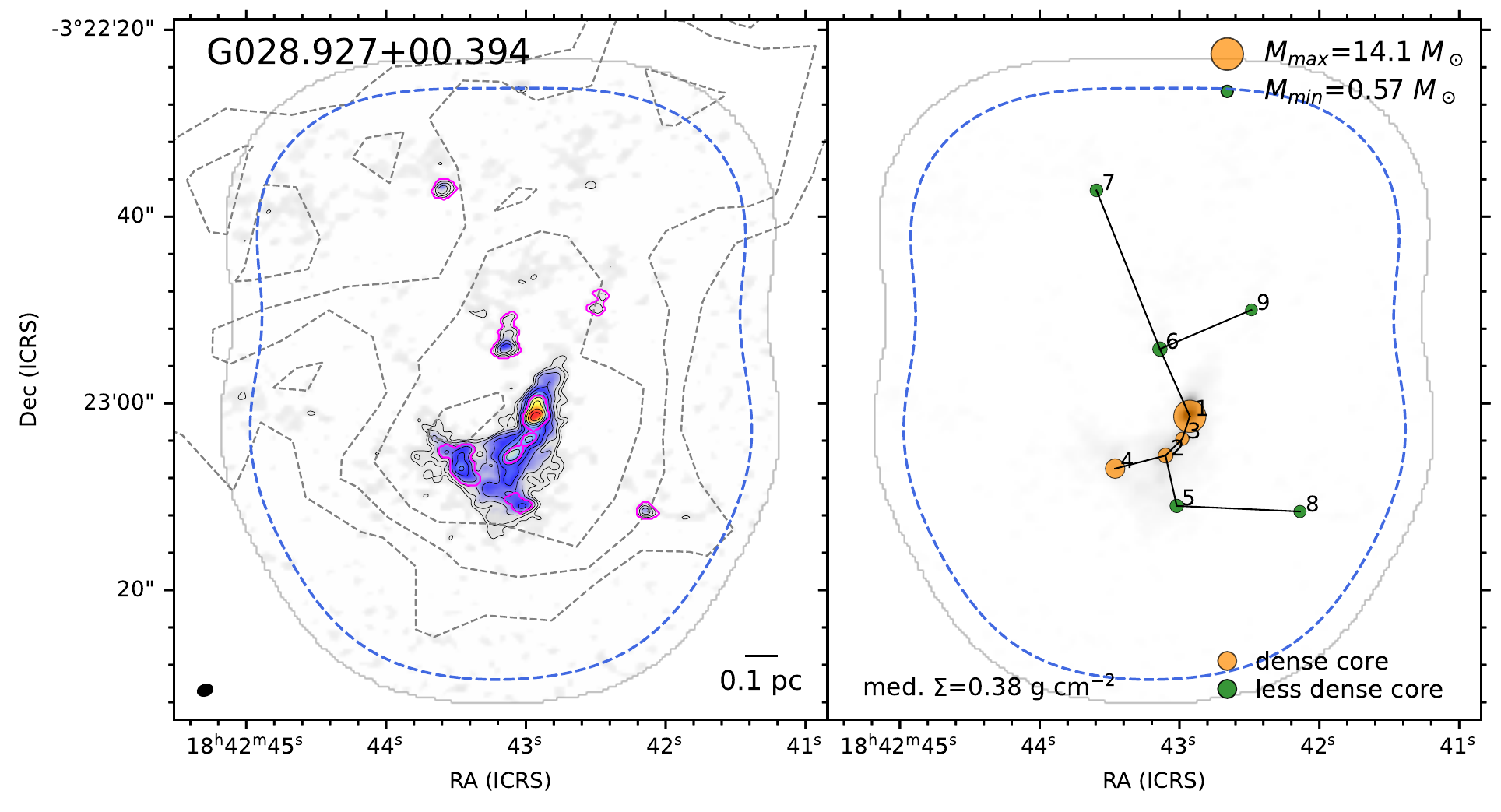}{0.75\textwidth}{}}\vspace{-3em}
    \gridline{\fig{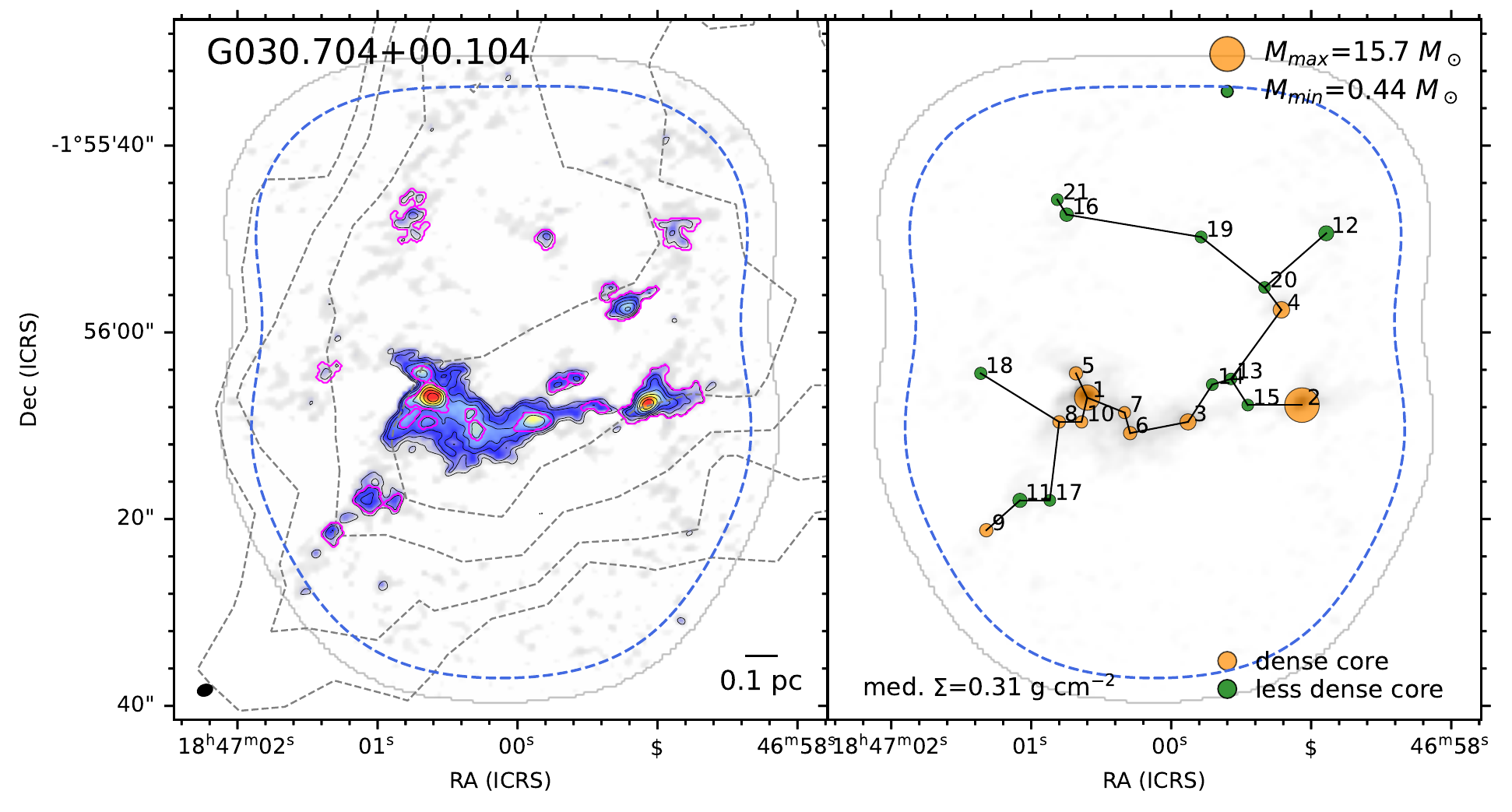}{0.75\textwidth}{}}\vspace{-3em}
    \gridline{\fig{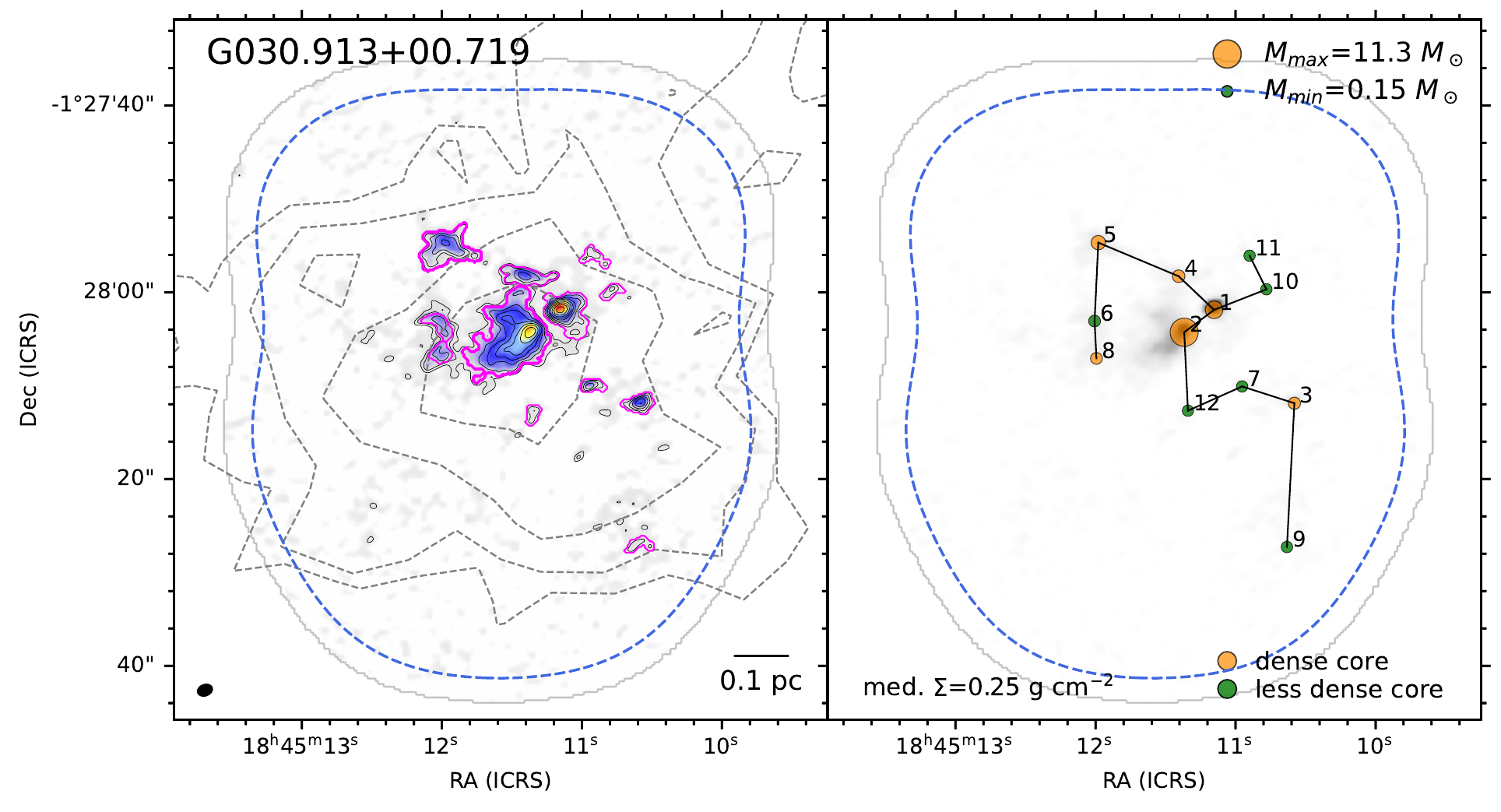}{0.75\textwidth}{}}\vspace{-3em}
    \caption{Same as Figure~\ref{fig:ashes_cont_1}.}
    \vspace{-5pt}
\end{figure*}

\begin{figure*}
    \gridline{\fig{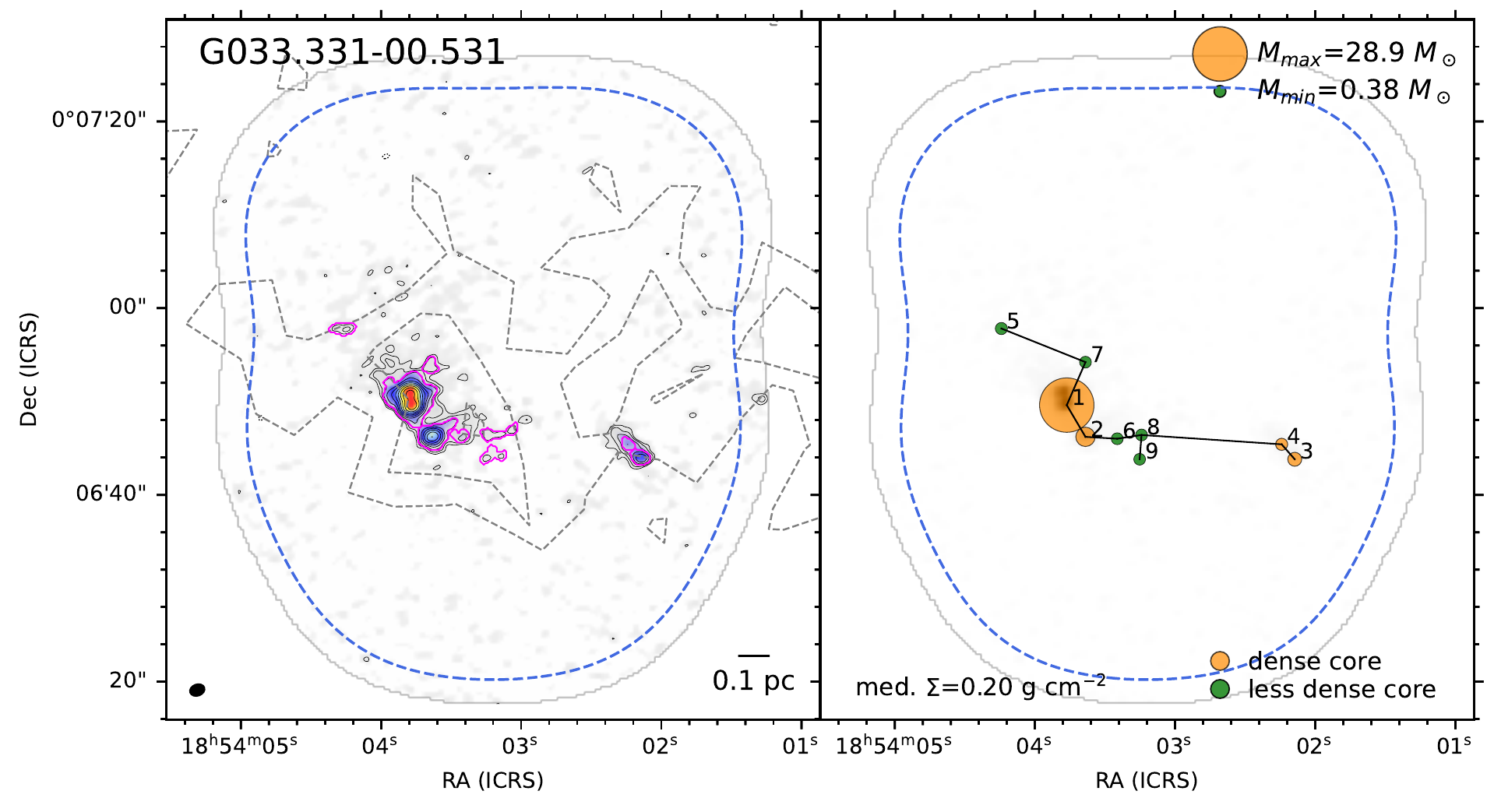}{0.75\textwidth}{}}\vspace{-3em}
    \gridline{\fig{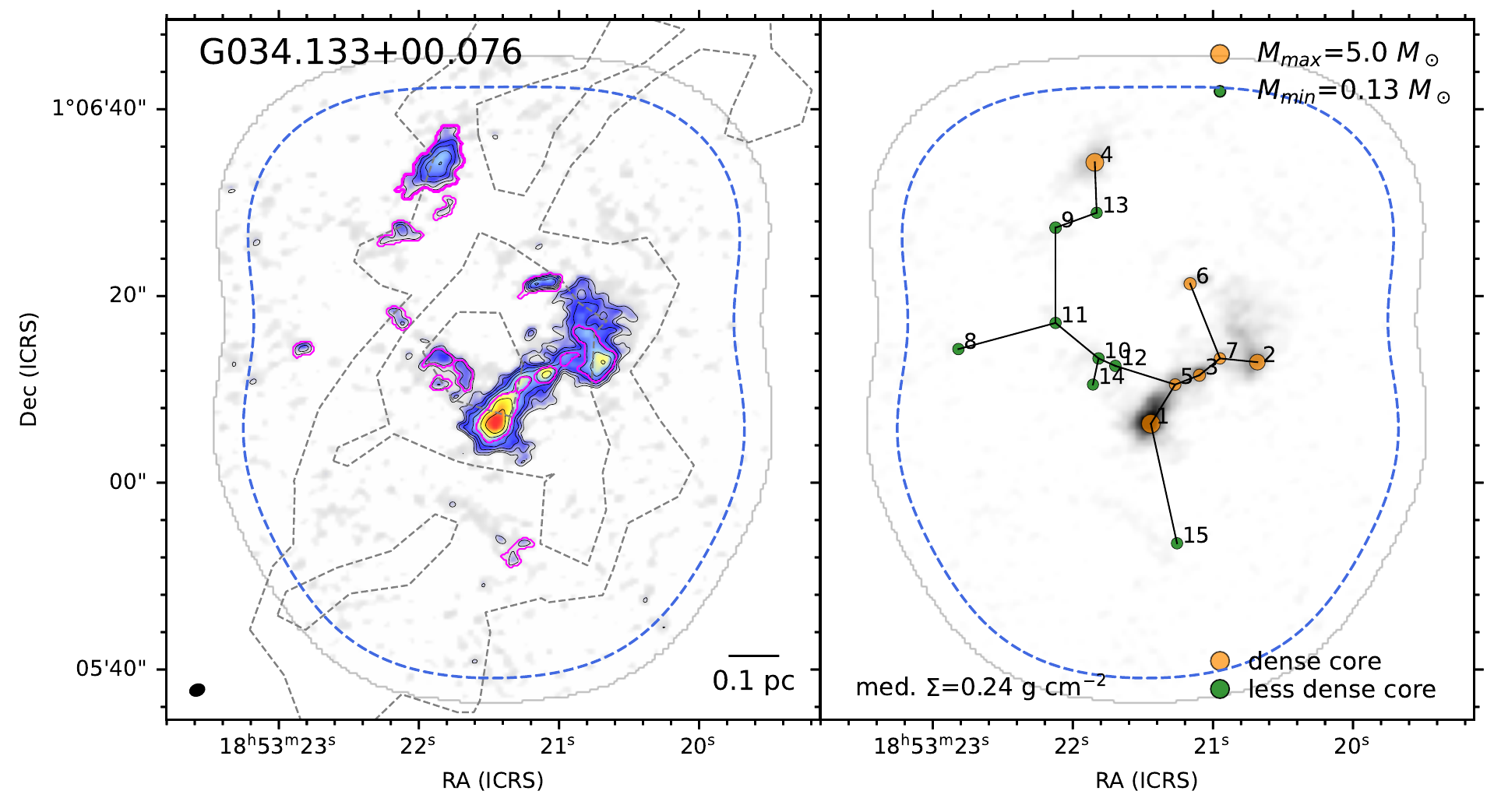}{0.75\textwidth}{}}\vspace{-3em}
    \gridline{\fig{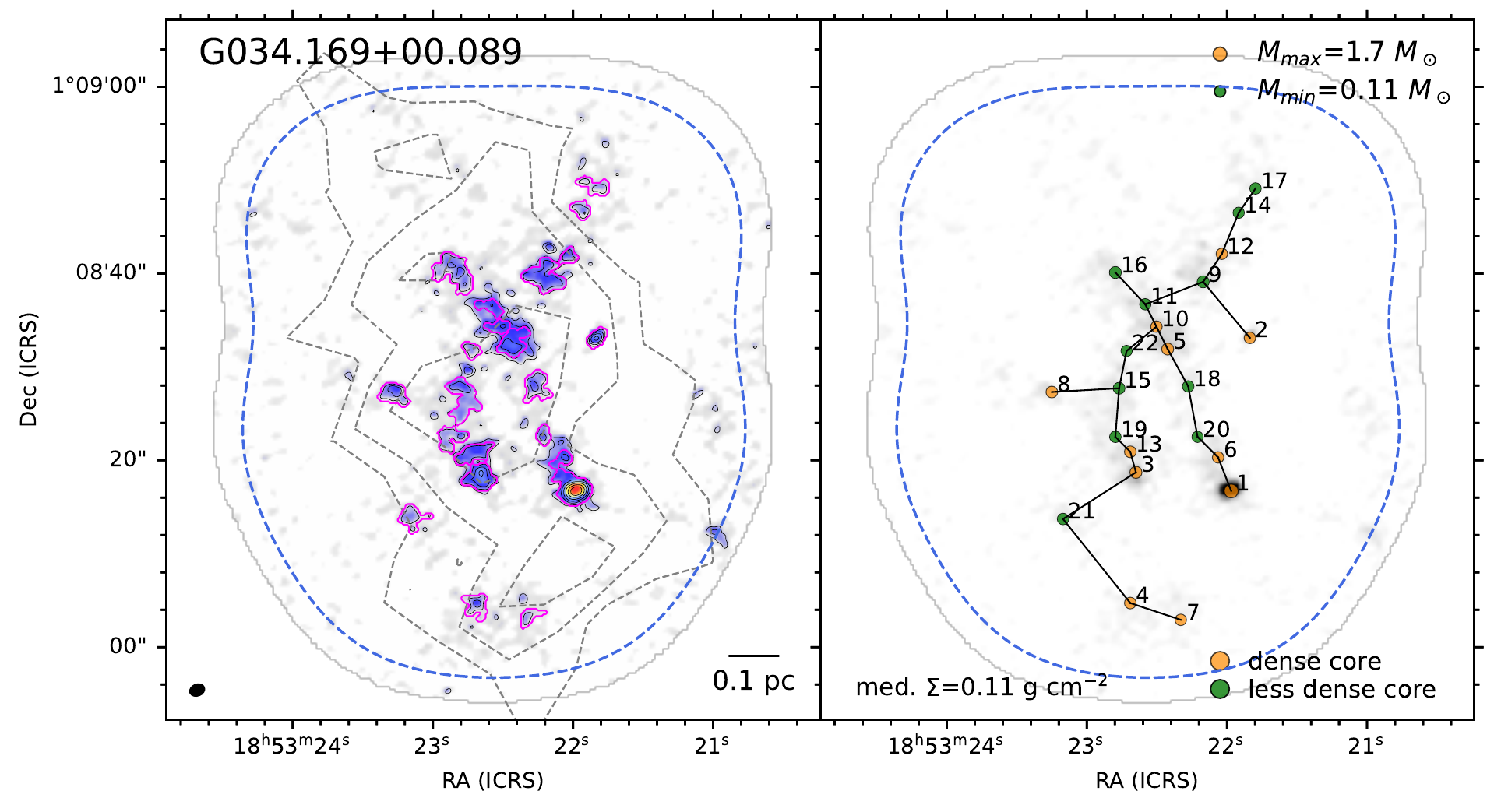}{0.75\textwidth}{}}\vspace{-3em}
    \caption{Same as Figure~\ref{fig:ashes_cont_1}.}
    \vspace{-9pt}
\end{figure*}

\begin{figure*}
    \gridline{\fig{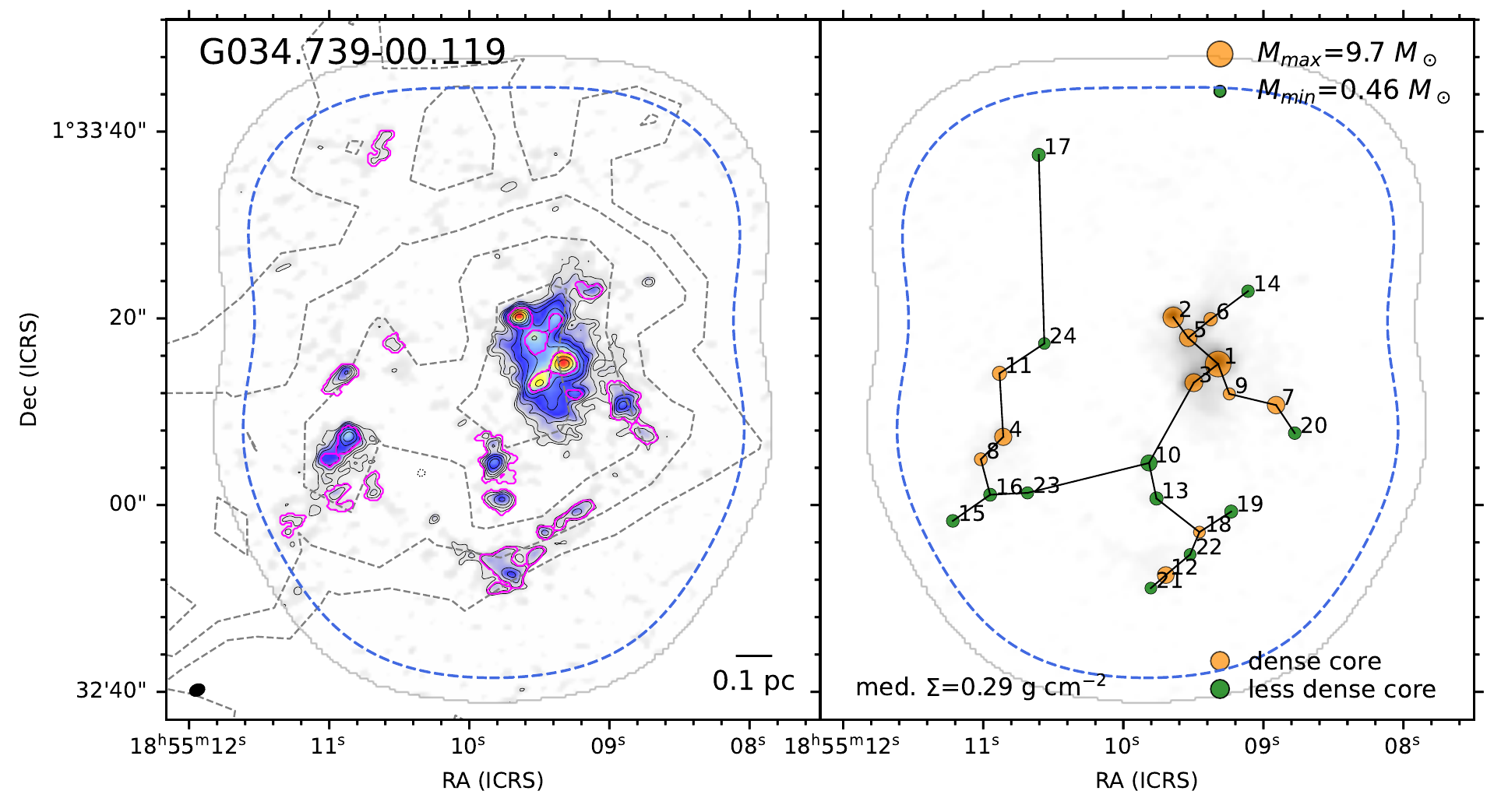}{0.75\textwidth}{}}\vspace{-3em}
    \gridline{\fig{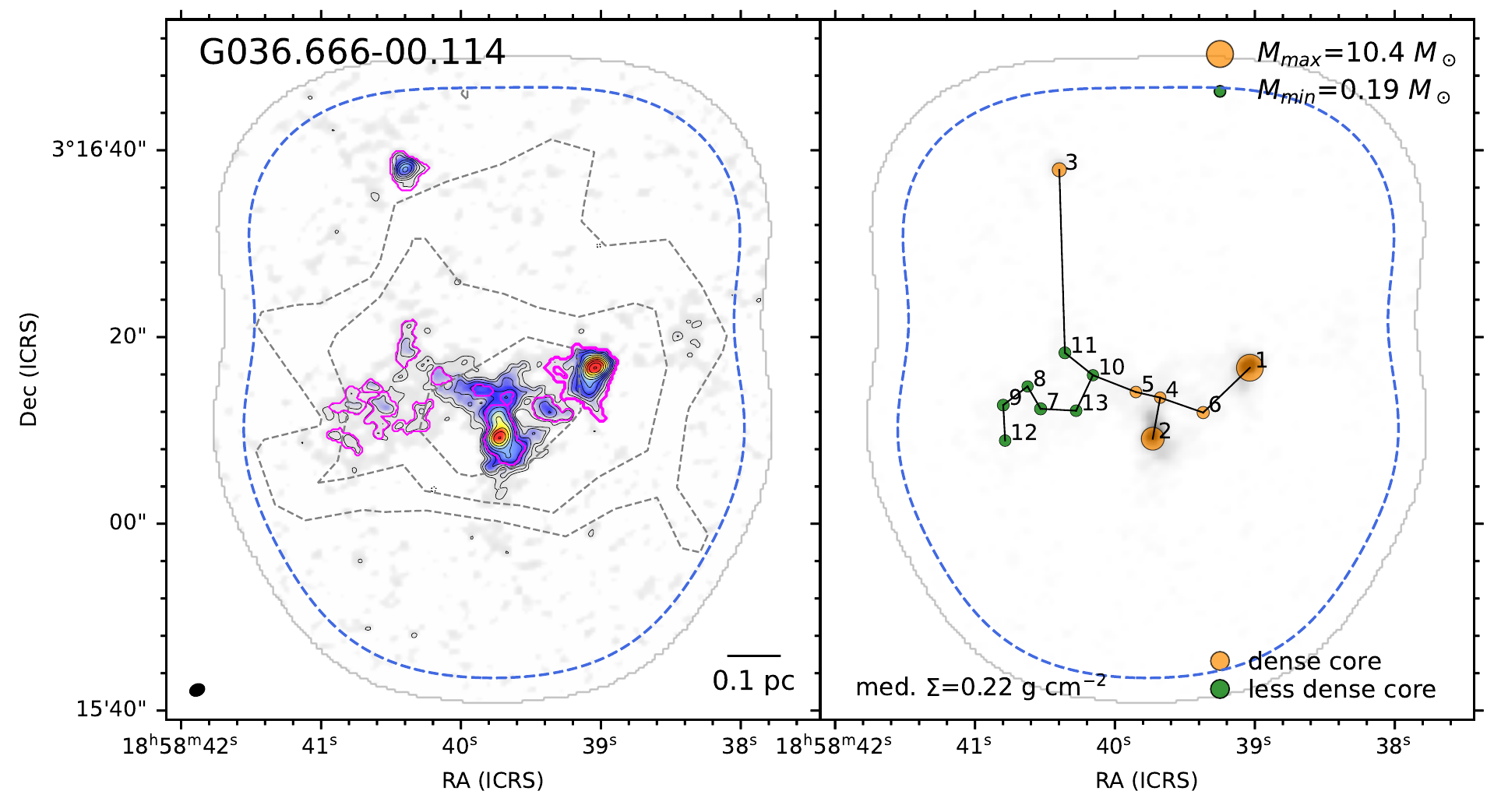}{0.75\textwidth}{}}\vspace{-3em}
    \gridline{\fig{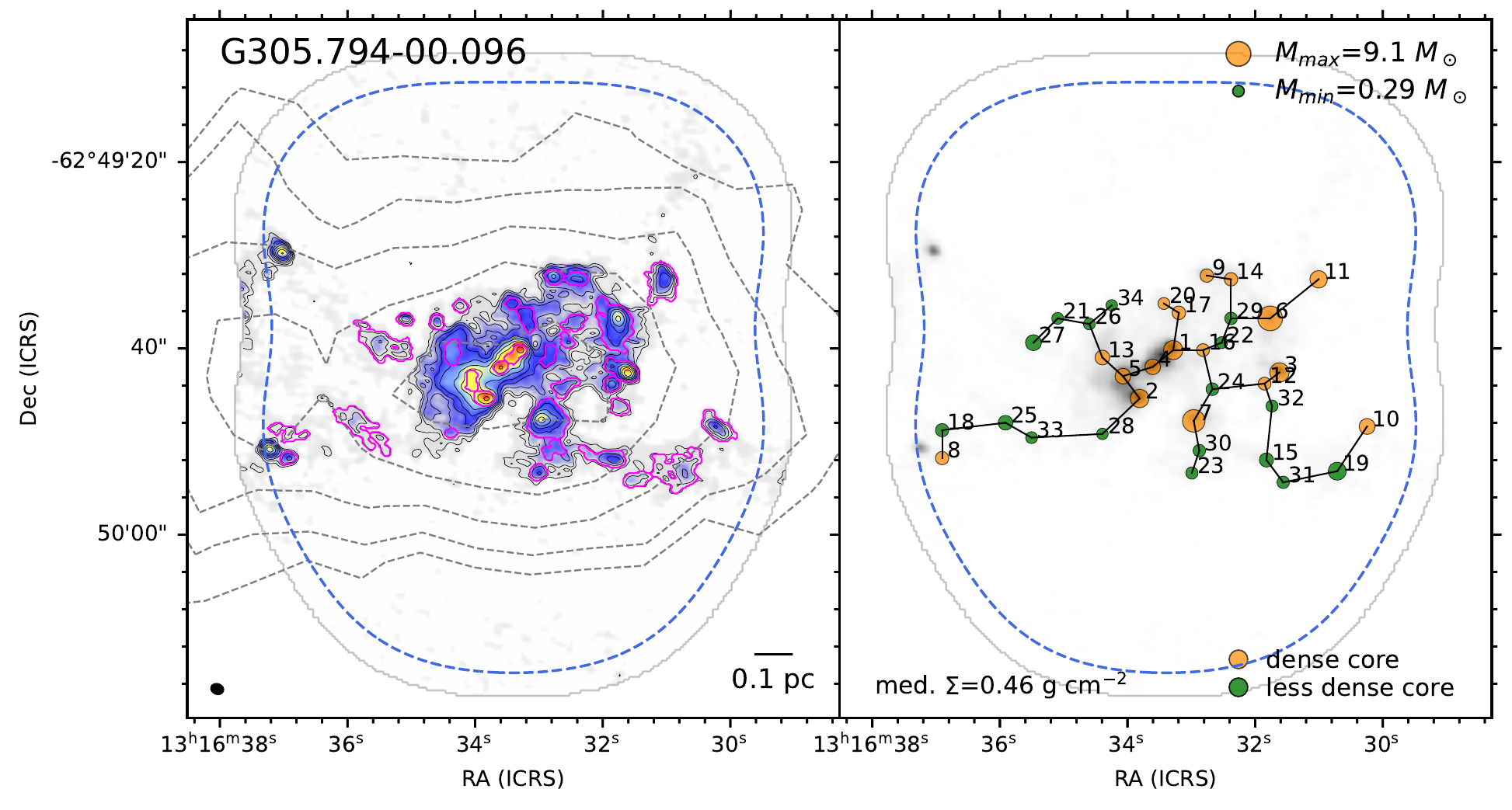}{0.75\textwidth}{}}\vspace{-3em}
    \caption{Same as Figure~\ref{fig:ashes_cont_1}.}
    \vspace{-7pt}
\end{figure*}

\begin{figure*}
    \gridline{\fig{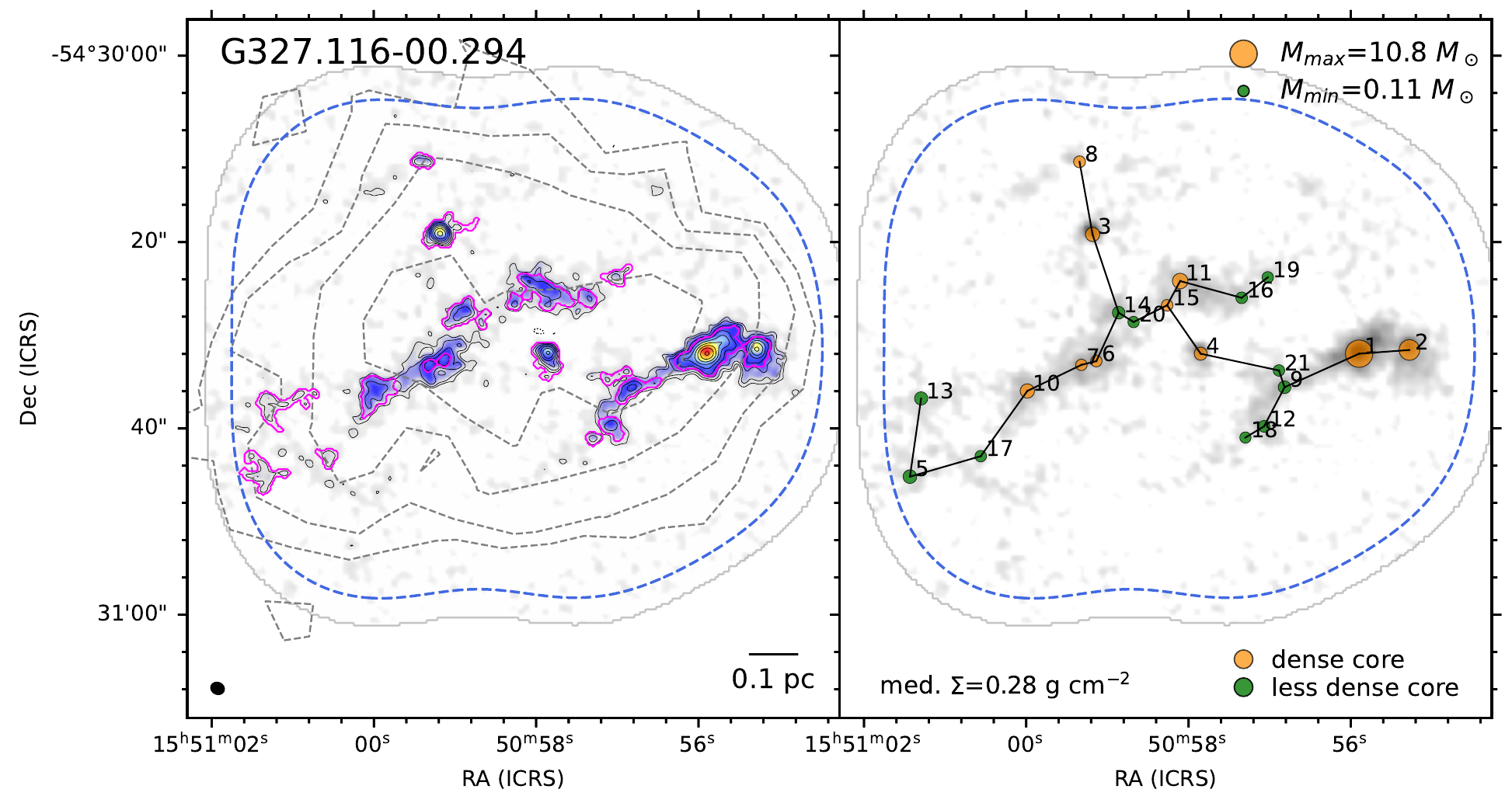}{0.75\textwidth}{}}\vspace{-3em}
    \gridline{\fig{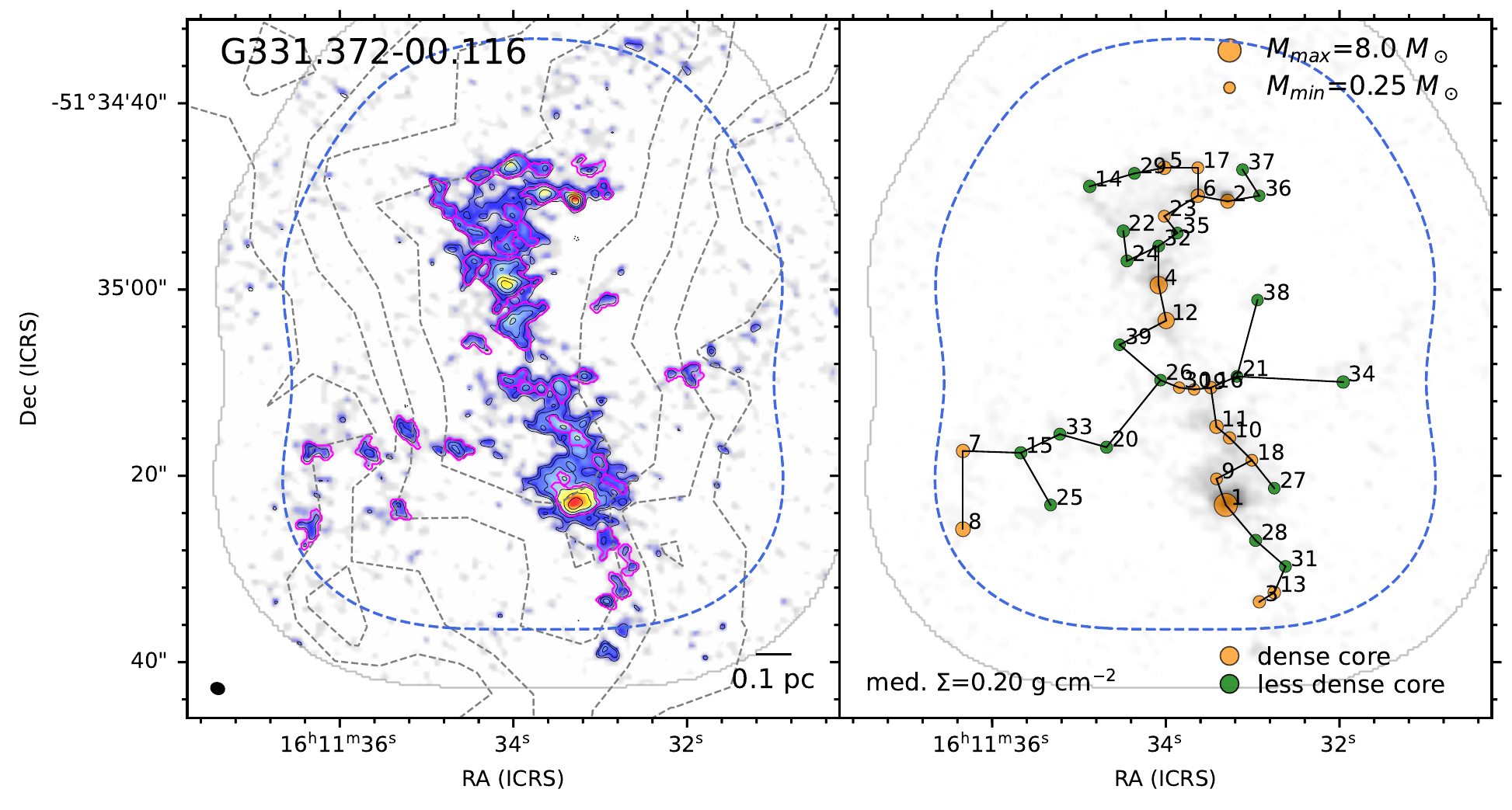}{0.75\textwidth}{}}\vspace{-3em}
    \gridline{\fig{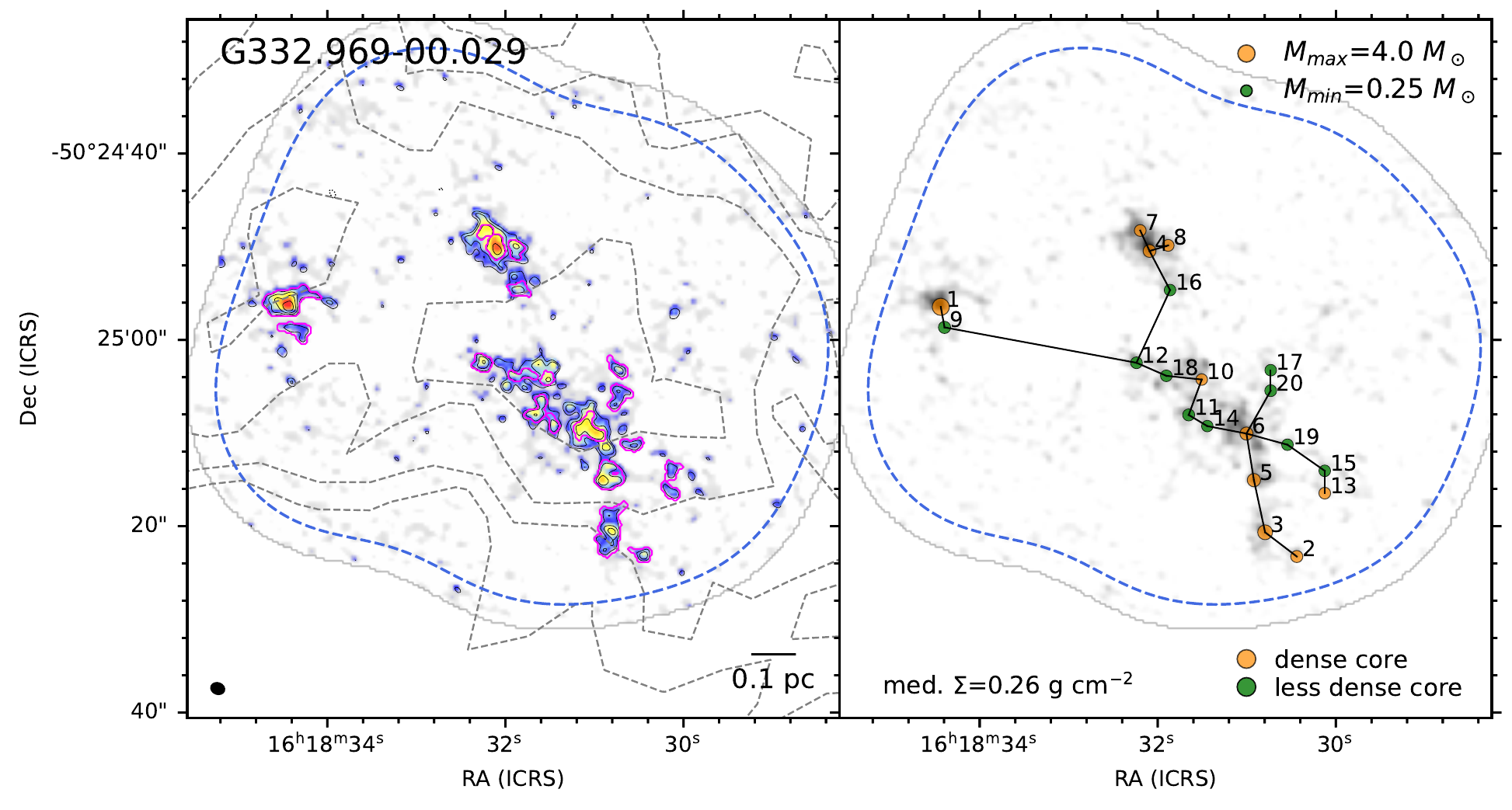}{0.75\textwidth}{}}\vspace{-3em}
    \caption{Same as Figure~\ref{fig:ashes_cont_1}.}
    \vspace{-5pt}
\end{figure*}

\begin{figure*}
    \gridline{\fig{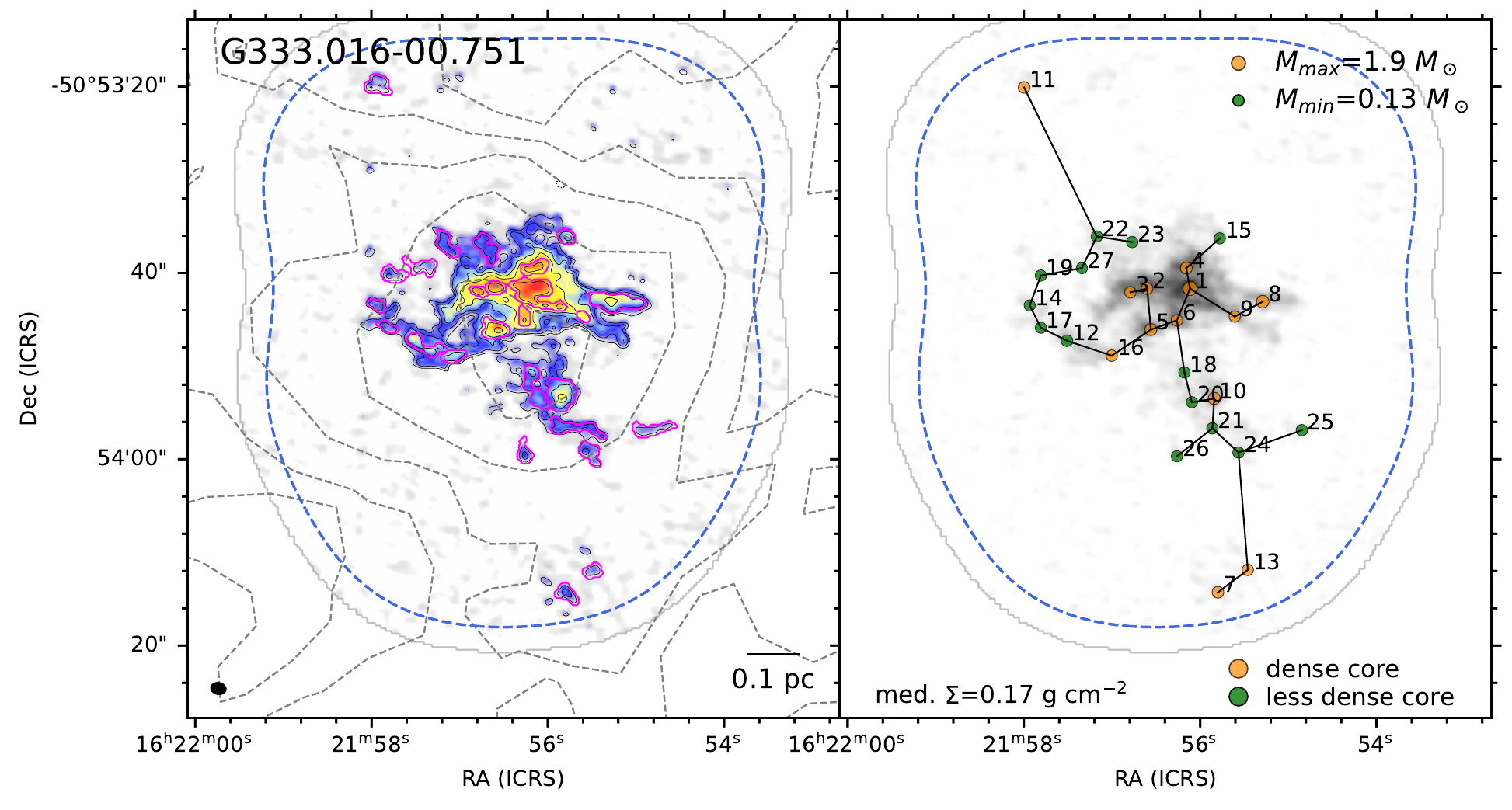}{0.75\textwidth}{}}\vspace{-3em}
    \gridline{\fig{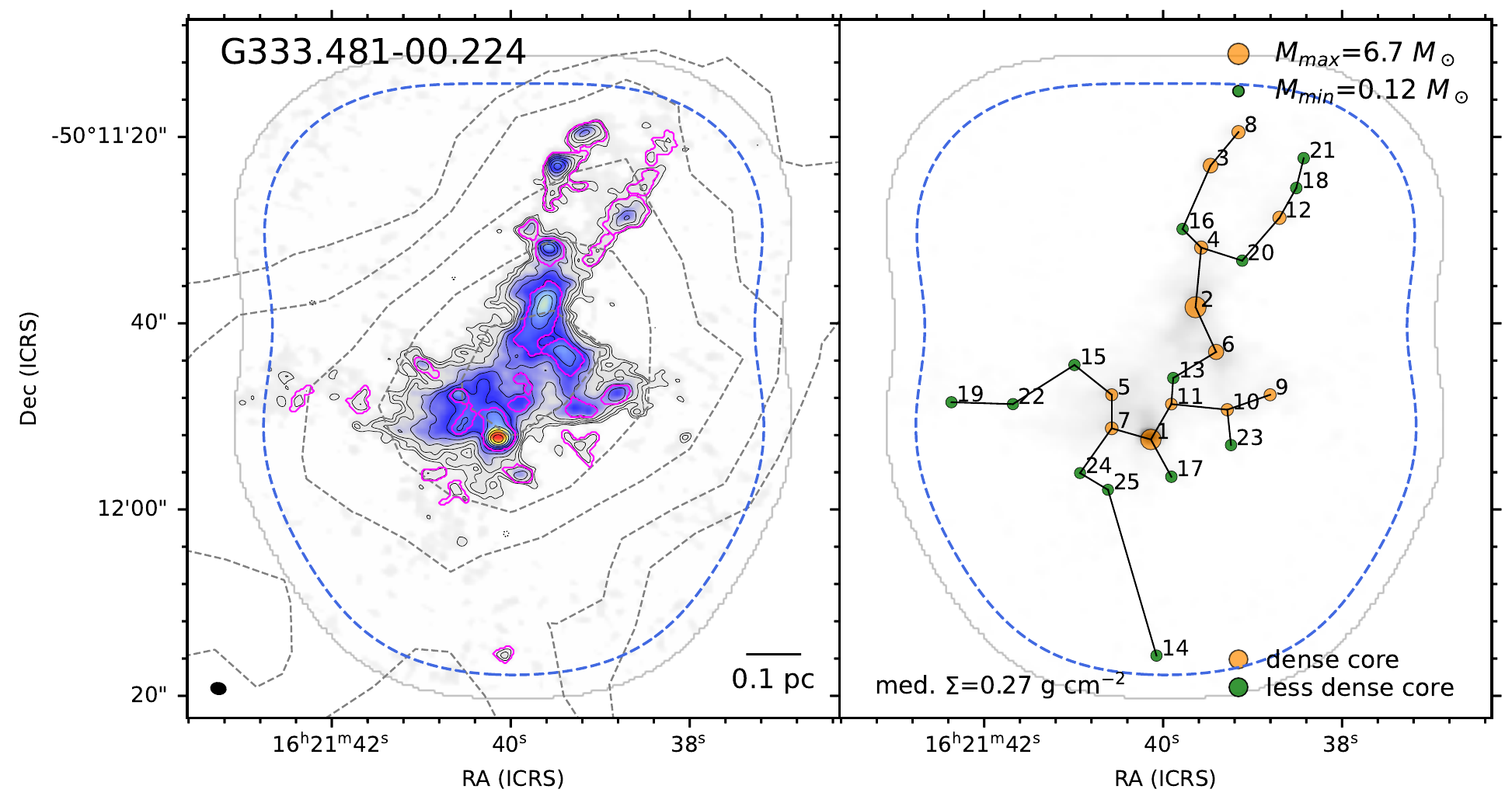}{0.75\textwidth}{}}\vspace{-3em}
    \gridline{\fig{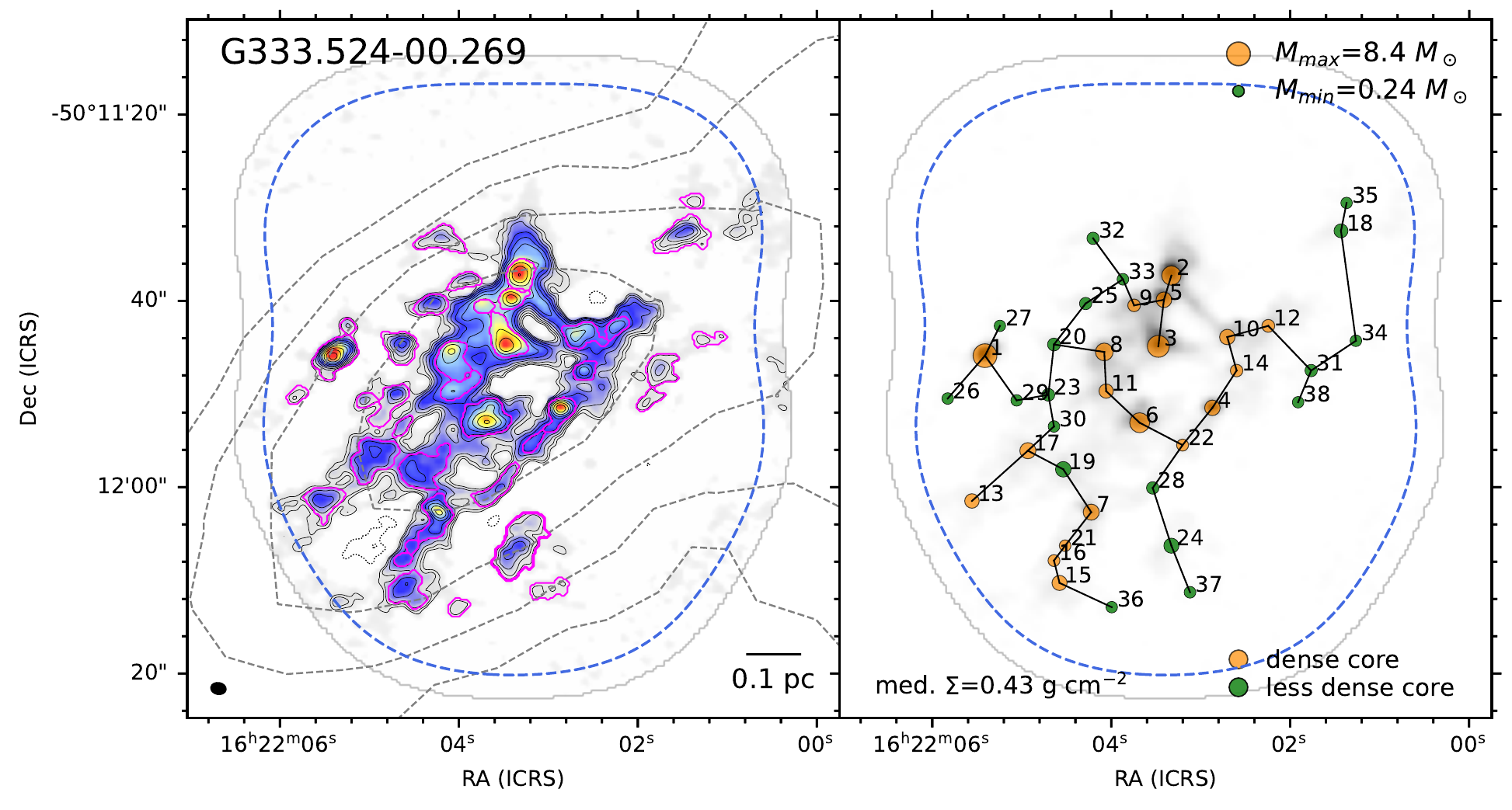}{0.75\textwidth}{}}\vspace{-3em}
    \caption{Same as Figure~\ref{fig:ashes_cont_1}.}
    \vspace{-5pt}
\end{figure*}

\begin{figure*}
    \gridline{\fig{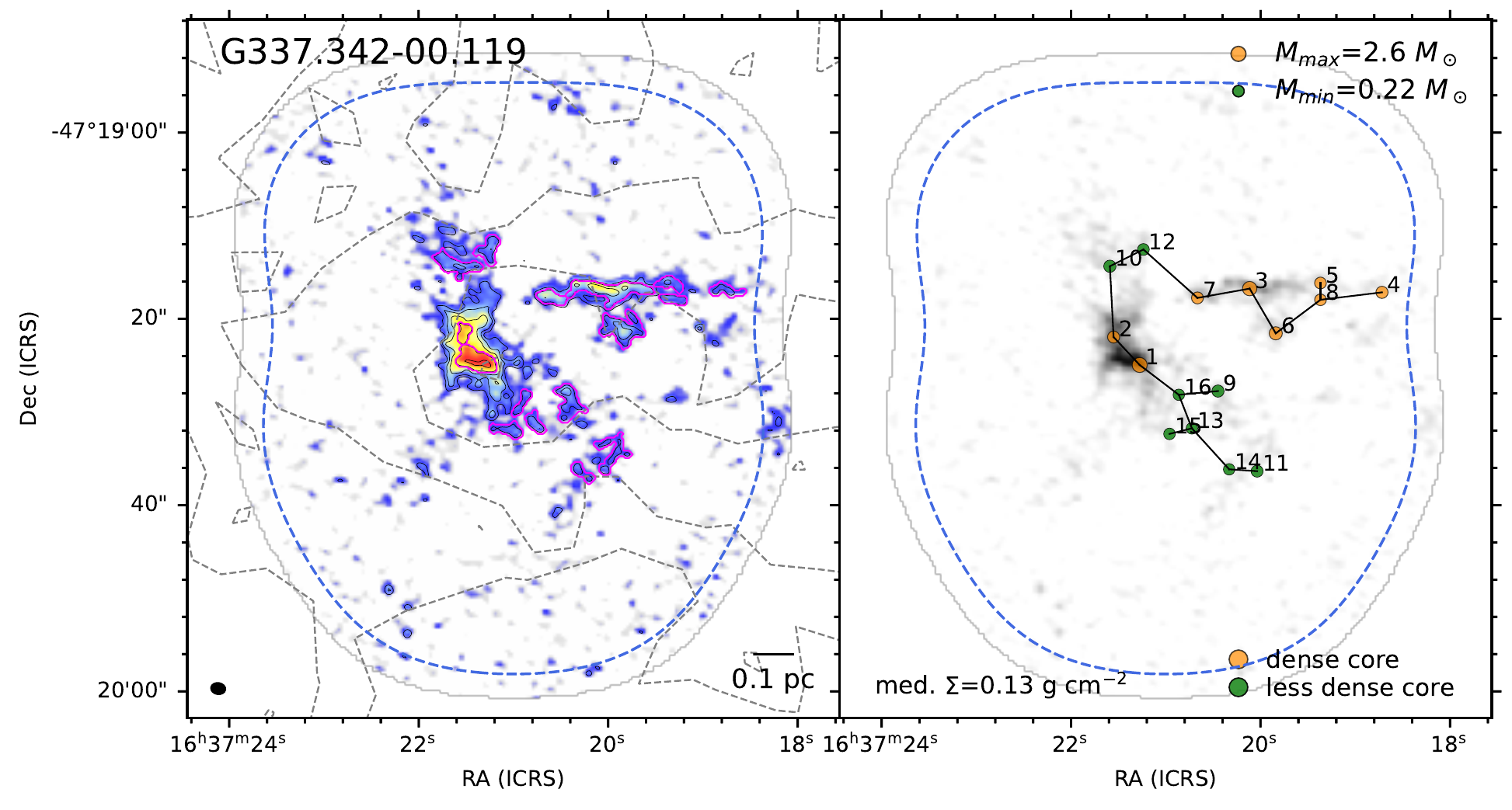}{0.75\textwidth}{}}\vspace{-3em}
    \gridline{\fig{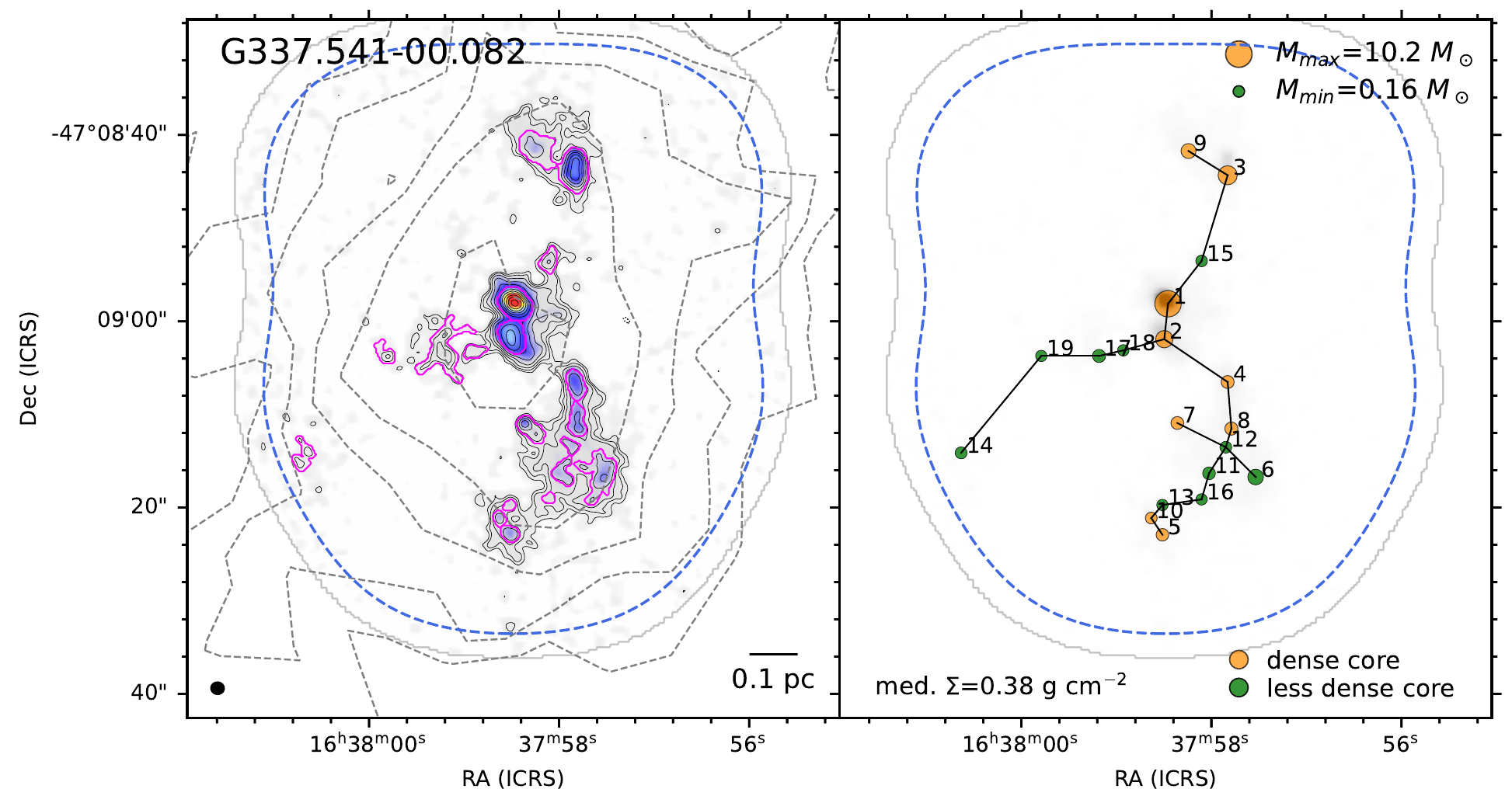}{0.75\textwidth}{}}\vspace{-3em}
    \gridline{\fig{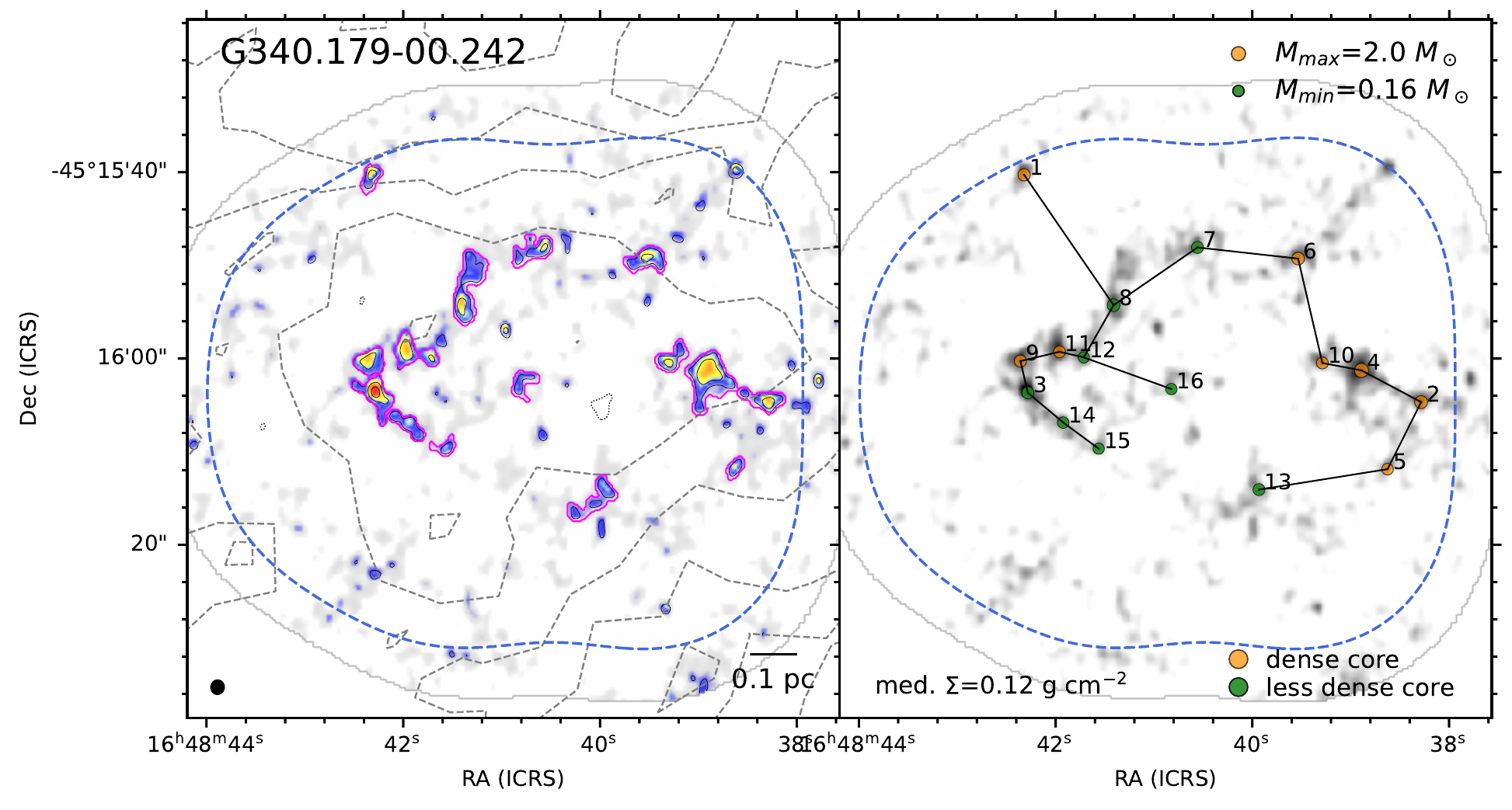}{0.75\textwidth}{}}\vspace{-3em}
    \caption{Same as Figure~\ref{fig:ashes_cont_1}.}
    \vspace{-5pt}
\end{figure*}

\begin{figure*}
    \gridline{\fig{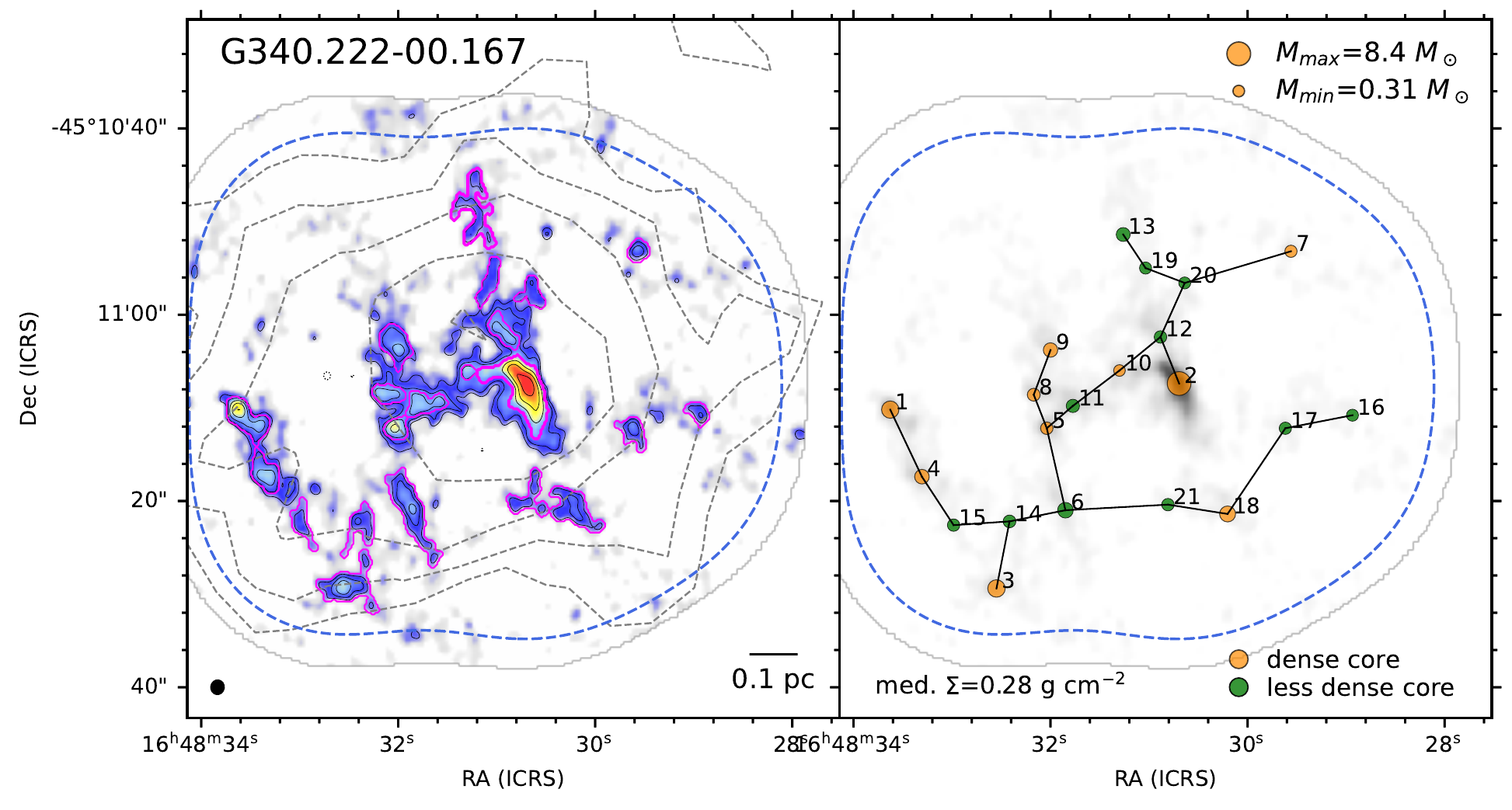}{0.75\textwidth}{}}\vspace{-3em}
    \gridline{\fig{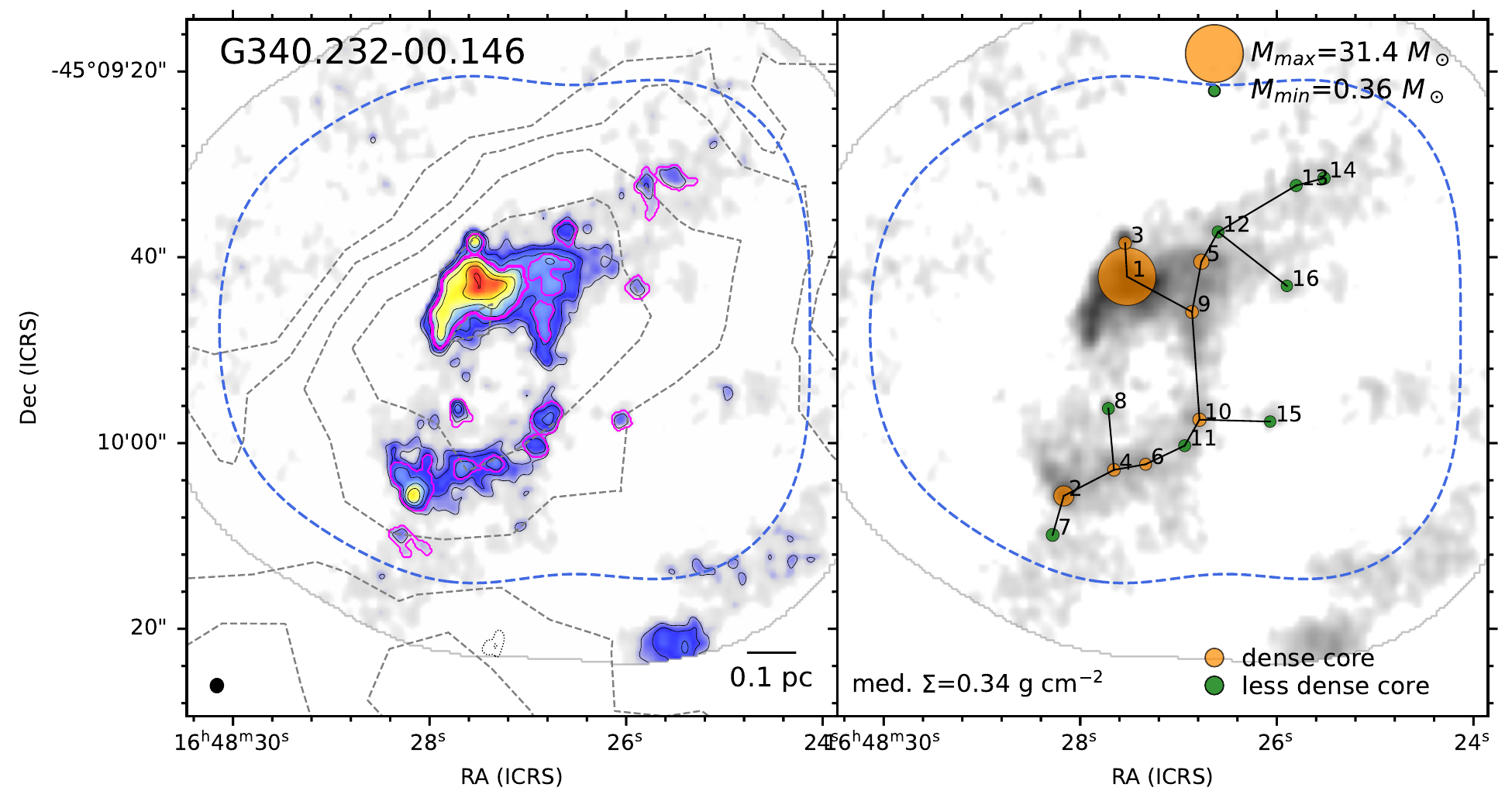}{0.75\textwidth}{}}\vspace{-3em}
    \gridline{\fig{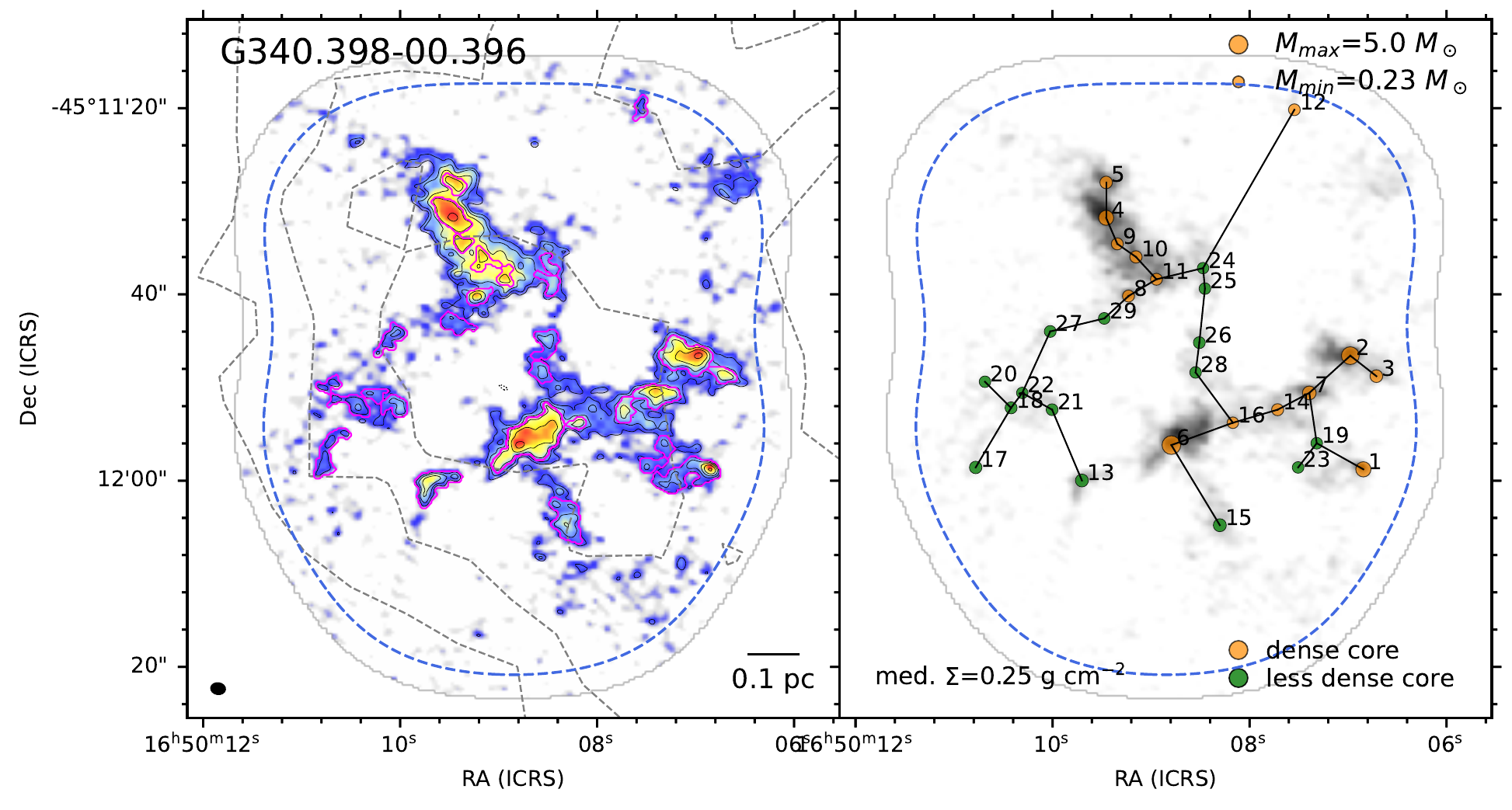}{0.75\textwidth}{}}\vspace{-3em}
    \caption{Same as Figure~\ref{fig:ashes_cont_1}.}
    \vspace{-5pt}
\end{figure*}

\begin{figure*}
    \gridline{\fig{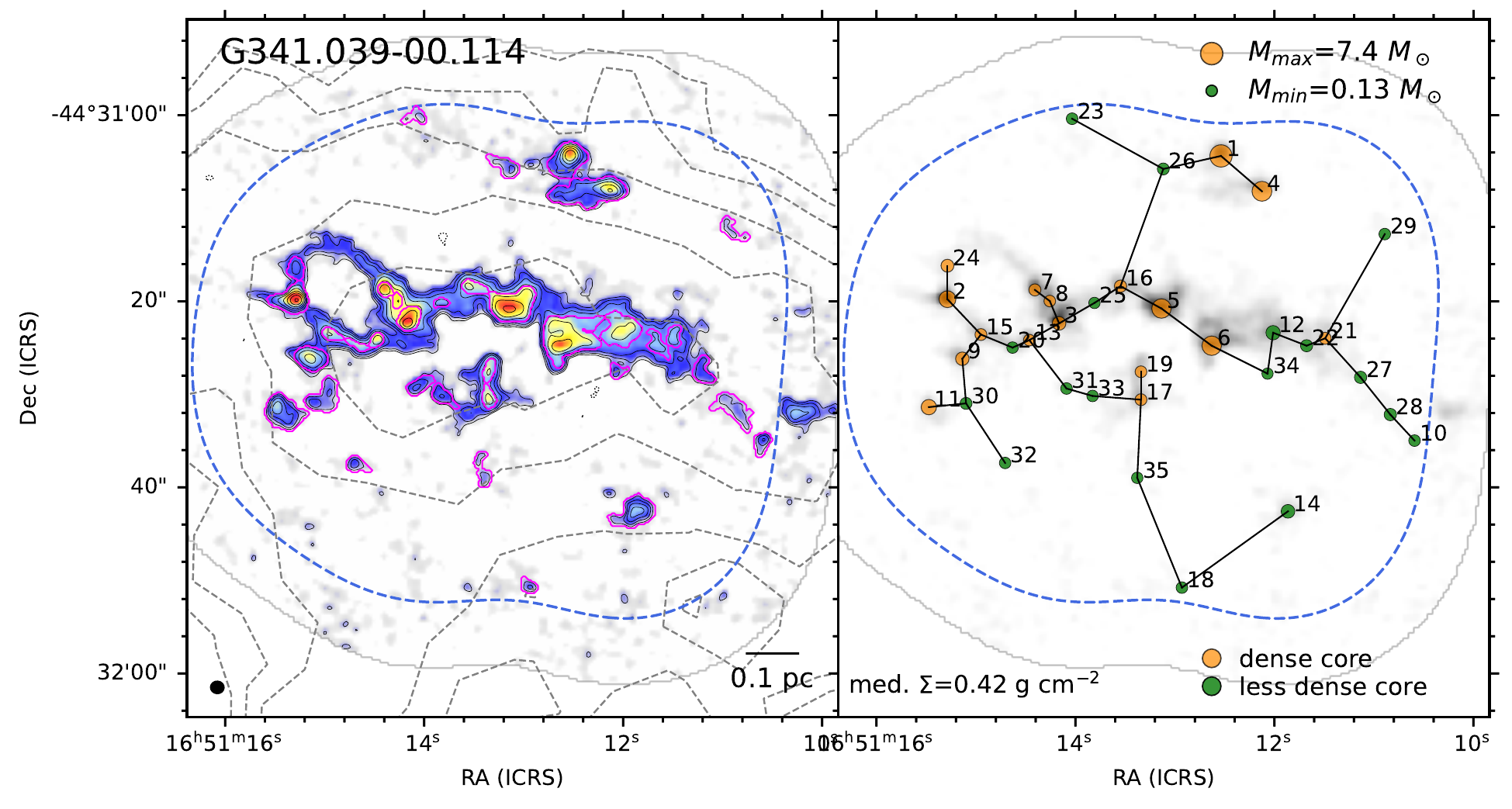}{0.75\textwidth}{}}\vspace{-3em}
    \gridline{\fig{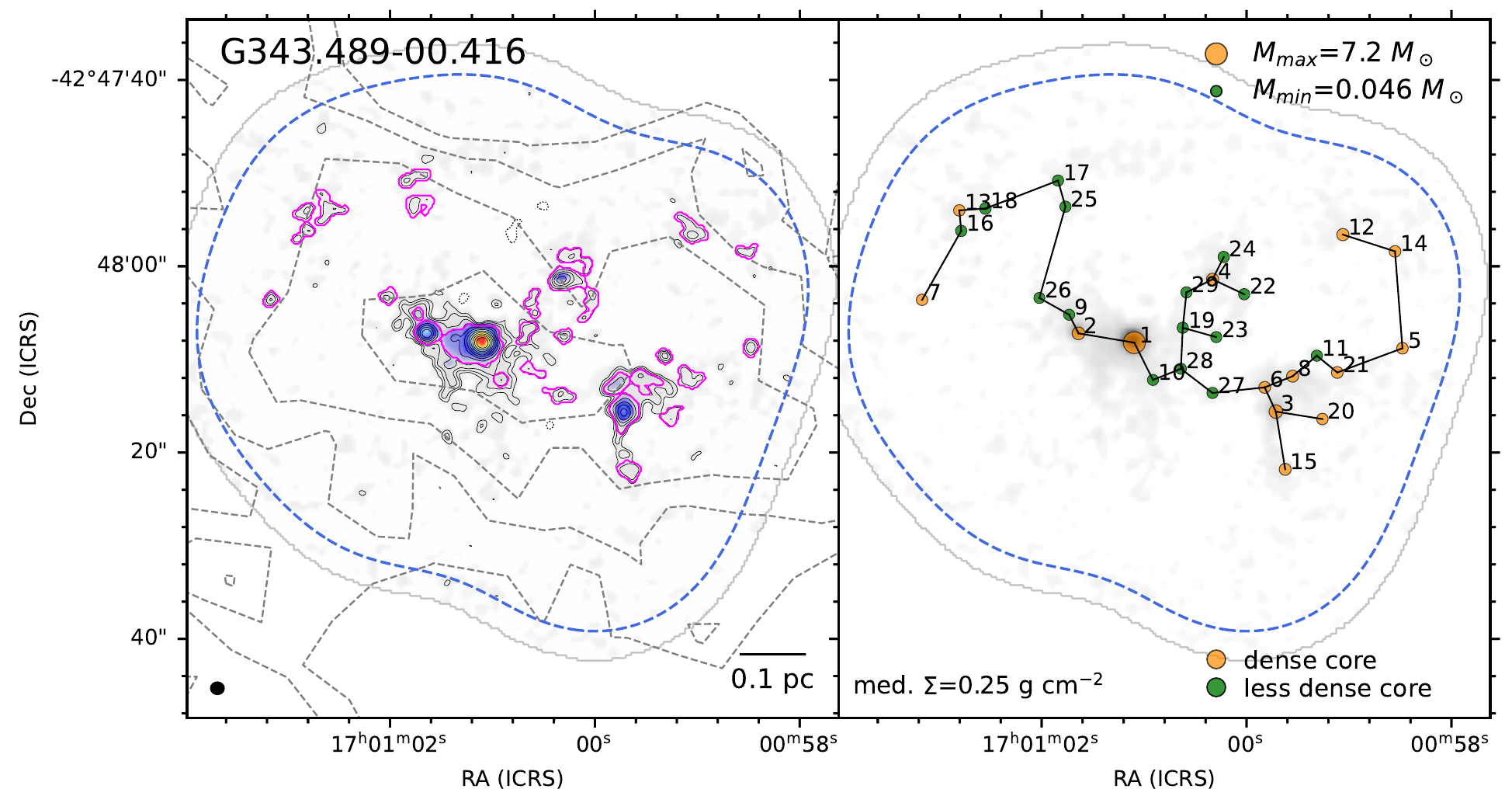}{0.75\textwidth}{}}\vspace{-3em}
    \gridline{\fig{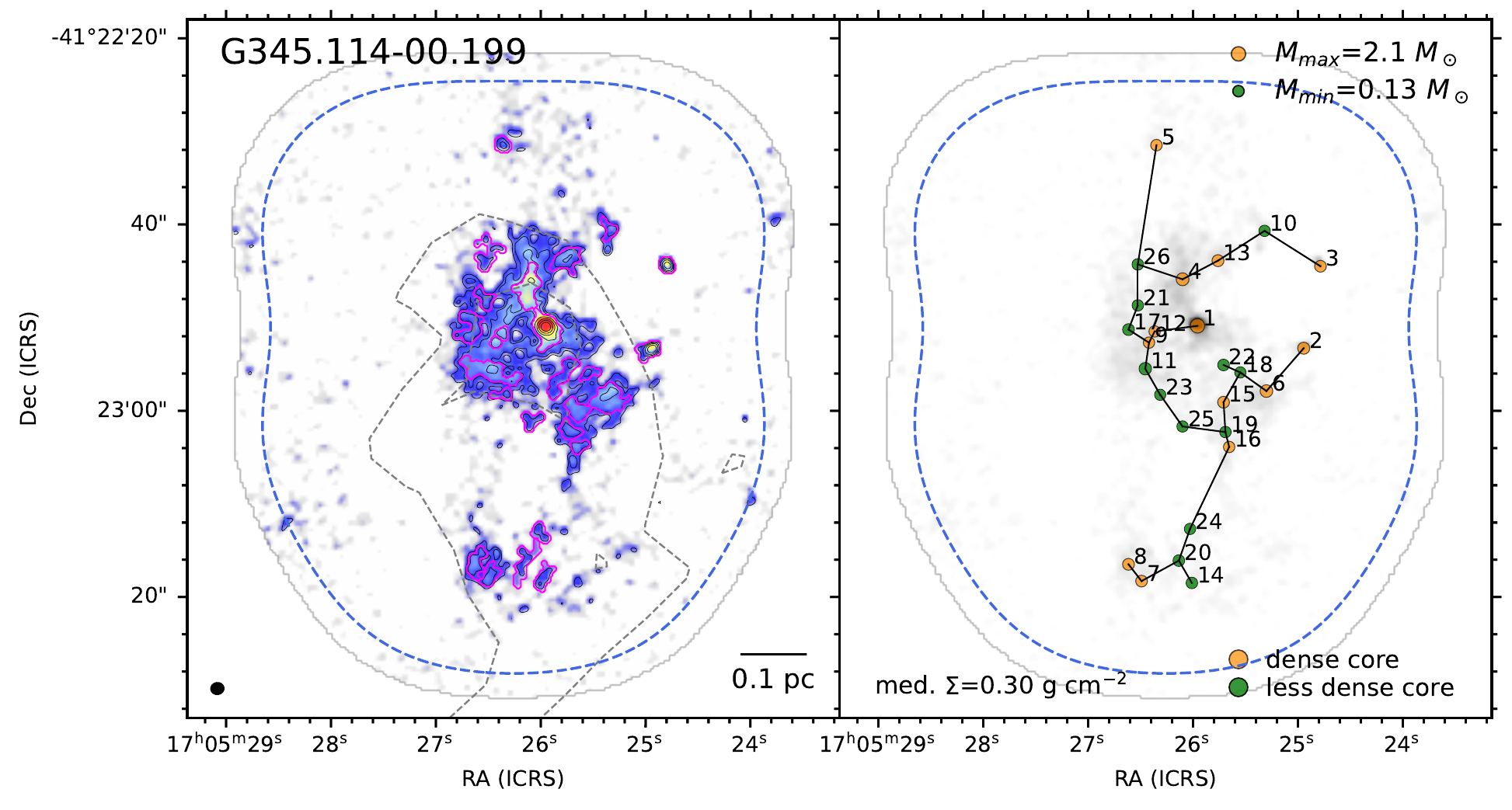}{0.75\textwidth}{}}\vspace{-3em}
    \caption{Same as Figure~\ref{fig:ashes_cont_1}.}
    \vspace{-5pt}
    \label{fig:Appendix_cont_last}
\end{figure*}


\begin{figure}
    \centering
    \includegraphics[width=15cm]{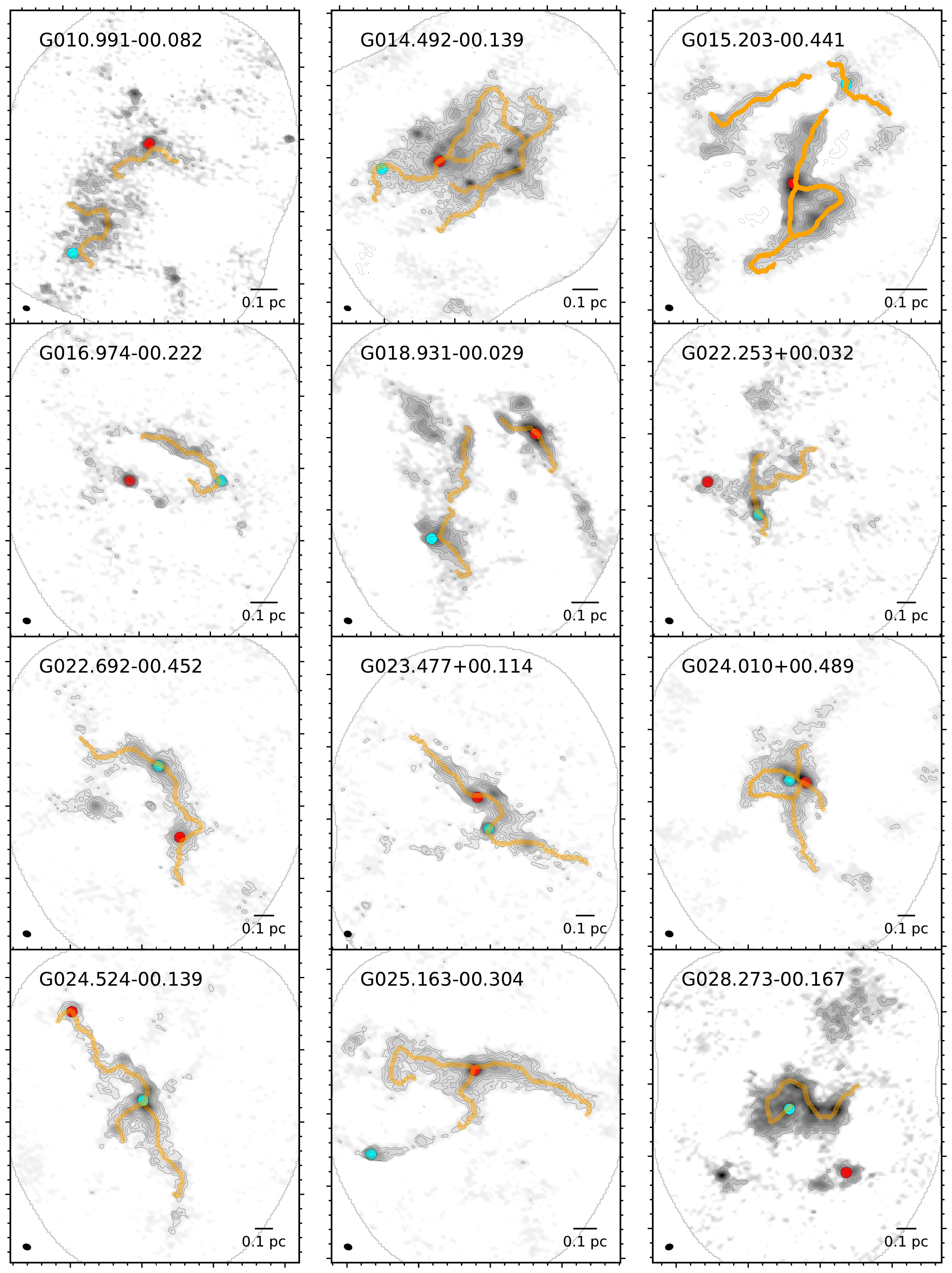}
    \caption{Same as Figure~\ref{fig:fil_G24} but presenting first 12 clumps.}
    \label{fig:Appendix_fil_1}
\end{figure}
\begin{figure}
    \centering
    \includegraphics[width=15cm]{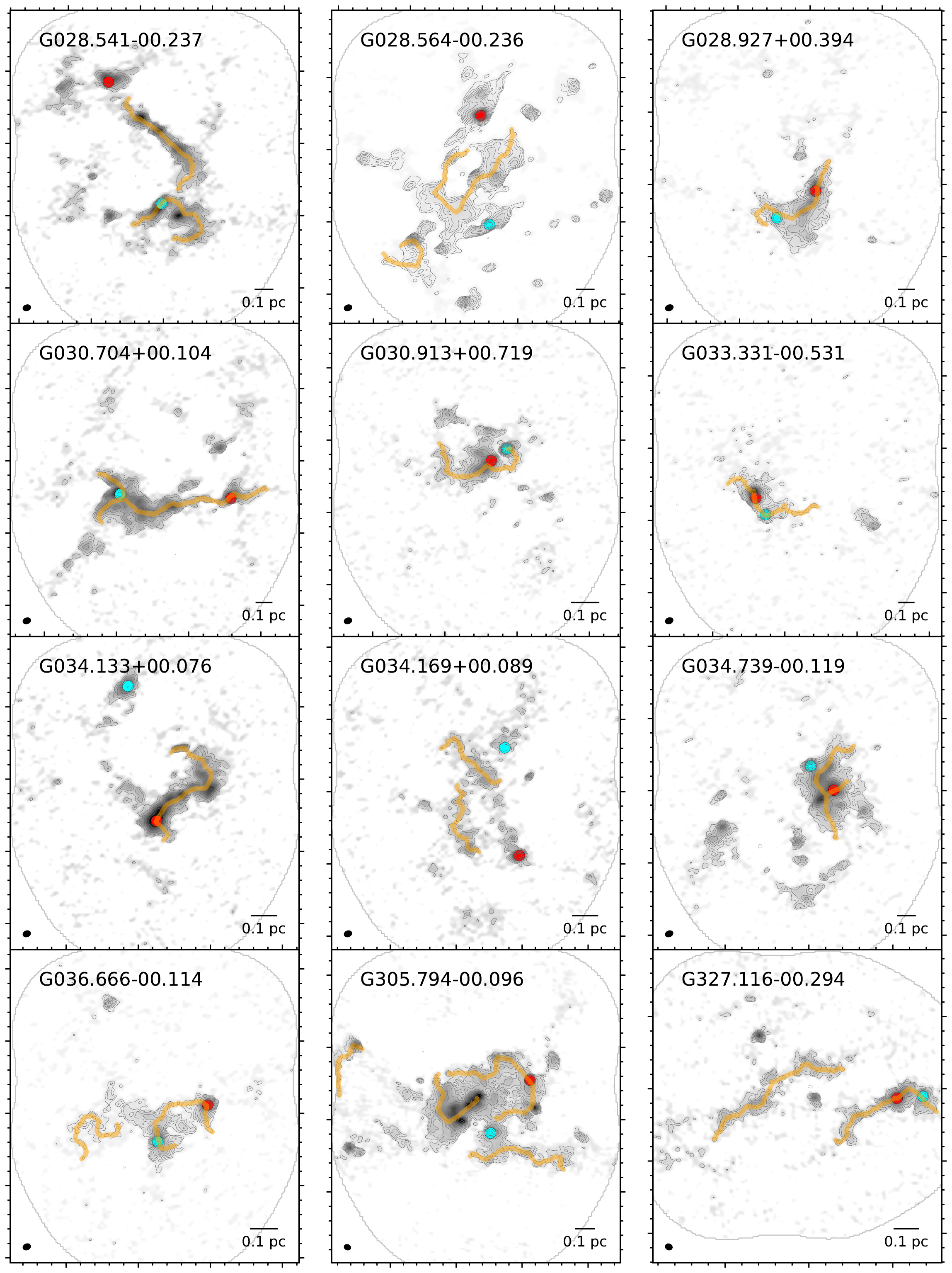}
    \caption{Same as Figure~\ref{fig:fil_G24} but presenting following 12 clumps.}
\end{figure}

\begin{figure}
    \centering
    \includegraphics[width=15cm]{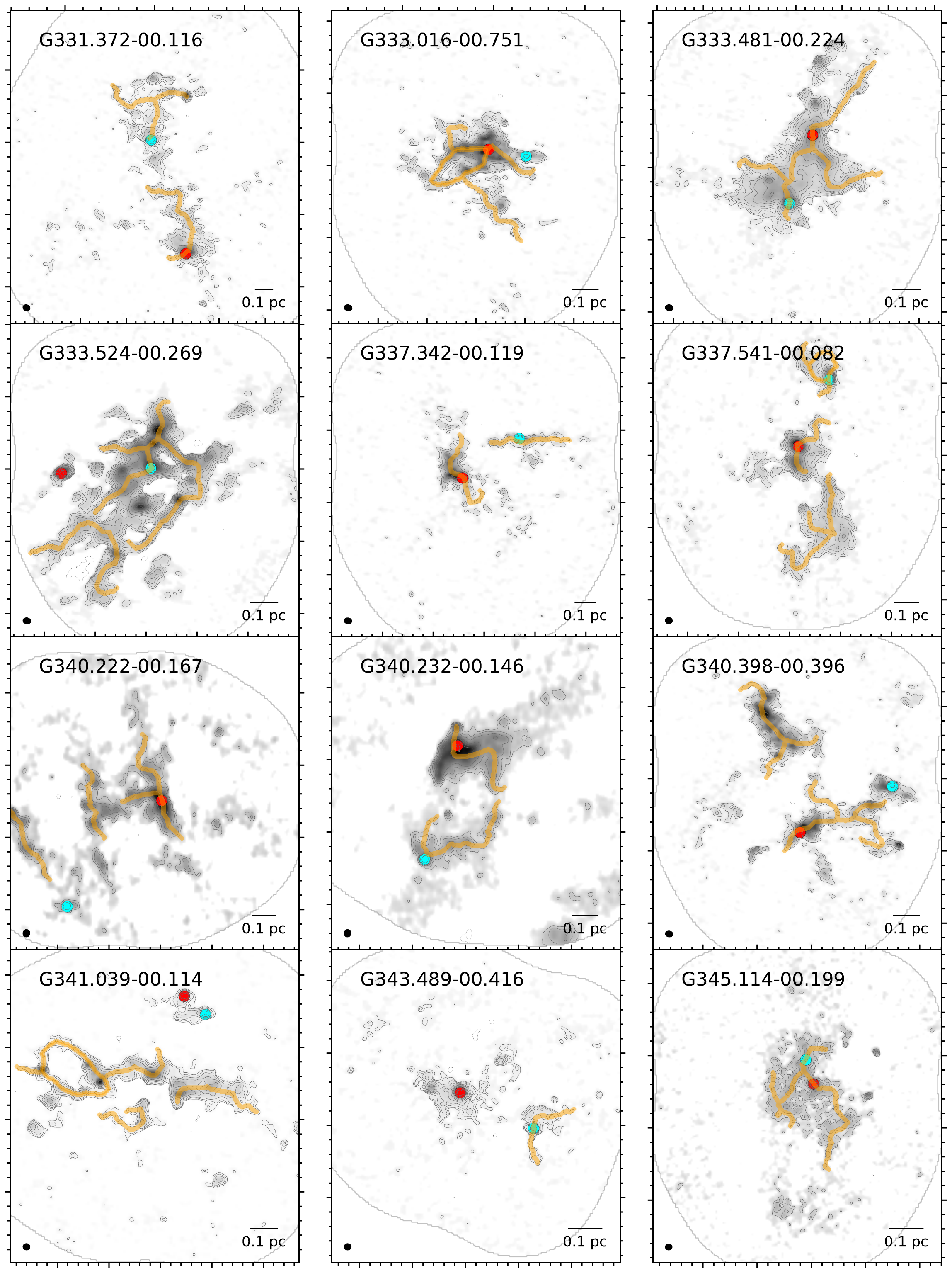}
    \caption{Same as Figure~\ref{fig:fil_G24} but presenting the remaining 11 clumps.}
    \label{fig:Appendix_fil_last}
\end{figure}

\end{document}